\documentclass[12pt,a4paper,epsfig]{article}
\usepackage{amsfonts}
\usepackage{epsfig}
\usepackage{mathtools,slashed}
\usepackage{color}

\definecolor{azulon}{rgb}{0.,0.,0.8}
\definecolor{verde}{rgb}{0.1,0.5,0.1}
\definecolor{rojo}{rgb}{0.5,0.1,0.1}

\newcommand \arcsinh{\mathop{\rm arcsinh}\nolimits}
\newcommand \sech{\mathop{\rm sech}\nolimits}

\title{{\bf The Jackiw-Rebbi model.\\  Bose/Fermi one-loop Kink mass shifts. }}

\author{ J. Mateos Guilarte
\\  {\normalsize {\it IUFFyM , Instituto de Fisica Fundamental y Matematicas}}\\{\normalsize {\it
Universidad de Salamanca, SPAIN}}
\date{}
}
\begin{document}
\maketitle
\section*{ABSTRACT}
Scalar and spinor Kink fluctuations are summarized in the Jackiw-Rebbi model. Canonical quantization is conventionally achieved by promoting the coefficients of the scalar and spinor field expansions to bosonic/fermionic creation/annihilation operators. One-particle states for bosons, respectively for fermions, are supplied by the scalar eigenfunctions of the Schr$\ddot{\rm o}$dinger, respectively, the eigenspinors of the Dirac Hamiltonian, evaluated either on the ground state or the Kink background. One-loop kink mass shifts are computed induced either by bosons or fermions. Both bound and scattering states contribute when the quanta are trapped by or scattered off the Kink. Zero modes of Kink fluctuations bosonic and fermionic do not modify the Kink mass at one-loop order. The Dashed-Hasslacher-Neveu prescription, derived in the canonical quantization framework, is generalized to the fermionic realm. Alternatively, quantization à la Feynman of the Jackiw-Rebbi model in the spaces of Kink  scalar and spinor fluctuations is developed. It is ensured that Berezin integration over the Grassmann coefficients of the Kink  spinor fluctuations leads to identical scenario as that revealed by including Fermi statistics in the DHN paradigm, at least with respect to the computation of the Kink Casimir energy.

\vspace{1cm}

\tableofcontents

\section{Introduction}

The Jackiw-Rebbi model is a relativistic Quantum Field Theory system in $(1+1)$-dimensional Minkowski space-time. The dynamics is governed by one action encompassing scalar and spinor fields with a quartic scalar self-interaction and a Yukawa scalar spinor interaction. Despite of being a simple Field Theory it is far from being only of academic interest in two completely different branches of Science. Firstly, in Condensed Matter Physics, the Jackiw-Rebbi model offered a theoretical framework
to explain  subtle phenomena arising in some linear chains of long macromolecules as polyacetilen and/or graphene nanotubes, see \cite{Jackiw}-\cite{Schrieffer}-\cite{Niemi1986}-\cite{JakubskyJPA14}. The main feature due to  the self-interaction of the scalar field is the existence of topological kinks, or, topological defects in the linear chain. The Yukawa scalar/spinor interaction couples the topological kink to spinor fluctuations in a way analyzed in these papers.
A new unexpected phenomenon emerged with far reaching consequences: the fractionization of the Fermi number. From the physical side quasi-particls carrying a fraction of the electron charge were predicted and experimentally found, breaking a principle experimentally 
supported from Millikan's era.

Secondly, in Mathematical Physics the paper of Jackiw and Rebbi built a bridge between QFT and sophisticated modern areas in high mathematics as index theorems, characteristic classes, etcetera, see also \cite{Fresneda} where the connection with index theorems, specifically the eta invariant, was explored. These findings aroused interest in studying in physical settings Fermions in presence of topological defects as kinks, see e.g. \cite{Vachaspati}-\cite{Bazeia}-\cite{Graham}. Of course, this idea is fruitful also for studying the interplay of fermions and higher dimensional topological defects like Abrikosov-Nielsen-Olesen vortices and Yang-Mills-Higgs magnetic monopoles, see \cite{Niemi1986}, \cite{Rubakov} and \cite{Rubakov1}.

In this paper we shall focus in the analysis of the mutual influence between Bosons, Fermions and Kinks in the Jackiw-Rebbi model. At the classical level we deal with scalar and spinor fluctuations that, after quantization, become respectively Bosonic and Fermionic quanta. Kinks, however, are solitary waves in the classical domain and remain to be so in the quantum realm at least 
if the couplings are weak. Our main goal here is to study how bosons and fermions affect kinks in the semiclassical expansion up to one-loop order, see e.g. \cite{Coleman}. Restriction to only scalar fluctuations has been  a topic widely developed in the Literature. The foundational paper is the Reference \cite{DHN}. We also mention the review \cite{FK} as an early important contribution to this theme. Attempts to include in the analysis evaluation of the one-loop  kink mass shifts due to fermionic fluctuations started later, at the end of the nineties, but were centered in supersymmetric kink systems. In the series of works \cite{Rebhan}-\cite{Shifman}-\cite{Graham}-\cite{Wimmer} a very delicate issue was addressed and beaten to death: does the classical Bogomolny bound of supersymmetric kinks receive any quantum correction ?  The answer is no if the supersymmetric field theory exhibits ${\cal N}=2$
extended supersymmetry and the \lq\lq quantum\rq\rq central charge of the SUSY algebra is identical to its classical counterpart.

Our playground, however, is the Jackiw-Rebbi model which do not ehibits supersymmetry and our main goal is to study how the fermions 
affect the kink in the semiclassical approximation. This endeavour requires a previous detailed study of the fermionic kink fluctuations that we shall develop following previous works in References \cite{Vachaspati}-\cite{Bazeia} in similar models encompassing fermions and kinks. We shall take profit of previous computations of one-loop kink mass shifts in two-dimensional 
relativistic field theoretical systems containing only scalar fields/bose quanta. The procedure will follow the lines explored in References \cite{AIJMG}
and \cite{GAGTS} where the Dashen-Hasslacher-Neveu paradigm has been applied to several models with only one and/or two scalar fields.
 The main new ingredient is that Fermi statistics adds to the kink classical mass one small amount of energy  at one-loop order. The efect behaves in the opposite sense as compared to bosonic one-loop kink mass shifts that are negative. Another difference lies in the fact that in the spectrum of the Jackiw-Rebbi model there are Fermions and anrti-Fermions but no bosonic anti-particles because the scalar field is real. Thus, the number of unoccupied states in the Fermionic Fock space is \lq\lq twice\rq\rq the identical number in the Bosonic Fock space. 

Fortunately, the spectrum of the Dirac Hamiltonian of the Jackiw-Rebbi model in the kink background may be identified as the square root of the spectrum of a diagonal matrix Schr$\ddot{\rm o}$dinger operator with consecutive transparent P$\ddot{\rm o}$sch-Teller potentials. Thus the DHN procedure is easy to apply but also, dealing with transparent PT potentials, the Cahill-Comtet-Glauber formula, see \cite{CCG}, also works. Of course, having in mind Fermi statistics and doubling. In fact J. Casahorran computed the fermionic one-loop kink mass shifts in the SUSY extensions of the sine-Gordon and $\phi^4_2$ models using this strategy in Reference \cite{Casa}.

Feynman quantization procedure provides one alernative way to derive the Fermionic DHN and CCG formulas for computing one-loop kink mass shifts. The Feynman functional integral definibg Fermion correlation functions is based on Berezin integrals over Grassmann variables. We shall see that this approach confirms the results obtained by the more transitaded canonical quantization path.

The organizarion of the paper is as follows: in Section \S.2 we describe the classical Jackiw-Rebbi model, summarize the canonical quantization of the system and identify the particle spectrum. Section \S.3 
is devoted to the presentation of the kink fluctuations both bosonic and fermionic. In this task we follow essentially the analysis
developed in \cite{Guilarte}. In Section \S.4 the novelties in this work are exhibited. After reconsidering the semi-classical expansion, both in the vacuum and kink sectors, the one-loop bosonic kink mass shift is briefly reviewed and the back-reaction of the kink to the impact of fermions at one-loop level
is unveiled. The tool for achieving this computation is the fermionic versions of the Cahill-Comtet-Glauber and Dashen-Hasslacher-Neveu 
procedures allowing the incorporation of fermions in this machinery. Section \S.5 is devoted to develop the quantization of the Jackiw-Rebbi model à la Feynman. Fermions in kink backgrounds are described in terms of a functional integral which is a Berezin integral over Grassmann variables. The subsequent effective action computed at semiclassical level includes the part of fermionic one-loop kink mass shifts which may be termed as fermionic Kink Casimir energy. No mass renormalization of ultraviolet divergences is discussed in this Feynman-Berezin scheme. Finally an outlook on future research is offered in Section \S.6 .

\section{The Jackiw-Rebbi model in $\mathbb{R}^{1,1}$ Minkowski space-time}

Let us consider a QFT of relativistic fermions and bosons restricted to move on a line. The \lq\lq classical\rq\rq dynamics is governed by the action, (\ref{JRact}): 
\begin{eqnarray}
S_{JR}[\phi, \Psi]&=& \int_{\mathbb{R}^{1,1}} \, d^2x \, \left\{ \frac{1}{2}\partial_\mu \phi\partial^\mu \phi - \frac{\lambda^2}{2}(\phi^2-1)^2 + i \bar{\Psi}\gamma^\mu \partial_\mu\Psi -g \bar{\Psi} \phi \Psi \right\} \label{JRact} \\ \bar{\Psi}&=& \Psi^\dagger \gamma^0 \quad , \quad [\phi]=1 \, \, \, , \, \, \, [\Psi]=L^{-1/2} \, \, \, , \, \, [\lambda]=L^{-1}=[g] \nonumber
\end{eqnarray}
encompassing a quartic self-interaction of the bosons plus a Yukawa coupling between fermions and bosons. In the natural system of units where the Planck constant and the speed of light in vacuum are set to one, $\hbar=c=1$, the dimensions of the fields and couplings are shown below the action above.
The Jackiw-Rebbi Hamiltonian $H = H_{FB}+H_{B}$, (\ref{JR}), is in turn obtained via a Legendre transformation :
\begin{eqnarray}
 H_{FB} &=&  \int\, dx \, \Psi^\dagger(t,x)\left\{-i \alpha \frac{\partial}{\partial x} \right\} \Psi(t,x) +\int \, dx \, \Psi^\dagger(t,x)\left\{ g \phi(t,x) \beta\right\}\Psi(t,x)  \nonumber \\ H_B&=&\frac{1}{2} \int \, dx \, \left\{\Pi^2(x)+ \left(\frac{\partial\phi}{\partial x}\right)^2+\lambda^2\left(\phi^2(t,x)-1\right)^2 \right\} \label{JR}
\end{eqnarray}
The Dirac matrices $\alpha=\sigma_2$ and $\beta=\sigma_1$ are chosen like in \cite{Jackiw}. Here, $\sigma_1$ and $\sigma_2$ are Pauli matrices
and this choice corresponds to the Clifford algebra
\begin{eqnarray*}
&& \gamma^0=\sigma_1 =\beta\, \, , \, \, \gamma^1=i \sigma_3 \quad , \quad \gamma^5=\gamma^0\gamma^1=\-i \sigma_2=-i \alpha \\
&&  \{\gamma^\mu,\gamma^\nu\}= 2 g^{\mu\nu}\quad , \quad g^{\mu\nu}={\rm diag}(1,-1) \, \, , \, \, \mu, \nu=0,1
\end{eqnarray*}

The Klein-Gordon and Dirac fields are maps from the Minkowski space respectively to the field of the reals and the fundamental irreducible representation of the ${\bf Spin}(1,1;\mathbb{R})$ group:
\[
\phi(t,x) = \,\, \mathbb{R}^{1,1} \, \, \longrightarrow \, \, \mathbb{R} \, \, , \, \, \, \, \Psi(t,x)=\left(\begin{array}{c} \psi_1(t,x) \\  \psi_2(t,x)  \end{array}\right): \, \, \mathbb{R}^{1,1} \, \, \longrightarrow \, \, {\it irr}_P {\bf Spin}(1,1; \mathbb{R}) \nonumber \, ,
\]
i.e., the transformations generated by $\frac{1}{4}[\gamma^0,\gamma^1]$, and characterized by the Lorentz boost parameter $\chi$, acts on one spinor in the form
\[
 S_L[\chi]=e^{\frac{\chi}{4}[\gamma^0,\gamma^1]}=\left(\begin{array}{cc} \cosh \frac{\chi}{2} & i \sinh \frac{\chi}{2} \\ -i \sinh \frac{\chi}{2} & \cosh \frac{\chi}{2}\end{array}\right) \,  , \, \, \cosh \chi=\frac{1}{\sqrt{1-v^2}} \, \, , \, \,   \Psi_L(t,x)=S_L[\chi]\Psi(t,x) \nonumber
\]

The Euler-Lagrange (classical) field equations read:
\begin{eqnarray}
&& \Box \phi+ 2 \lambda^2 \phi(t,x)(\phi^2(t,x)-1)+g \bar{\Psi}(t,x)\Psi(t,x)=0 \label{hfieldeq}\\
&& i \gamma^\mu \frac{\partial \Psi}{\partial x^\mu}- g \phi(t,x)\Psi(t,x)=0 \label{ffieldeq} \, \, .
\end{eqnarray}
This system of coupled non-linear PDE's is very difficult to solve. The situation is better -and pertinent to the posterior canonical quantization of the system- if some static solution of the system of the form $(\phi_S(x), \Psi_S=0)$ is discovered:
\[
-\frac{d^2 \phi_S}{d x^2}+2 \lambda^2 \phi_S(x)(\phi_S^2(x)-1)=0
\]
In that case, one may search for solutions close to these static solutions
which linearize the (\ref{hfieldeq}-\ref{ffieldeq}) PDE system: 
 \begin{eqnarray}
 &&  \phi(t,x)=\phi_S(x)+\eta(t,x) \, \, \Rightarrow \, \, \Box\eta+2\lambda^2(3\phi_S^2(x)-1)\eta(t,x)={\cal O}(\eta^2) \label{hfieldeql}\\
 && i \gamma^\mu \frac{\partial\Psi}{\partial x^\mu}-g \phi_S(x)\Psi(t,x)={\cal O}(\eta\Psi)\label{ffieldeql}
 \end{eqnarray}
Since the terms in (\ref{hfieldeql}-\ref{ffieldeql}) with no derivatives of the fields are time inedependent it is convenient to solve the linear system via Fourier transform in time:
\begin{equation}
\eta(t,x)=\int_{-\infty}^\infty \frac{d \omega_B}{2\pi} \, e^{i \omega_B t}\eta(x;\omega_B) \quad , \quad \Psi(t,x)=\int_{-\infty}^\infty \, \frac{d\omega_F}{2\pi} e^{i \omega_F t} \Psi(x;\omega_F) \nonumber
\end{equation}
The linear PDE system (\ref{hfieldeql}-\ref{ffieldeql}) becomes equivalent to the spectral problem (\ref{scham}-\ref{dham}):
\begin{eqnarray}
&& h_{\rm Sch}\eta(x;\omega_B)=\Big[ -\frac{d^2}{d x^2}+2 \lambda^2(3 \phi_S^2(x)-1)\Big]\eta(x;\omega_B)=\omega_B^2 \eta(x;\omega_B) 
\label{scham} \\
&& h_{\rm D}\Psi(x;\omega_F)=\Big[ -i \alpha \frac{d}{d x}+g \beta \phi_S(x)\Big]\Psi(x;\omega_F)=\omega_F \Psi(x;\omega_F) \label{dham}
\end{eqnarray}
where $h_{\rm Sch}$ and $h_{\rm D}$ are respectively the quantum mechanical Schr$\ddot{\rm o}$dinger and Dirac operators in the background created by the $\phi_s$ classical solution.

The standard canonical quantization procedure is based on finding the eigenwave functions of the Schr$\ddot{\rm o}$dinger operator and the eigenspinors of the Dirac operator. The space of stationary states of the \lq\lq free\rq\rq $H\equiv (h_{\rm Sch},h_{\rm D})$ are then taken as the one-particle states and quantum transitions within the Fock space states can be evaluated by considering the interaction Hamiltonians as perturbations. 

\subsection{From scalar/spinor ground state fluctuations to Bose/Fermi quanta}

The simplest solutions of the classical field equations are the two homogeneous, independent of $t$ and $x$, minima of the scalar potential energy whereas $H_{BF}$ is minimized by $\Psi_V=0$:
\begin{equation}
\phi_S(t,x)^\pm=\phi_V^\pm =\pm 1 \quad ,  \quad \Psi_S(x)=\Psi_V=0 \nonumber
\end{equation} 
Choice of one of these two configurations, e.g. $(\phi_S(t,x)^+=+1, \Psi_S(t,x)=0)$, as the ground state,  spontaneously breaks the  $\phi \to -\phi$ symmetry and the linear Klein-Gordon and Dirac equations become:
\begin{eqnarray}
 \frac{\partial^2 \phi}{\partial t^2}&=&\frac{\partial^2 \phi}{\partial x^2}- 4 \lambda^2 \phi(t,x) \label{eqKG} \\
 \left(\begin{array}{cc} i \frac{\partial}{\partial t} & 0 \\ 0 & i \frac{\partial}{\partial t}\end{array}\right) \cdot\left(\begin{array}{c} \psi_1(t,x) \\ \psi_2(t,x)\end{array}\right)&=& \left(\begin{array}{cc} 0 &-\frac{\partial}{\partial x}+g \\ \frac{\partial}{\partial x}+g & 0 \end{array}\right)\cdot\left(\begin{array}{c} \psi_1(t,x) \\ \psi_2(t,x)\end{array}\right) \label{eqDir}.
\end{eqnarray} 
Because there are no time and space dependent terms in the PDE operators in the linear equations (\ref{eqKG}) and (\ref{eqDir}) it is convenient to search for the general solution relying on  Fourier transform integrals, see (\ref{scawavpac}-\ref{spinwavpac})-\ref{spinwavpacp}){\footnote{We denote back $\eta(t,x)$ as $\phi(t,x)$ to follow the conventional notation.}}
\begin{eqnarray}
\phi(t,x)&=&  \int_{\mathbb{H}^+} \frac{d k}{\sqrt{2\pi}\sqrt{2 \omega_B(k)}} \left[a(k)e^{-i\, \omega_B(k)\, t+i k x} + a^*(k) e^{i \, \omega_B(k)t-i k x}\right] \label{scawavpac}\\
\Psi(t,x)&=&\sqrt{g}\int_{\mathbb{H}^+} \frac{d k}{\sqrt{2\pi}\sqrt{2\omega_F(k)}} \left[b(k) u(k)e^{-i \omega_F(k) t+i k x} + c^*(k) v(k) e^{i \omega_F(k)t-i k x}\right] \label{spinwavpac}\\ \Psi^\dagger(t,x)&=&\sqrt{g}\int_{\mathbb{H}^+} \frac{d k}{\sqrt{2 \pi} \sqrt{2 \omega_F(k)}} \left[b^*(k) u^\dagger(k)e^{i \omega_F(k) t-i k x} + c(k) v^\dagger(k) e^{-i \omega_F(k)t+i k x}\right] \label{spinwavpacp}
\end{eqnarray}
where the integration is performed over the upper branches $\mathbb{H}_B^+$ and $\mathbb{H}_F^+$ of the hyperbolas: $\omega_B^2=k^2+4\lambda^2, \omega_B=+\sqrt{k^2+4 \lambda^2}$, and $\omega_F^2=k^2+g^2, \omega_F=+\sqrt{g^2+k^2}$.

In order to be the spinor expansion (\ref{spinwavpac}) the general solution of (\ref{eqDir}), $u(k)$ and $v(k)$ must be respectively the eigenspinors of the $2\times 2$-matrices with eigenvalues $\omega_F$ {\footnote{Note however that the spectral equation for $u$ is the same as the spectral equation for $v$ if $(\omega_F,k)$ is replaced by $(-\omega_F,-k)$. Notice also that it is possible to come back to $(\omega_F;k)$ provided that $g$ will be transmutted to $-g$, the essential property of antimatter.}}:
\begin{eqnarray*}
&& \left( \begin{array}{cc} 0 & g-i k \\ g+i k & 0 \end{array}\right)\cdot \left(\begin{array}{c} u_{1}(k) \\ u_{2}(k) \end{array}\right)= +\sqrt{k^2+g^2} \left(\begin{array}{c} u_{1}(k) \\ u_{2}(k) \end{array}\right) \\ && \left( \begin{array}{cc} 0 & -g-i k \\ - g+i k & 0\end{array}\right)\cdot \left(\begin{array}{c} v_{1}(k) \\ v_{2}(k) \end{array}\right)= + \sqrt{k^2+g^2} \left(\begin{array}{c} v_{1}(k) \\ v_{2}(k) \end{array}\right) \\ && u(k)= +\left(\frac{\sqrt{k^2+g^2}}{2 g}\right)^{1/2}\cdot\left( \begin{array}{c} 1 \\ \frac{g+i k}{\sqrt{k^2+g^2}}\end{array}\right) \quad , \quad  v(k)= \left(\frac{\sqrt{k^2+g^2}}{2 g}\right)^{1/2}\cdot\left( \begin{array}{c} \frac{- g-i k}{\sqrt{k^2+g^2}} \\ 1 \end{array}\right)
\end{eqnarray*}
which satisfy the standard orthonormality conditions:
\[
u^\dagger(k)u(k)=\frac{\omega_F}{g}=v^\dagger(k)v(k) \, \, , \, \, \, \bar{u}(k)u(k)=1=-\bar{v}(k)v(k)
\] 
together with $u^\dagger(k)v(-k)= 0$. 

In the canonical quantization procedure the Fourier coefficients of the scalar field are promoted to creation and annihilation bosonic operators satisfying the commutationon rules:
\begin{equation}
[ \hat{a}(k_1),\hat{a}^\dagger(k_2)]=\delta(k_1-k_2) \quad , \quad [ \hat{a}(k_1),\hat{a}(k_2)]=0= [ \hat{a}^\dagger(k_1),\hat{a}^\dagger(k_2)]
\nonumber
\end{equation}
The ground state with no meson particles at all is annihilated by all the destruction bosonic operators
\begin{equation}
\hat{a}(k) \vert 0 \rangle_B =0 \, , \quad \forall k \nonumber
\end{equation}
Meson multiparticle states have the form (\ref{bfss}):
\begin{equation}
\prod_{j=1}^N\, [\hat{a}^\dagger(k_j)]^{n_j} \vert 0 \rangle_B= \vert n_1n_2 \cdots n_N \rangle \, \, , \quad n_j \in \mathbb{N} \, \, , \quad \sum_{j=1}^N n_j=N \label{bfss}
\end{equation}
and span the basis of the Bosonic Fock space, obtained via symmetric tensor product. 

Simili modo, the canonical quantization of the Dirac field courses via anticommutation relations (\ref{facr1}-\ref{facr2})
\begin{equation}
\{\hat{b}^\dagger(k_1), \hat{b}(k_2) \}=\delta(k_1-k_2) \quad , \quad \{\hat{c}^\dagger(k_1), \hat{c}(k_2) \}=\delta(k_1-k_2) \label{facr1}
\end{equation}
\begin{equation}
\{\hat{b}(k_1), \hat{b}(k_2) \}=0= \{\hat{c}(k_1), \hat{c}(k_2) \}  \label{facr2}
\end{equation}
Likewise the fermionic ground state  (\ref{fgs}) with neither electrons nor positrons is annihilated by all the
fermionic destruction operators
\begin{equation}
\hat{b}(k) \vert 0 \rangle_F=0=\hat{c}(k) \vert 0 \rangle_F \, \, , \quad \forall k \label{fgs}
\end{equation}
Electron/positron{\footnote{We shall refer to the Fermi quanta in the Jackiw-Rebbi model as electrons/positrons by analogy with QED. In the JR system there is no electric charge. The N$\ddot{\rm o}$ther's invariant associated to the $\mathbb{U}(1)$ symmetry is the Fermi number. }} multiparticle states (\ref{melec}-\ref{pelec}) form the basis of the Fermionic Fock space built from the one-particle states via antisymmetric tensor product:
\begin{equation}
\prod_{j=1}^N\, [\hat{b}^\dagger(k_j)]^{n^-_j} \vert 0 \rangle_F= \vert n^-_1n^-_2 \cdots n^-_N \rangle \, \, , \quad n^-_j = 0 \,{\rm or} \, 1 \, \, , \quad \sum_{j=1}^N n^-_j=N \label{melec}
\end{equation}
\begin{equation}
\prod_{j=1}^N\, [\hat{c}^\dagger(k_j)]^{n^+_j} \vert 0 \rangle_F= \vert n^+_1n^+_2 \cdots n^+_N \rangle \, \, , \quad n^+_j = 0 \, {\rm or} \, 1 \, \, , \quad \sum_{j=1}^N n^+_j=N \label{pelec}
\end{equation}
Quantization by anticommutators forces the antisymmetry of the multiparticle states and thus one state can only be either unoccupied, $n_j^\pm=0$, or occupied by one Fermion, $n_j^\pm=1$. Note that $n^+_j=1$ and $n^-_j=1$ is simultaneously possible describing one state 
with one electron and one positron, both with momentum $k_j$. 

To finish this Section we compute the most important observable in the Fermionic Fock space, the quantum Dirac free Hamiltonian:
\begin{equation}
\hat{H}_D=\int_{-\infty}^\infty \, dx \, \hat{\Psi}^\dagger(x,t) (i\frac{\partial}{\partial t})\hat{\Psi}(x,t) \label{ffen} \, \, .
\end{equation}
Thus,
\begin{eqnarray*}
\hat{H}_D &=& g \int_{-\infty}^\infty dx \int_{-\infty}^\infty \int_{-\infty}^\infty \frac{dk^\prime dk}{2\pi} \cdot  \frac{\omega_F(k^\prime)}{2\sqrt{\omega_F(k^\prime)\omega_F(k)}}\times \\ && \times \left[ \hat{b}^\dagger(k^\prime)\hat{b}(k)u^\dagger(k^\prime)u(k)e^{i(\omega_F(k^\prime)-\omega_F(k))t}e^{-i(k^\prime-k)x}- \hat{c}(k^\prime)\hat{c}^\dagger(k)v^\dagger(k^\prime)v(k)e^{-i(\omega_F(k^\prime)-\omega_F(k))t}e^{i(k^\prime-k)x}\right] \, \, .
\end{eqnarray*}
There are also crossed terms of the form
\begin{eqnarray*}
&& g \int_{-\infty}^\infty dx \int_{-\infty}^\infty \int_{-\infty}^\infty \frac{dk^\prime dk}{2\pi} \cdot  \frac{\omega_F(k^\prime)}{2\sqrt{\omega_F(k^\prime)\omega_F(k)}}\times \\ && \times \left[ \hat{b}^\dagger(k^\prime)\hat{c}^\dagger(k)u^\dagger(k^\prime)v(k)e^{i(\omega_F(k^\prime)+\omega_F(k))t}e^{-i(k^\prime+k)x}- \hat{c}(k^\prime)\hat{b}(k)v^\dagger(k^\prime)u(k)e^{-i(\omega_F(k^\prime)+\omega_F(k))t}e^{i(k^\prime+k)x}\right] 
\end{eqnarray*}
but
\[
\int_{-\infty}^\infty \frac{ dx}{2\pi} e^{\pm i(k^\prime\mp k)x}=\delta(k^\prime \mp k) \, \, .
\]
In the crossed terms then arise the null factors:
\[
u^\dagger(k)v(-k)=0= v^\dagger(k)u(-k)
\]
The surviving terms prompts the answer
\begin{equation}
\hat{H}_D= \int_{-\infty}^\infty \, dk \, \sqrt{k^2+g^2} \left(\hat{b}^\dagger(k)\hat{b(k)}-\hat{c}(k)\hat{c}^\dagger(k)\right) \label{enfsf}
\end{equation}
The crucial point is that use of the anticommutation relations renders the free Dirac Hamiltonian to the form
\begin{equation}
\hat{H}_D= \int_{-\infty}^\infty \, dk \, \sqrt{k^2+g^2} \left(\hat{b}^\dagger(k)\hat{b(k)}+\hat{c}^\dagger(k)\hat{c}(k)-\lq\lq \delta"(0)\right)\label{enfsfno} \, \, .
\end{equation}
Therefore, the free Dirac Hamiltonian is the sum of a normal ordered fermionic operator plus one ultra singular negative energy suplied by the Fermionic vacuum $\vert 0\rangle_F$ (the Dirac sea) 
\[
\hat{b}(k) \vert 0 \rangle_F= \hat{c}(k) \vert 0 \rangle_F=0 \quad , \quad \forall k\in \mathbb{R} \, .
\]
Observe that the sign of the energy of the Fermionic vacuum, where all the Fermionic states are unoccupied, is the opposite of the energy of the bosonic vacuum, where all the bosonic states are unnocupied.

A more precise meaning of the distribution Dirac $\lq\lq \delta"(0)$ at its most dangerous point may be given in the spirit of zeta function regularization, see \cite{Kirsten}. In QFT it is common practice to normalize the system in a very large but finite inteval of length $L$, 1D box, 
and impose periodic boundary conditions on the spinor fields $\Psi(x,t)=\Psi(x+L,t)$. The allowed momenta become discretized and the anti-commutation relations are also modified:
\[
k_n=\frac{2\pi}{L}\cdot n \, \,  , \, \,  n\in \mathbb{Z} \quad , \quad \{\hat{b}^\dagger(k_{n_1}), \hat{b}(k_{n_2})\}=\delta_{n_1 n_2}
=\{\hat{c}^\dagger(k_{n_1}), \hat{c}(k_{n_2})\}
\]
where $\delta_{n_1 n_2}$ is now the Kronecker delta. Under these assumptions
\[
\int_{-\infty}^\infty \, dk \, \sqrt{k^2+g^2} \, \, \lq\lq \delta"(0)\, \, \longrightarrow \, \, \frac{ 4\pi^2}{L^2}\sum_{n\in \mathbb{Z}} \, \sqrt{n^2+g_L^2} \, \, \delta_{nn} \quad , \quad g_L=\frac{L}{2\pi} g
\]
if $1<< L <+\infty$. Recall that
\[
\zeta_{\cal E}(s,g_L^2\vert 1)= \sum_{n\in\mathbb{Z}}\, \frac{1}{(n^2+g_L^2)^s} \quad , \quad s\in \mathbb{C}
\]
 is one Epstein zeta meromorphic function. The series above is  only convergent for ${\rm Re}>1$. The Epstein function is then obtained by analytic continuation of the convergent series to the whole complex $s$-plane finding that the only singularities are isolated poles.  We regularize the expectation value of the Dirac Hamiltonian at the vacuum asigning to it the value of the zeta function at a regular point of $s$:
 \[
 \mu^{-(1+2s)}\cdot {}_F\langle 0 \vert \hat{H}_D^{-s} \vert 0 \rangle_F= - \mu^{-(1+2s)}\left(\frac{4\pi^2}{L^2}\right)^{-s}\cdot \sum_{n\in\mathbb{Z}} \, \frac{1}{(n^2+g_L^2)^s}= \mu^{-(1+2s)}\left(\frac{4\pi^2}{L^2}\right)^{-s}\cdot \zeta_{\cal E}(s,g_l^2\vert 1) \, .
 \]
A parameter $\mu$ of physical dimensions of inverse length is introduced to keep the dimensions correct.

The Mellin transform and the summation Poisson formula
\[
\zeta_{\cal E}(s,g_L^2\vert 1)=\frac{1}{\Gamma(s)}\int_0^\infty \, d\beta \, \beta^{s-1} \, \sum_{n\in\mathbb{Z}}\, e^{-\beta(n^2+g_L^2)} \qquad , \qquad \sum_{j=-\infty}^\infty e^{-\beta j^2} = \sqrt{\frac{\pi}{\beta}} \sum_{j=-\infty}^\infty e^{-\frac{\pi}{\beta} j^2}
\]
allow to express the Epstein zeta function in the form
\begin{equation}
\zeta_{\cal E}(s,g_L^2\vert 1)=\sqrt{\pi}\frac{\Gamma(s-\frac{1}{2})}{\Gamma(s)} g_L^{\frac{1}{2}-s}+2\pi^s \frac{g_L^{\frac{1-2s}{2}}}{\Gamma(s)}+\sum_{n\in\mathbb{Z}} n^{s-\frac{1}{2}} K_{\frac{1}{2}-s}(2\pi g_L n^2) \label{fcczr}
\end{equation}
where the $K_\nu(z)$ functions are modified Bessel functions. As expected the physical value $s=-\frac{1}{2}$ is a pole of the Epstein zeta function and proper renormalization should be implemented to compute the expectation values in the Fermionic Fock space of the normal ordered Dirac Hamiltonian. To achieve that task the computation of the zeta regularized vacuum expectation value of the Dirac Hamiltonian may be used as an intermediate step.

\section{From scalar/spinor Kink fluctuations to Bose/Fermi quanta in presence of Kinks}

Besides the homogeneous static solutions this system admits also static but space dependent stable solutions that, via Lorentz transformations, are travelling waves centered at $x=a$ with Kink shape:
\begin{eqnarray*}
&& -\frac{d^2 \phi}{d x^2}\pm 2 \lambda^2 \phi(x)(\phi^2-1)=0 \, \, \, \Leftarrow \, \, \, \phi_K^\pm(\frac{x-v t-a}{\sqrt{1- v^2}})= \pm \tanh [\lambda( \frac{x-vt-a}{\sqrt{1-v^2}})] \\ && -i \alpha \frac{d \Psi_K}{d x}+\beta(g \phi_K +i g \alpha )\Psi_K=0 \, \, \Leftarrow \, \, \, \Psi_K=0 \, \, .
\end{eqnarray*} 
Note that the spinor field solution is still taken null even in presence of the kink.

Defining non dimensional space-time coordinates $\tau= \lambda t$, $y=\lambda x$ and considering small fluctuations 
\[
\phi(\tau,y)\simeq \phi_K(y)+ \eta(\tau,y) \quad , \quad \Psi(\tau,y)=0+ \sqrt{g} \,\psi(\tau,y)
\]
on the Kink classical background the expansion above is still solution of the field equations if the linear system of coupled PDE's
hold:
\begin{eqnarray*}
&& \left( \frac{\partial^2}{\partial \tau^2}-\frac{\partial^2}{\partial y^2}+4-\frac{6}{\cosh^2 y}\right)\phi(\tau,y)= {\cal O}(\eta^2) \\ && \left(i\frac{\partial}{\partial \tau}-i \sigma_2 \frac{\partial}{\partial y}+\nu \sigma_1 \phi_K(y)\right)\psi(\tau,y)={\cal O}(\eta\psi )
\end{eqnarray*}
We stress that: (1) We choose $\Psi_K=0$ as Fermionic ground state and thus we neglected the Fermionic backreaction on the Kink at the classical level. (2) We introduce the important non dimensional parameter $\nu=\frac{g}{\lambda}$ that measures the strength of the Yukawa versus the scalar self-interaction coupling. (3) Again we abuse of notation in the linearized equations by writing $\phi(\tau,y)$ instead of $\eta(\tau,y)$ . Because there are no $\tau$-dependent terms in both operators it is natural the search of solutions via $\tau$-Fourier transform integrals:
\begin{eqnarray*}
&& \phi(\tau,y)= \int_{-\infty}^\infty \, d\tau \, e^{i \Omega_B \tau} \,  f_{\Omega_B}(y) \quad ,\quad \Omega_B=\frac{\omega_B}{\lambda}\\ && \psi(\tau,y)=  \int_{-\infty}^\infty \, d\tau \, e^{i \Omega_F \tau} \, \psi_{\Omega_F}(y) \qquad, \qquad \Omega_F=\frac{\omega_F}{\lambda}
\end{eqnarray*}
such that the general solution of the linearized equations requires the solution of two quantum mechanical spectral problems,
one for a P$\ddot{\rm o}$sch-Teller/Schr$\ddot{\rm o}$dinger operator the other one for a Dirac operator in a Kink potential background.

 \subsection{From scalar Kink fluctuations to Higgs boson/Kink weak interactions}
 Starting with the scalar/Bose case, the quantum mechanical spectral problem (\ref{pthams}) governing the scalar Kink fluctuations  reads:

 \begin{equation}
 h_{PT}f_{\Omega_B}(y)=\left(-\frac{d^2}{ d y^2}+4 -\frac{6}{\cosh^2 y}\right) f_{\Omega_B}(y)=\Omega_B^2 f_{\Omega_B}(y)  \label{pthams} 
\end{equation}
 Fortunately the eigenvalues and eigenfunctions of this one-particle Hamiltonian are well known, we follow the References \cite{Morse}-\cite{Drazin}-\cite{Abramowitz} to summarize their properties.  The discrete spectrum posseses two bound state eigenfunctions whose corresponding eigenvalues s are respectively $\Omega_B^2=0$ and $\Omega_B^2=3$., namely:
 
  1. \underline{Zero mode}
 \begin{equation}
 \Omega_0=0 \qquad , \qquad f_0(y)= \frac{1}{\cosh^2 y} \nonumber 
 \end{equation}
 This eigenfunction responds to the spontaneous breaking of the translational symmetry of the model by the Kink.
 
2. \underline{The shape fluctuation mode}
 \begin{equation}
 \Omega_{\sqrt{3}}^2=3 \qquad , \qquad f_{\sqrt{3}}(y)=\frac{\sinh y}{\cosh^2 y} \nonumber
 \end{equation}
 The next eigenfunction in the discrete spectrum corresponds to vibrations of the Kink, rather than translations,  with a frequency of $\sqrt{3}$ and it is referred to as shape mode because it is accompanied by variations in the Kink shape.
 
 Above these two bound fluctuation modes the eigenstates with energies over the threshold of the continuous spectrum $\Omega_B^2(0)=4$ arise.

3. \underline{Scattering states: the continuous spectrum}
 
\begin{equation}
\hspace{-0.3cm}\Omega^2_B(q)= q^2 +4 \, \,  , \quad f(y;q)= e^{i q y}(3 \tanh^2 y -3 i q \tanh y-(1+q^2)) = e^{i q y} P_2(\tanh y;q) \nonumber
\end{equation}
where $q^2=\frac{k^2}{\lambda^2}$ and $P_2(z;q)$ are Jacobi polynomials of order $2$.
It is remarkable that the scattering involved is transparent, i.e, the reflection amplitude $r(q)$ is zero and the modulus of the transmission amplitude is one:
\[
t(q)=\frac{1-i q}{1+i q}\cdot \frac{2-i q}{2+ i q}  \, \, .
\]
Imposing Periodic Boundary Conditions on the continuous spectrum eigenfunctions in a very long interval of length $L$,
$f(-\lambda \frac{L}{2};q)=f(\lambda \frac{L}{2};q)$, the spectral condition incorporates the phase shifts due to the kink presence:
\[
q_n + \delta(q_n)= 2\pi n \, \, \, \, \Rightarrow \, \, \, \, \delta(q_n)=2\left({\rm arctan}\frac{1}{q_n}+{\rm arctan}\frac{2}{q_n}\right) \quad , \, \, n\in \mathbb{Z} 
\]
and the spectral density is modified:
\[
\rho_B(q)=\frac{1}{2\pi}\left(\lambda L+ \frac{d\delta}{d q}\right) \, \, \, \Leftarrow \, \, \, \frac{d \delta}{dq}=\lim_{\Delta q\to 0} \frac{\delta(q_n+\Delta q)-\delta(q_n)}{\Delta q} \, \, , \, \, \Delta q=\frac{2\pi}{\lambda L} \, \, .
\]

The scalar field expansion (\ref{scafield}) is obtained as a linear superposition 
in terms of the eigenfunctions of the one-particle operator (\ref{pthams})
\begin{eqnarray}
\phi(\tau,y)&=&\lim_{\varepsilon \to 0}\left(A_0 e^{-i \varepsilon \tau}+ A_0^* e^{i \varepsilon \tau}\right) f_0(y) +\left( A_{\sqrt{3}}e^{-i \sqrt{3}\tau}+A^*_{\sqrt{3}}e^{i\sqrt{3} \tau}\right)\cdot f_{\sqrt{3}}(y)\nonumber \\ &+& \int \, \frac{dq}{2\Omega_B(q)}\left( A(q)e^{-i \sqrt{q^2+4} \tau}f(y;q)+ A^*(a) e^{i \sqrt{k^2+4} \tau}f^*(y;q) \right) \label{scafield}
\end{eqnarray} 

Canonical quantization courses as usual replacing the complex coefficients of the spectral expansion by quantum operators satifying commutative quantization relations:
\[
[\hat{A}_0, \hat{A}_0^\dagger ]=1 ,\quad , \quad [\hat{A}_{\sqrt{3}},\hat{A}^\dagger_{\sqrt{3}}]=1 \quad , \quad [\hat{A}(q_1),\hat{A}^\dagger(q_2)]=\delta(q_1-q_2)
\]
The differences with respect to  the Bosonic Fock space in the vacuum sector are three: (1) There is one state, the zero mode; where a Boson is bounded to the Kink center travelling with  it at no cost of energy. (2) The shape mode is one state in the Fock space where one Boson is trapped by the Kink giving rise to one Kink excited state characterized by its vibration frequency. (3) There are many states where the Higgs quanta are scattered off the Kink but the outgoing particles escape out of the Kink center as plane waves distorted by Jacobi polynomias of order 2.

\subsection{From spinor Kink fluctuations to Fermion/Kink weak interactions}

\begin{itemize} 

\item \underline{Spinor Kink fluctuations}

Spinor Kink fluctuations are determined from the spectral problem of the one-particle Kink-Dirac Hamiltonian (\ref{specfkf}) :
\begin{eqnarray}
&& h_{DK}\psi(y; \Omega_F)=\Omega_F \psi_(y;\Omega_F) \qquad , \qquad \psi(y;\Omega_F)=\left(\begin{array}{c}\psi_1(y;\Omega_F) \\ \psi_2(y; \Omega_f) \end{array}\right) \label{specfkf}\\
 && h_{DK}= \left(\begin{array}{cc} 0 & -\frac{d}{dy}+ \nu \tanh y  \\ \frac{d}{dy}+\nu \tanh y  & 0 \end{array}\right) \, \, ,\,\, \ \nu=\frac{g}{\lambda} \, \, \, , \, \, \, [\psi_1(y;\Omega_F)]=[\psi_2(y;\Omega_F)]=1 \nonumber \, .
\end{eqnarray}
Instead of solving directly the spectral problem  (\ref{specfkf}) we notice that the
 square of the Dirac-Kink operator is a diagonal matrix of P$\ddot{\rm o}$sch-Teller-Schr$\ddot{\rm o}$dinger operators. In the case when $g$ is a positive integer multiple of $\lambda$ such that
 $\nu=N$, the square of the Dirac/Kink operator encompasses in the diagonal two PTS consecutive operators belonging
 to the hierarchy of transparent PT operators. The potential wells are determined respectively by $N(N+1)$ and $N(N-1)$. times 
 $\sech^2 y$.
 Moreover, defining the first-order differential operator $d_N=\frac{d}{d y}+N \tanh y$, the Darboux-Crumm factorization method
\begin{eqnarray*}
 h_{DK}^2 &=& \left(\begin{array}{cc} -\frac{d^2}{d y^2}+ N^2-\frac{N(N+1)}{\cosh^2 y} & 0 \\ 0 & -\frac{d^2}{d y^2}+ N^2- \frac{N(N-1)}{\cosh^2 y}\end{array}\right) \\ &=&\left(\begin{array}{cc} d_N^\dagger d_N & 0 \\ 0 & d_{N}d_{N}^\dagger \end{array}\right) = \left(\begin{array}{cc} d_N^\dagger d_N & 0 \\ 0 & d_{N-1}^\dagger d_{N-1}+2 N -1 \end{array}\right) 
\end{eqnarray*}
 may be succesfully applied to find the spectrum. Note that the two second order differential operators in the diagonal are isospectral.

1. \underline{Spinor zero modes}
Immediately one normalizable and one non normalizable spinor waves zero modes are recognizedas living respectively
in the kermels of $d_N$ or $d_N^\dagger$. One finds:
\[
 h_{DK} \left(\begin{array}{c} \psi_1^{(0)}(y) \\ 0 \end{array}\right)=0 \, \, \Rightarrow \, \, \psi_1^{(0)}(y)=\frac{1}{\cosh^N y} \, \, \, , \, \, \,  h_{DK} \left(\begin{array}{c} 0 \\ \psi_2^{(0)}(y) \end{array}\right)=0 \, \, \Rightarrow \, \, \psi_2^{(0)}(y)=\cosh^N y
\]
Needless to say changing from Kink to anti-Kink the normalizable and non-normalizable zero modes are exchanged.

2. \underline{Bounded spinor vibrational modes. Oscillating Fermi-anti-Fermi pair}\\ \underline{ bounded around the Kink center.}

In order of increasing energy we list $N-1$ proper bound states which correspond to vibrating spinor fluctuation modes. The eigenvalues are well known{\footnote{In this case there is one half-bound state just at the threshold of the continuous spectrum with $l=N$. These \lq\lq half-states\rq\rq do not exist if $\nu \notin \mathbb{N}^*$. }}:
 
  \underline{Eigenvalues}:
\[
\Omega_F^{(l)}(\kappa_l;N)=\sqrt{(2N-l)l}=\sqrt{N^2-\kappa_l^2} \, \, , \, \, l=1,2, \cdots, N-1 \quad , \quad \kappa_l=N-l
\]
 
The eigenspinors are found using the Crumm-Darboux method:

\underline{Eigenspinors} $\psi^{(l)}(y)= \left(\begin{array}{c} \psi_1^{(l)}(y)\\  \psi_2^{(l)}(y)\end{array}\right)$
\begin{eqnarray*}
\psi_1^{(l)}(y)&=&\prod_{j=0}^{N-1} \left(-\frac{d}{dy}+(N-j)\tanh y\right) e^{-(N-l) y}
\\  \psi_2^{(l)}(y)&=&\prod_{j=1}^{N-1} \left(-\frac{d}{dy}+(N-j)\tanh y\right) e^{-(N-l) y}
\end{eqnarray*}
The product should be undestood as product of differential operators where the order matters. The $j$-terms are placed from left to right in increasing order
Note that the bound state labelled by $l$ in the upper diagonal component and the bound state 
labelled by $j=l-1$ in the lower diagonal component of the square of the Dirac operator share identical eigenvalues. 

3. \underline{Spinor plane waves distorted by Kinks}

It remains to describe the spinor fluctuations scattered off Kinks, i.e., not bounded around the Kink center. Of course, these fluctuations belong to the continuous spectrum of $h_{DK}$ (\ref{dkss}):
\begin{equation}
h_{DK} \psi(y;q)= \Omega_F(q) \psi(y,q) \qquad , \qquad \Psi(y;q)=\sqrt{g}\left(\begin{array}{c} \psi_1(y;q)\\ \\ \psi_2(y;q)\end{array}\right)
\label{dkss}
\end{equation}

The eigenvalues are:

\underline{Eigenvalues}: $\Omega_F^2(q)=q^2+N^2 \, \, \Rightarrow \, \, \Omega_F(q)=+\sqrt{q^2+N^2}$

whereas the eigenspinors may also be computed using the factorization method:

\underline{Eigenspinors}
\begin{eqnarray*}
 \psi_1(y;q)&=& \prod_{j=0}^{N-1} \left(-\frac{d}{dy}+(N-j)\tanh y\right)\cdot e^{i q y}
\\  \psi_2(y;q)&=& \prod_{j=1}^{N-1} \left(-\frac{d}{dy}+(N-j)\tanh y\right)\cdot e^{i q y}
 \label{scatwf2}
\end{eqnarray*}

The scattering transition amplitudes are read from the spinor scattering waves and are:
\[
t^{1)}(q)= \prod_{l=0}^{N-1}\, \frac{N-l -i q}{N-l + i q} \quad \, \, ,  \quad \, \,  t^{2)}(q)= \prod_{l=1}^{N-1}\, \frac{N-l -i q}{N-l + i q}
\]
which means that the reflection amplitudes are null: $r^{(1)}(q)=0=r^{(2)}(q)$, the reason for being called transparent potentials.
Thus, the poles of the transition amplitudes in the imaginary axis of the $q$-complex plane correspond to the shape mode bound states.
Like scalar scattering waves, also spinor scattering  waves are characterized by their phase shifts, obtained as the eigenvalues of the S-matrix, and/or spectral densities. Considering the system defined on a finite interval of very large length $L$ with PBC the spinor waves phase shifts induced passing
through the kink are
\[
\delta_F^{(1)}(q)= 2 \sum_{j=1}^N {\rm arctan} \frac{j}{q} \quad , \quad \delta_F^{(2)}(q)= 2 \sum_{j=1}^{N-1} {\rm arctan} \frac{j}{q}
\]
We thus write now the spectral densities of the scattering through the Kink wells experienced respectively by the scalar fluctuations and  the upper and lower components of the spinor fluctuations : 
\begin{eqnarray*}
&& \rho_B(q)= \frac{\lambda L}{\pi}-\frac{1}{\pi}\left( \frac{1}{1+q^2}+\frac{2}{4+q^2}\right) \quad , \quad 
 \rho_F(q)=\rho_F^{(1)}(q)+\rho_F^{(2)}(q) \\
&& \rho_F^{(1)}(q) =\frac{g L}{2\pi}-\frac{1}{\pi}\sum_{j=1}^{N-1} \frac{j}{j^2+q^2}+\frac{N}{N^2+q^2} \quad , \quad  \rho_F^{(2)}(q)=\frac{g L}{2\pi}-\frac{1}{\pi}\sum_{j=1}^{N-1} \frac{j}{j^2+q^2}
\end{eqnarray*}
Note that besides the precise characterization of the continuous spectrum in terms of the wave number $q$ the bound states resurface as poles in the spectral density.

\item  \underline{Complex conjugate spinor Kink fluctuations}
The Dirac classical field is complex and it is necessary to analyze behave spinors under complex conjugate transformations in orden to study complex conjugate spinor fluctuations. Because the complex conjugate field is $\Psi^C(t,x)=\sigma_1\Psi^*(-t,-x)$ ,
the one-particle complex conjugate Dirac Hamiltonian becomes: 
\begin{equation}
h_D^C(\mu)= \sigma_1 h_D(\mu)\sigma_1 =\left(\begin{array}{cc} -\mu & -\frac{d}{d y}-N \tanh y \\ -\frac{d}{d y}+N \tanh y & \mu\end{array}\right) \label{condir} \, \, .
\end{equation}
 A light chiral mass $0<\mu<< \nu^2$ has been introduced to provide an infrared cutoff to the zero modes.
 Denoting $\Psi^C(t,x)=\Phi(t,x)$ the spectral problem
 \[
 h_D^C(\mu) \Phi^{\Omega_F}(y)=-\Omega_F \Phi^{\Omega_F}
 \]
is equivalent to the ODE's system, if we restrisct $\nu$ to be a positive integer $N$:
\begin{equation}
(\frac{d}{d y}+N \tanh y)\phi_2^{\Omega_F}(y)= (\Omega_F-\mu) \phi_1^{\Omega_F}(y) \quad , \quad (-\frac{d}{d y}+N \tanh y)\phi_1^{\Omega_F}(y)= (\Omega_F+\mu) \phi_2^{\Omega_F}(y)
\label{posfluc} \quad .
\end{equation}
We list next the solutions of the system (\ref{posfluc}):
\begin{enumerate}

\item  \lq\lq Filled\rq\rq hole
\[
\Omega_F^{(-\mu)}=-\mu  \quad , \quad  \Phi^{-\mu}(y)=\sqrt{g}\left(\begin{array}{c}0 \\ \frac{1}{\cosh^N y} \end{array}\right)
\]
This solution is normalizable but because its energy is negative belongs to the Dirac sea. Therefore it will be occupied 
by one fermion after anticommutative quantization.

\item Non-normalizable shifted zero mode
\[
\Omega_F^\mu=\mu \quad , \quad \Phi^{\mu}(y)=\sqrt{g}\left(\begin{array}{c}\cosh^N y  \\ 0 \end{array}\right)
\]

\item \lq\lq Conjugate\rq\rq spinor vibrational modes. If $\phi^{(l)}(y)=\sqrt{g}\phi^{(l)}(y)=\left(\begin{array}{c}\phi^{(l)}_1(y) \\ \phi_2^{(l)}(y)\end{array}\right)$
\begin{eqnarray*}
\phi_1^{(l)}(y)&=&\prod_{j=1}^{N-1} (-\frac{d}{dy}+(N-j)\tanh y)e^{-(N-l)y}\\ \phi_2^{(l)}(y) &=& \prod_{j=0}^{N-1} (-\frac{d}{dy}+(N-j)\tanh y)e^{-(N-l)y}
\end{eqnarray*} 

\item Conjugate spinor scattering waves

\begin{eqnarray*}
\phi_1(y;q)&=& \prod_{j=1}^{N-1} (-\frac{d}{dy}+(N-j)\tanh y)e^{-i q y}\\ \phi_2(y;q)&=& \prod_{j=0}^{N-1}  (-\frac{d}{dy}+(N-j)\tanh y)e^{- i q y}
\end{eqnarray*} 

\end{enumerate}

In any case the $\Phi$ and $\Psi$ eigenspinors are mutually orthogonal:
\begin{eqnarray*}
&& \int_{-\infty}^\infty dy \, \Phi^{(l)\dagger}(y) \Psi^{(l)}(y)=\int_{-\infty}^\infty dy \, \Psi^{(l)\dagger}(y) \Phi^{(l)}(y)=0\\
&& \int_{-\infty}^\infty dy \, \Phi^\dagger(y;q) \Psi(y;q)=\int_{-\infty}^\infty dy \, \Psi^\dagger(y;q) \Phi(y;q)=0
\end{eqnarray*}
because the densities to be integrated are odd functions. The Kink spinor scattering fluctuations are not normalizable, like the spinor scattering fluctuations of the ground state, and special care is needed to deal with this problem, e.g., putting the system in a very long length interval and demanding periodic boundary conditions to the spinor waves, planes or distorted by the Kink.. 

\end{itemize}

Note that the eigen-spinors $\psi^{(l)}(y)$, $\psi(y;q)$ and $\phi^{(l)}(y)$, $\phi(y;q)$ depends only on one -component, the upper component in the first case, the lower in the second case. The reason is that the Dirac equation determines one from the other.

\subsection{Excited kinks in the Jackiw-Rebbi model}

The combined effect of the scalar and spinor modes of linear fluctuations of the kink is encoded in the time-dependent configurations
\begin{equation}
\left( \tanh y +\varepsilon  \, e^{i \sqrt{3}\, \tau} \frac{\tanh y}{\cosh y}, \varepsilon \, (a \, e^{i \sqrt{(2N-l)l}\, \tau} \, \psi^{(l)}(y)+ b^* \, \,e^{- i\sqrt{(2 N-l)l}\,\tau} \,\phi^{(l)}(y)) \right) \label{defkink}
\end{equation}
where $a$ and $b^*$ are complex Grassmann constants.
In this formula (\ref{defkink}) the scalar kink fluctuation of frequency $\sqrt{3}$  has been selected and added to the classical kink configuration. The effect is to change the kink profile that oscillates with frequency $\sqrt{3}$. Other two fluctuations are selected of the static kink configuration, in this case of spinor type. The addition of these configurations to the pure scalar kink profiles
do not modify the classical kink but gives rise to vibrations of the spinor field with frequency $\sqrt{(2N-l)l}$ in the background of this scalar topological defect. There are bounded spinor fluctuations of type $\psi^{(l)}(y)$ -electrons- and type $\phi^{(l)}(y)$ -positrons- catalyzed by the kink. Thus, one electron-positrom pair emerges and oscillates over the kink background. In the $N=2$ case  the frequency $\sqrt{3}$ by which the kink changes its shape due to the bounded scalar fluctuation is identical to the vibration frequency of the spinor fields.

\subsection{The coupling's ratio $\frac{g}{\lambda}=3$ }

We next describe \, the very interesting situations when the ratios between the Yukawa coupling and the scalar self-coupling is three: $g=3 \lambda$. We shall list first the two Kink shape mode spinors. The lower components are identical to the scalar Kink  fluctuation modes. The upper component spinor Kink fluctuations modes are:

\begin{enumerate}

\item \underline{Spinor Kink fluctuation zero mode}
\[
\Omega_F^{(0)}=0 \quad , \quad \frac{1}{\sqrt{g}}\Psi^{(0)}(y)=\left(\begin{array}{c} \frac{1}{\cosh^3 y} \\ 0 \end{array}\right)
\]

\item \underline{Spinor Kink fluctuation shape mode 1}

\begin{eqnarray*}
\Omega_F^{(1)}&=&+\sqrt{5+\mu^2} \qquad , \qquad 
\frac{1}{\sqrt{g}}\Psi^{(1)}(y)=\frac{1}{\cosh^2 y}\left(\begin{array}{c}\tanh y \\  1 \end{array}\right) \\   \Psi^{(1)T}(y)\Psi^{(1)}(y)&=&g \frac{\cosh 2y}{\cosh^6 y} \qquad , \qquad  N^{(1)}= \frac{1}{g}\int_{-\infty}^\infty \, dy \,  \Psi^{(1)T}(y)\Psi^{(1)}(y) = \frac{8}{5} \
\end{eqnarray*}
\begin{figure}[ht]
\centerline{
\includegraphics[height=3.6cm]{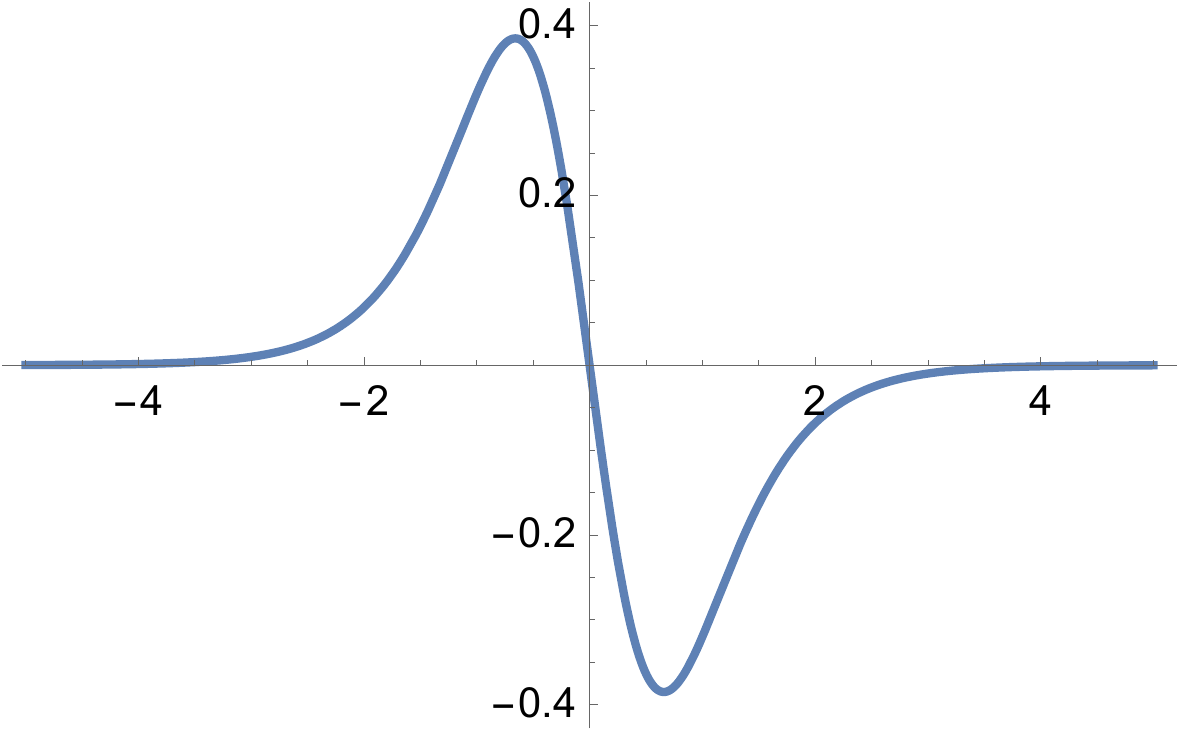} \hspace{0.1cm}\includegraphics[height=3.6cm]{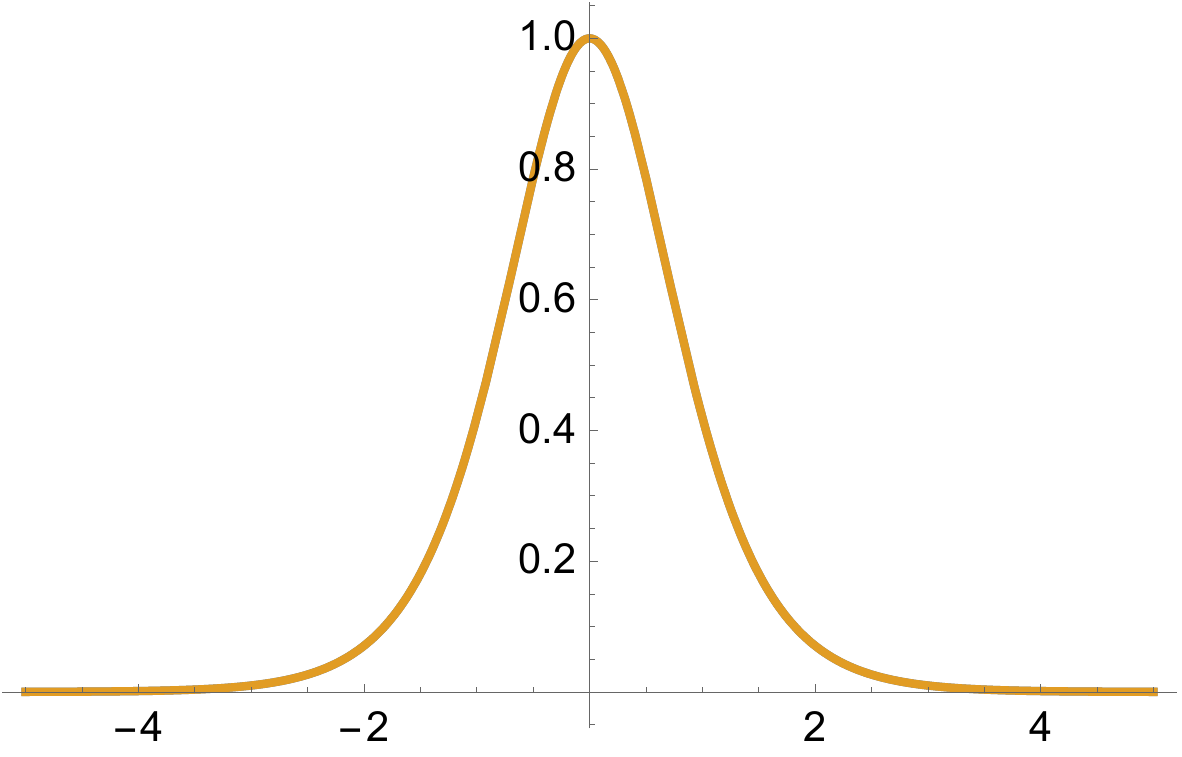} \hspace{0.2cm} \includegraphics[height=4cm]{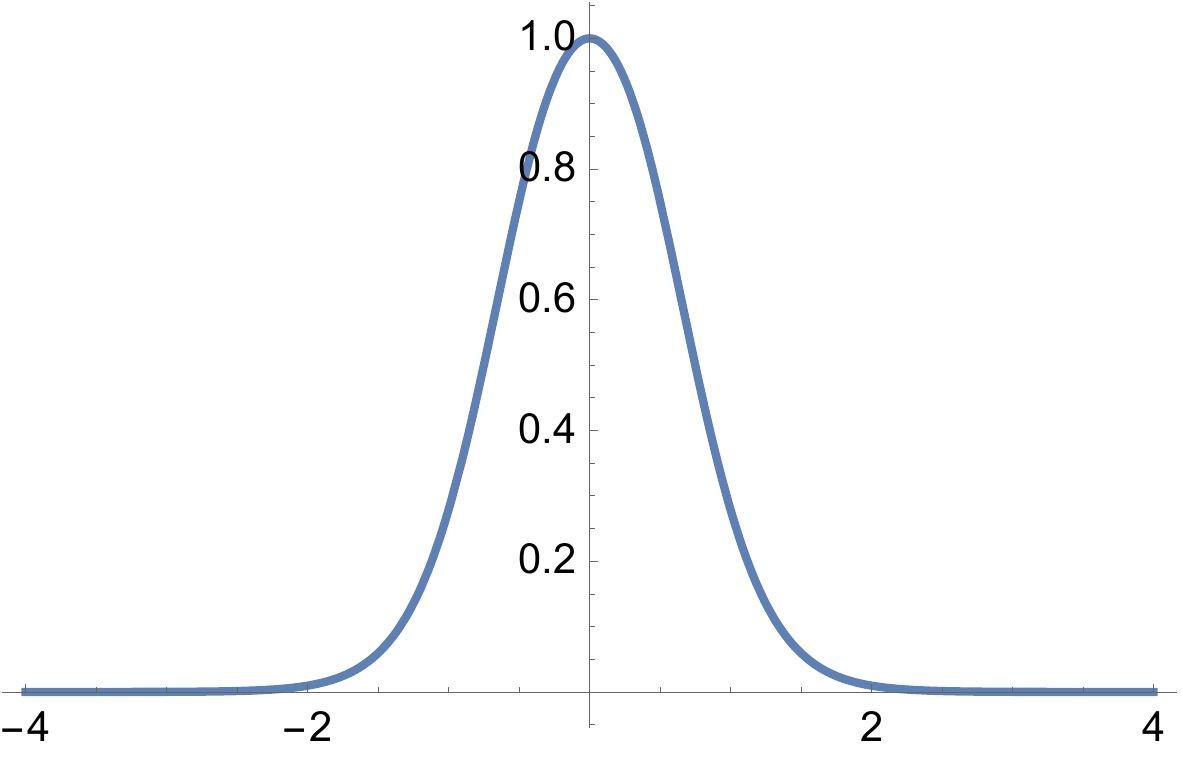}}
\caption{ Upper (left) lower (center) components of the shape 1 eigenspinor. (right) Probability density of the shape mode 1}
\end{figure}

\item \underline{ Spinor Kink fluctuation shape mode 2}

\begin{eqnarray*}
\Omega_F^{(2)}&=&+\sqrt{8+\mu^2} \qquad , \qquad \
\frac{1}{\sqrt{g}}\Psi^{(2)}(y)=\frac{1}{\cosh y}\left(\begin{array}{c} 4-\frac{5}{\cosh^2 y}\\  \tanh y \end{array}\right) \\ \frac{1}{g} \Psi^{(2)T}(y)\Psi^{(2)}(y)&=&\frac{1}{ \cosh^2 y}\left[(4-5 \sech[y]^2)^2+\tanh[y]^2\right]\, \,  , \quad N^{(2)}= \frac{1}{g}\int_{-\infty}^\infty \, dy \, \Psi^{(2)T}(y)\Psi^{(2)}(y)=6
\end{eqnarray*}
\begin{figure}[ht]
\centerline{
\includegraphics[height=3.6cm]{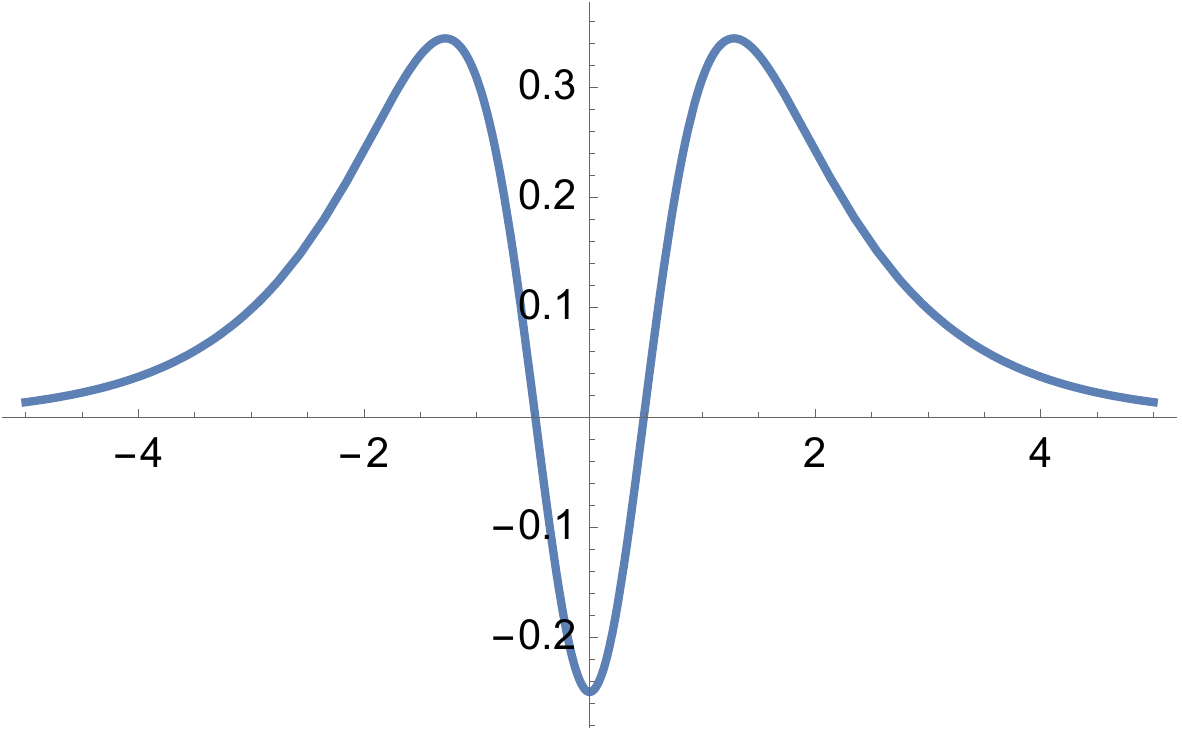} \hspace{0.1cm}\includegraphics[height=3.6cm]{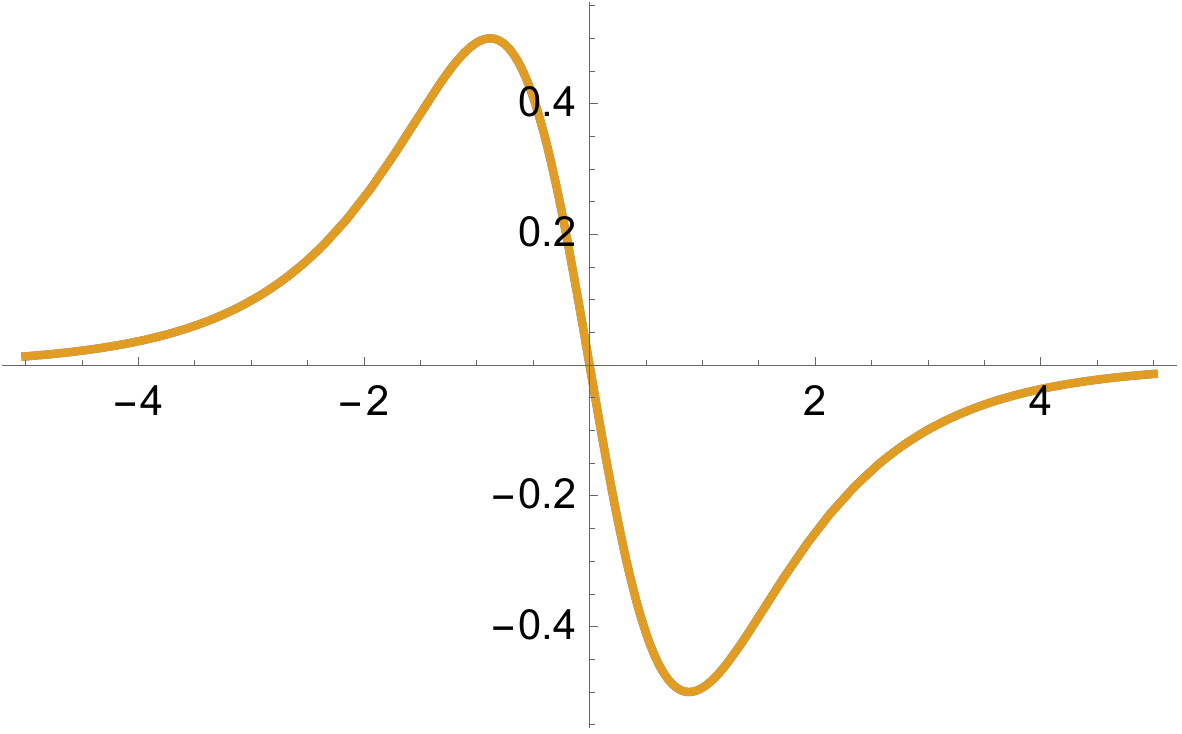} \hspace{0.2cm} \includegraphics[height=4cm]{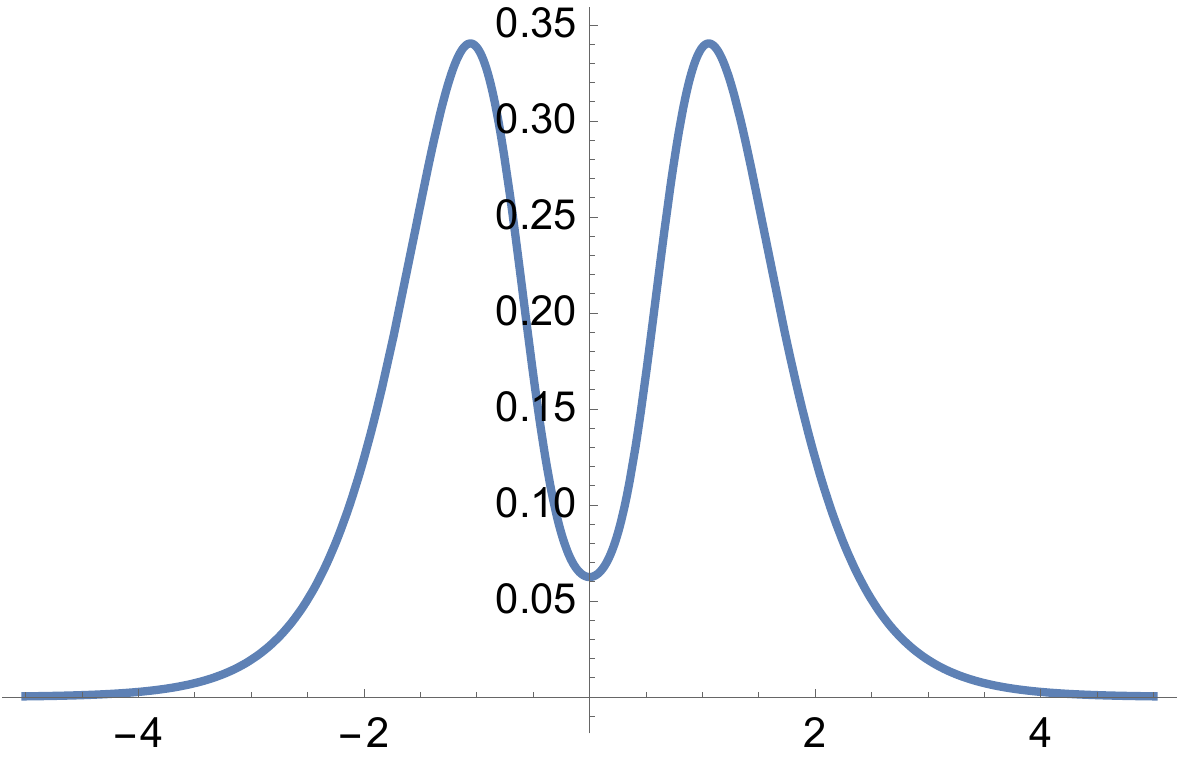}}
\caption{ Upper (left) lower (center) components of the shape 2 eigenspinor. (right) Probability density of the shape mode 2}
\end{figure}

Also one may consider the Lorentz invariant densities for the $a=1$ and $a=2$ shape modes{\footnote{Recall that $J_0^{(a)}(y)=\Psi^{(a)\dagger}(y)\Psi^{(a)}(y)$ is the $0$-component of a two-vector.}}:
\begin{eqnarray*}
g^{-1} \Psi^{(1)\dagger}(y)\gamma^0\Psi^{(1)}(y) &=& \sech^2 y \tanh y  \\ g^{-1}\Psi^{(2)\dagger}(y)\gamma^0 \Psi^{(2)}(y)&=& \sech^2 y(4-5 \sech^2 y) \tanh y\end{eqnarray*}
which are plotted in Figure 3.

\begin{figure}[ht]
\centerline{
\includegraphics[height=4.6cm]{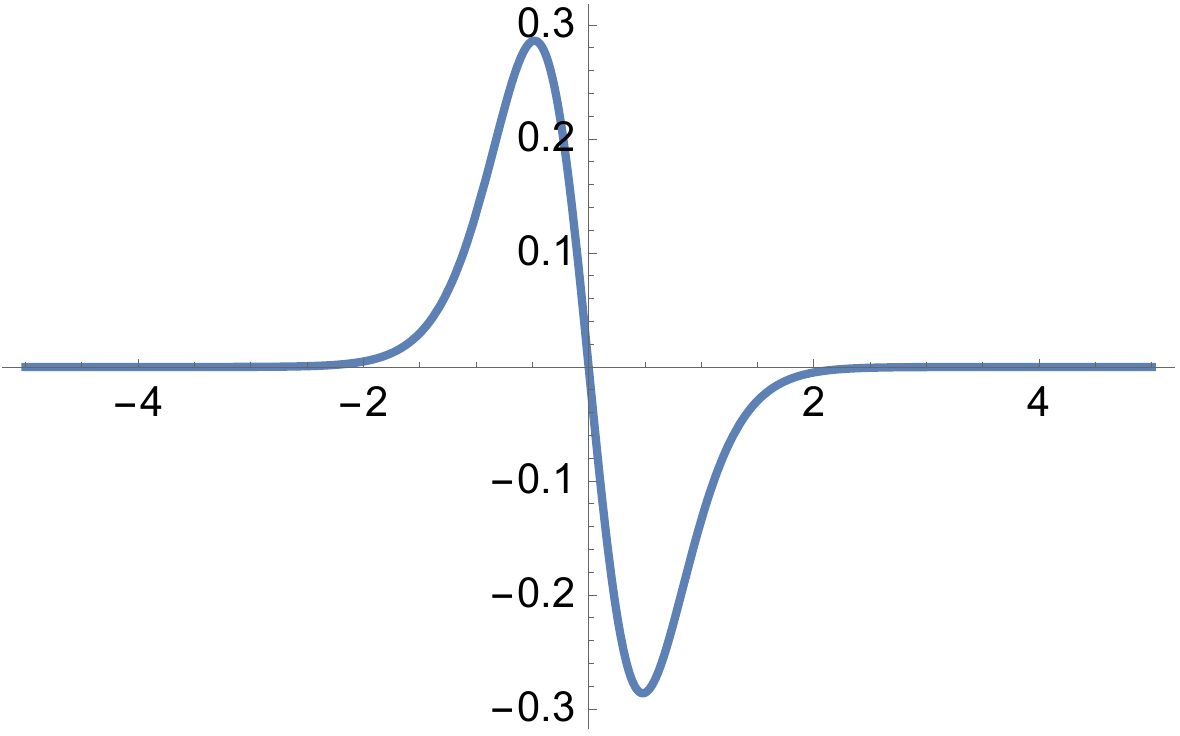} \hspace{0.5cm}\includegraphics[height=4.6cm]{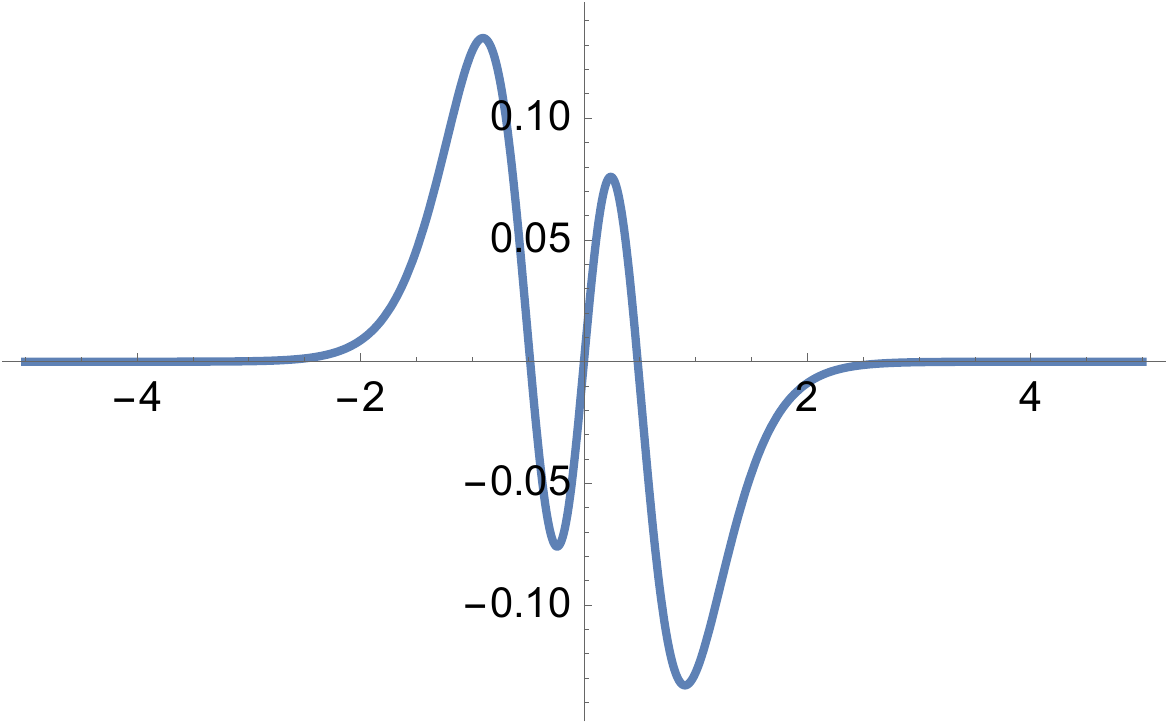} }
\caption{ Scalar density of the shape mode of frequency $\sqrt{5}$ (left) Scalar density of the frequency  $\sqrt{8}$ shape mode (right)}
\end{figure}

\item \underline{ Spinor distorted plane waves scattered off kinks }
We shall write the scattering eigenspinors using the the Crumm-Darboux-Infeld-Hull method: 

\begin{eqnarray*}
&& \Omega_F(q)=+\sqrt{q^2+9+\mu^2} \\
&& \Psi(y;q)=\sqrt{g}\, e^{i q y} \left(\begin{array}{c} i(3\sech^2 y(2 q+5 i \tanh y)+( q+3 i \tanh y)(-2-q^2+3 i q \tanh y) \\ (3\tanh^2 y -3 i q \tanh y -1-q^2) \end{array}\right)
\end{eqnarray*}

\begin{figure}[ht]
\centerline{
\includegraphics[height=4.6cm]{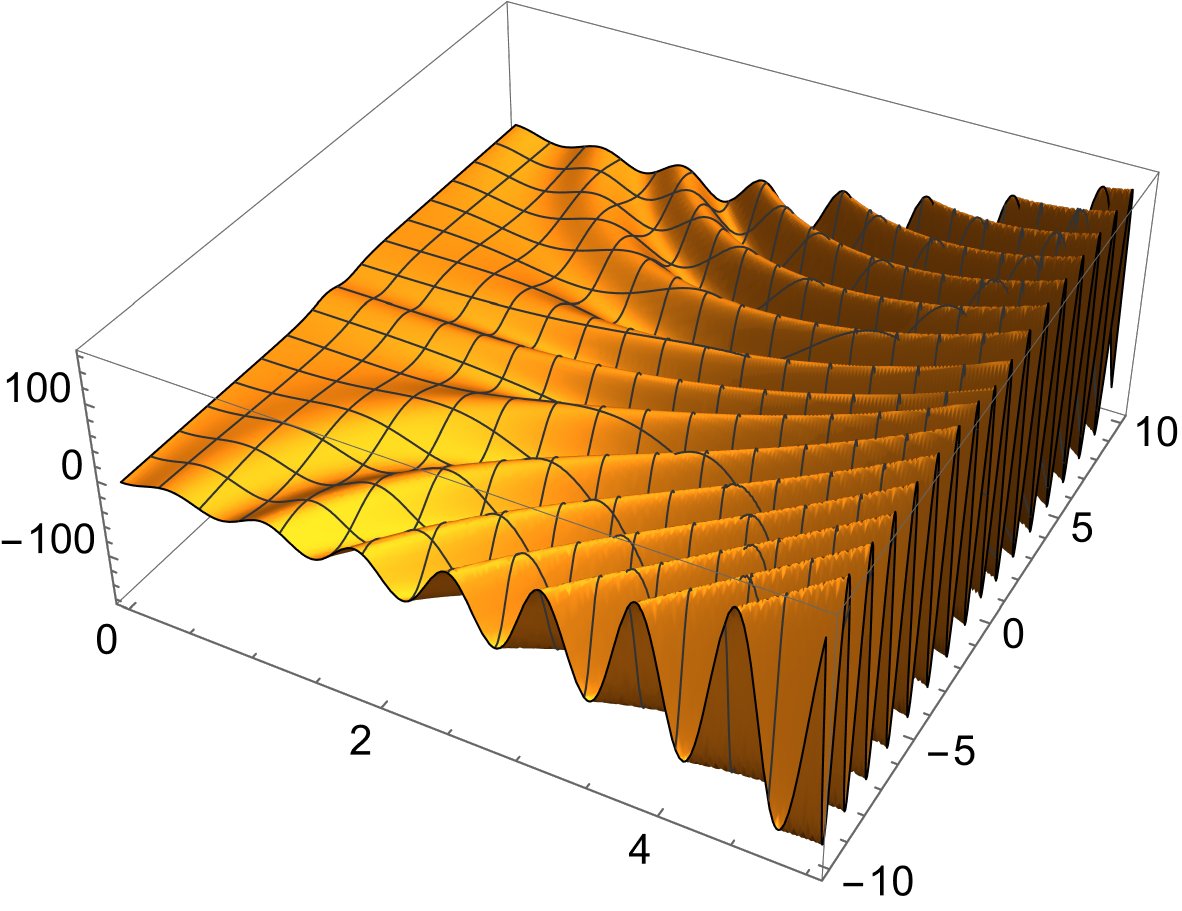} \hspace{0.1cm}\includegraphics[height=4.6cm]{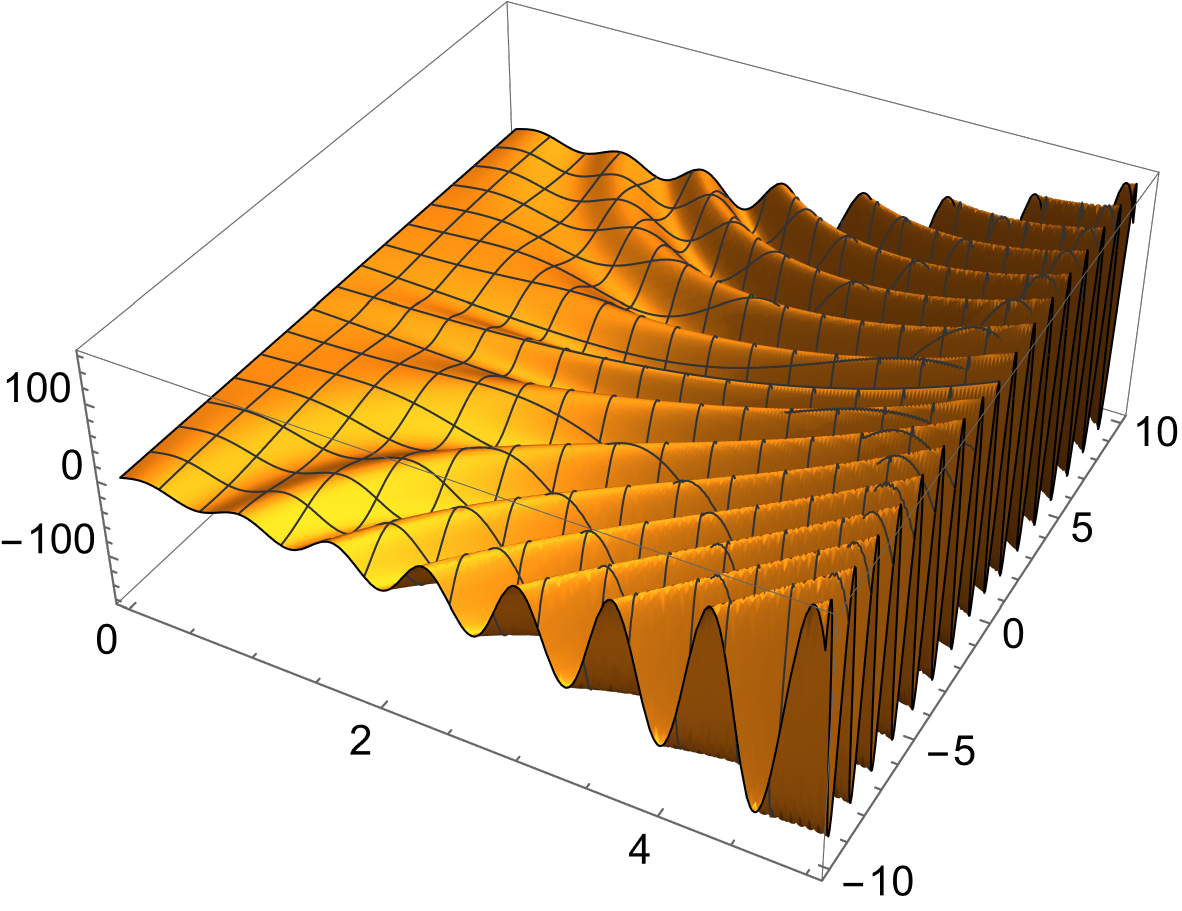} \hspace{0.2cm} \includegraphics[height=4.6cm]{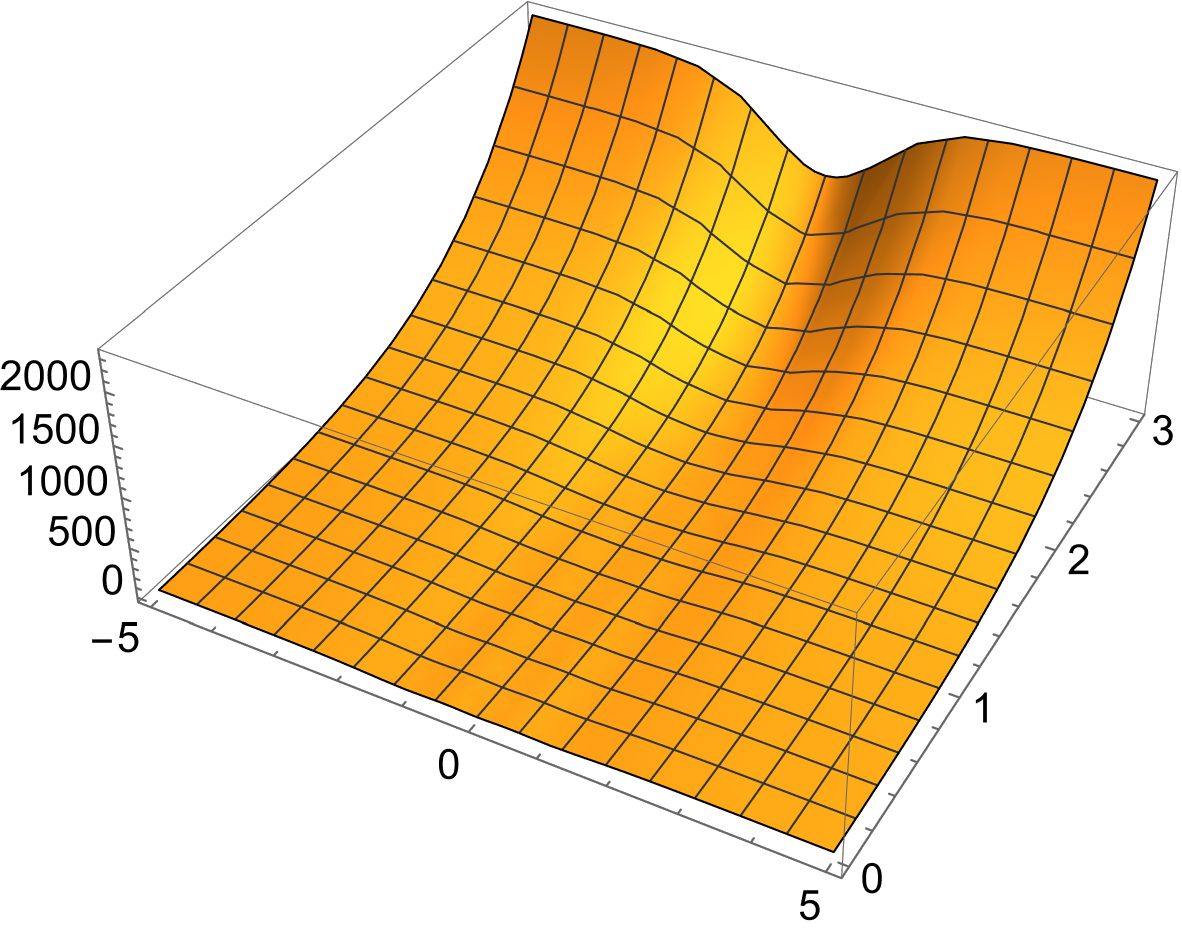}}
\caption{ 3D graphics of the upper component of scattering eigenspinors of energy $\Omega_F=\sqrt{9+q^2}$ as functions of $y$ and the wave number $q$. Real part (left) Imaginary part (center) Absolute value (right). It is difficult to observe in this Figure, that the real and imaginary parts of the upper component of the spinor scattering waves have different phases but they do. The difference varies with $q$.}
\end{figure}

\begin{figure}[ht]
\centerline{
\includegraphics[height=4.6cm]{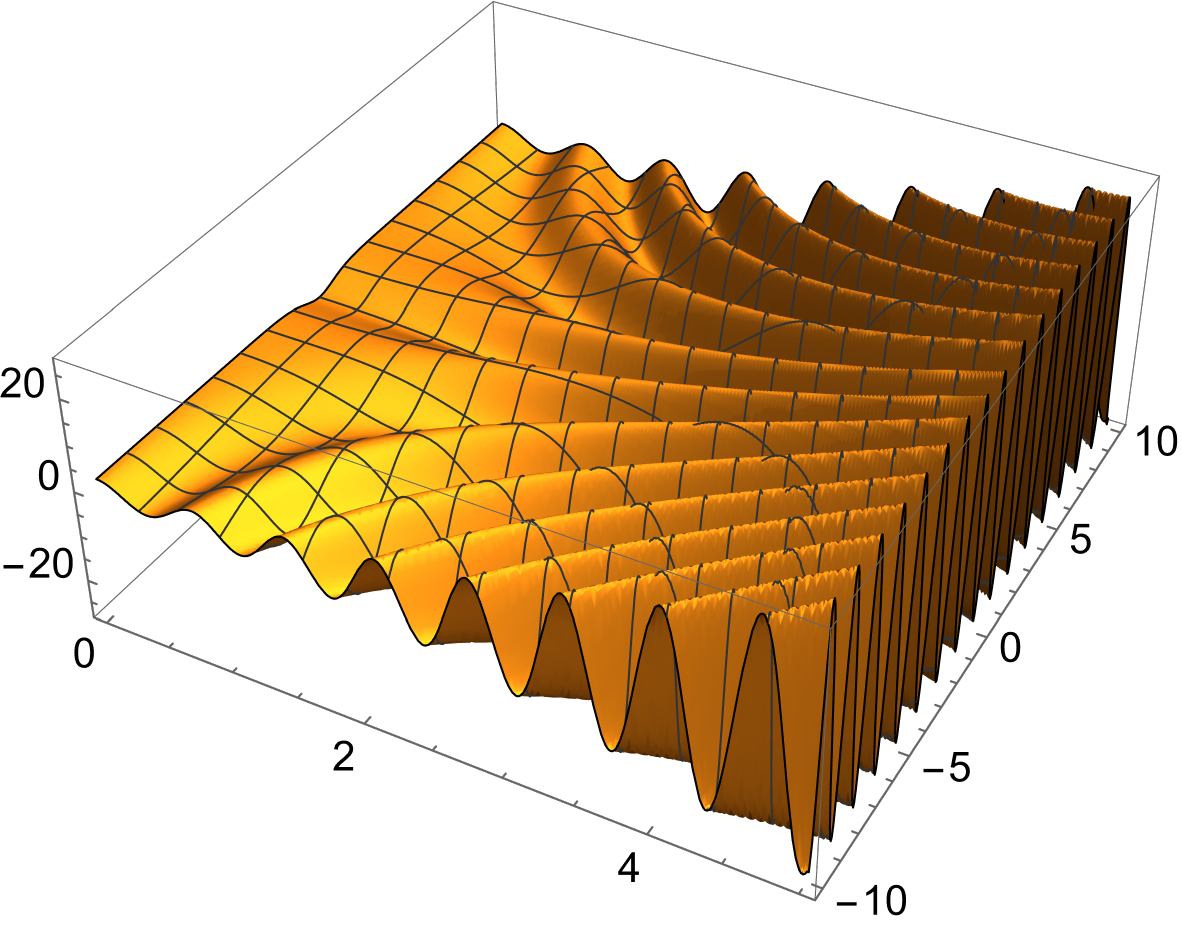} \hspace{0.1cm}\includegraphics[height=4.6cm]{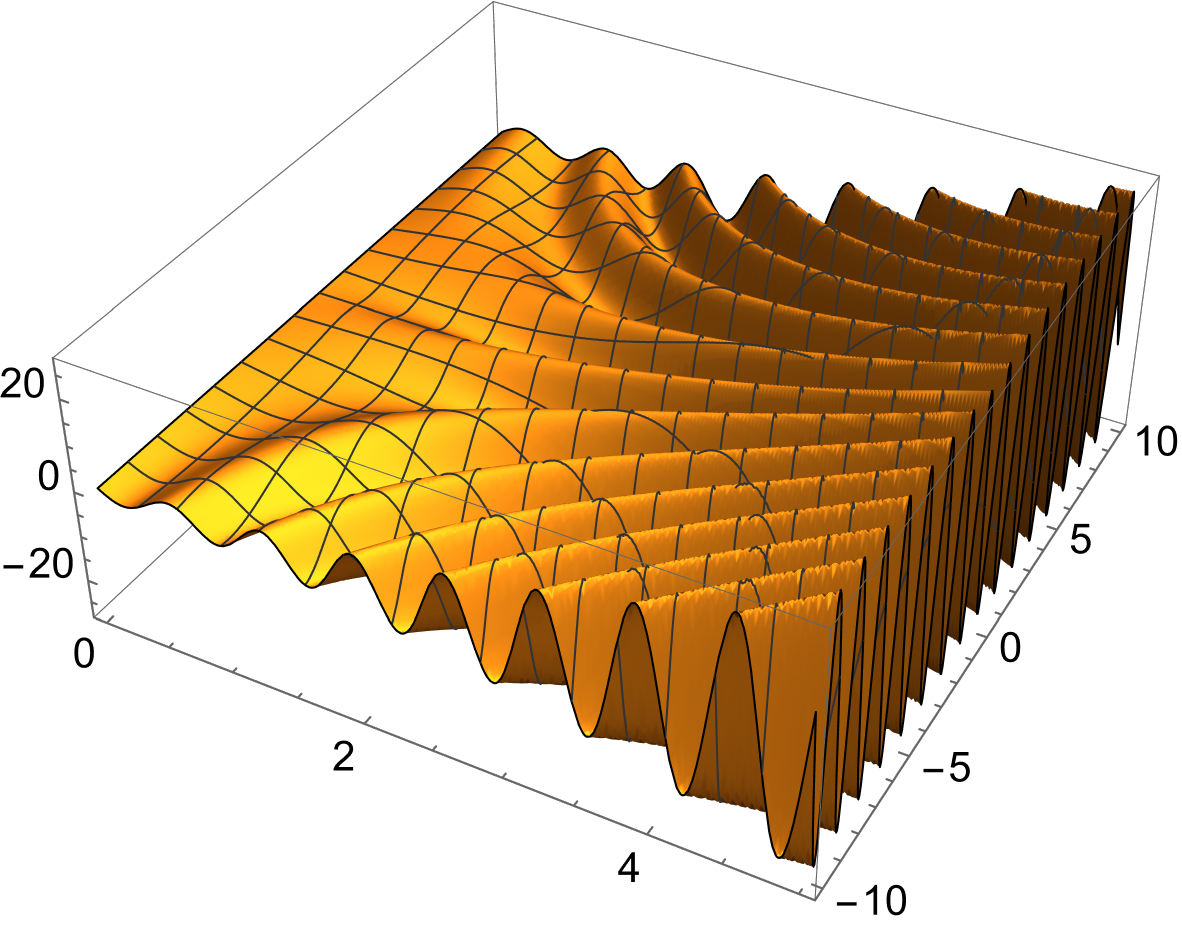} \hspace{0.2cm} \includegraphics[height=4.6cm]{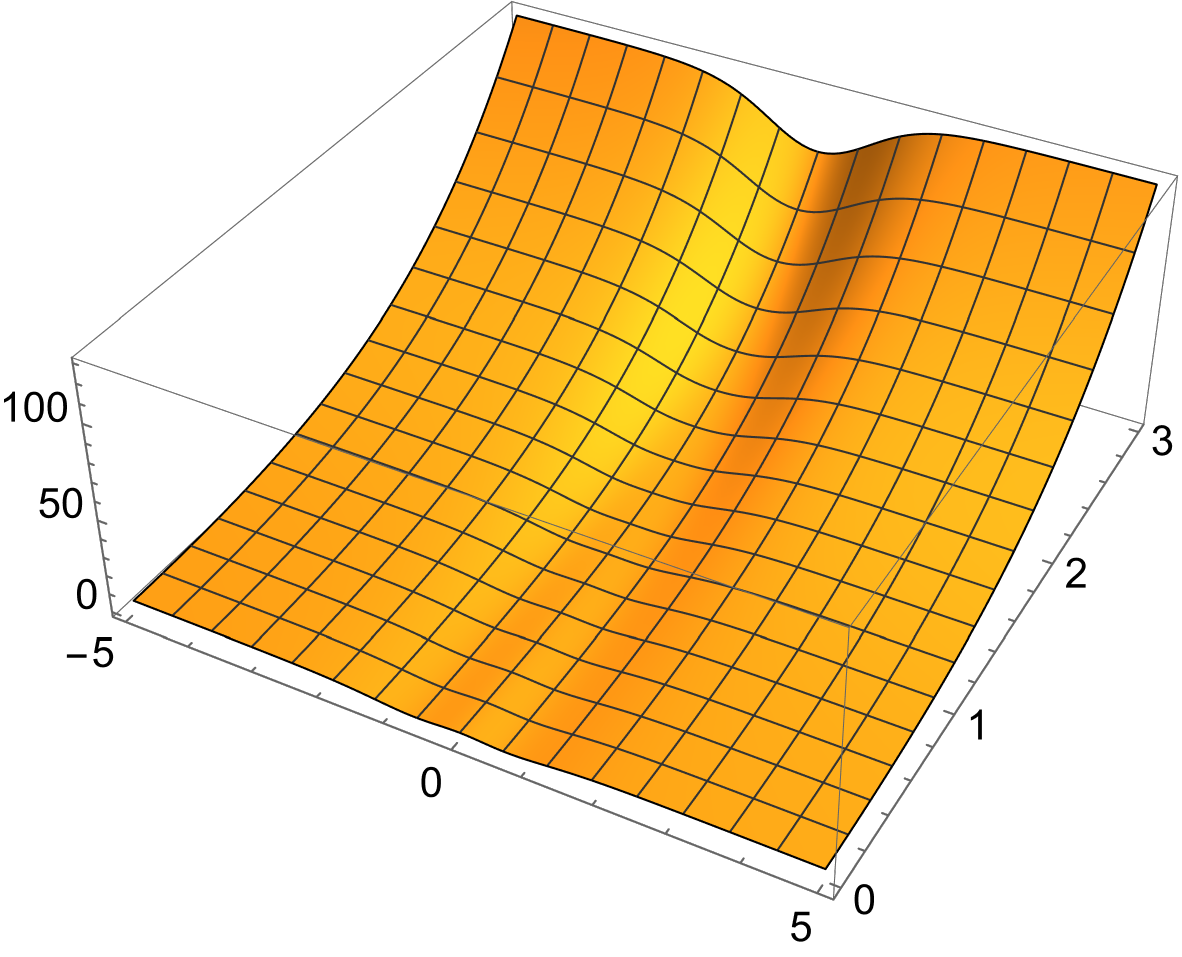}}
\caption{ 3D graphics of the lower component of scattering spinor waves as functions of $y$ and the wave number $q$. Real part (left) Imaginary part (center) Absolute value (right). Also the real and imaginary parts have different phases and the differences varies with $q$.}
\end{figure}

It remains to compute the relative density of probability of the scattering wave spinors with respect to spinor plane waves. If we denote 
as $ \Psi_0(y,q)= \sqrt{g} e^{i q y}\left(\begin{array}{c} 1 \\ 1 \end{array}\right)$ the probability density of free wave spinors, the quantity
\begin{eqnarray*}
&& \rho(y;q)-\rho_0(y;q)= \Psi^\dagger(y;q)\Psi(y;q)-\Psi_0^\dagger(y;q)\Psi_0(y;q)=\frac{1}{N^{1)}}\psi_1^*(y;q)\psi_1(y;q)+\\ && +\frac{1}{N^{(2)}}\psi^*_2(y;q)\psi_2(y;q)-2 \quad , \quad  N^{(1)}(q) =N^{(2)}(q)(9+q^2) \, \, , \quad  N^{(2)}(q)= (1+q^2)(4+q^2)
\end{eqnarray*}
measures the density of probability of spinor waves scattered off the Kink with respect to that of spinor waves propagating in vacuum.
Integration of the relative density of probability in a finite but very long interval of length $L$ gives the relative probability

\begin{eqnarray*}
&& P(L;q)-P_0(L;q) = \int_{-L/2}^{L/2} \, dy \, (\rho(y;q)-\rho_0(y;q)) =\\ && =\frac{240 \left(q^2+5\right)
   \sinh
   ^4\left(\frac{L}{2}\right)
   \text{csch}^3(L)-6 \tanh
   \left(\frac{L}{2}\right)
   \left(15
   \text{sech}^4\left(\frac{L}
   {2}\right)+2 q^4+16
   q^2+22\right)}{\left(q^2+1
   \right) \left(q^2+4\right)
   \left(q^2+9\right)}\\ &&-\frac{6 \tanh
   \left(\frac{L}{2}\right)
   \left(\tanh
   ^2\left(\frac{L}{2}\right)+
   q^2+1\right)}{\left(q^2+1
   \right) \left(q^2+4\right)}
\end{eqnarray*}
The limit of infinite length is given by the derivative of the total phase shift with respect to the wave number
\[
P(\infty; q)-P_0(\infty;q)=-2 \frac{d}{d q}\left(\delta^{(1)}(q)+\delta^{(2)}(q)\right))=-\frac{d }{d q}\left(\arctan (\frac{q}{3})+2 \arctan (\frac{q}{2})+\arctan (q)\right)
\]
such that the \lq\lq sum\rq\rq over all the momenta up to a cutoff is
\[
\Delta P(\Lambda)=\int_0^\Lambda \, dq \, \left(P(\infty;q)-P_0(\infty;q)\right)=-2\left(\arctan (\frac{\Lambda}{3})+2 \arctan (\frac{\Lambda}{2})+ 2 \arctan (\Lambda) \right)
\]
and pushing the cutoff to infinity we obtain: $\Delta P(\infty)=-5 \pi$.

\begin{figure}[ht]
\centerline{
\includegraphics[height=4.6cm]{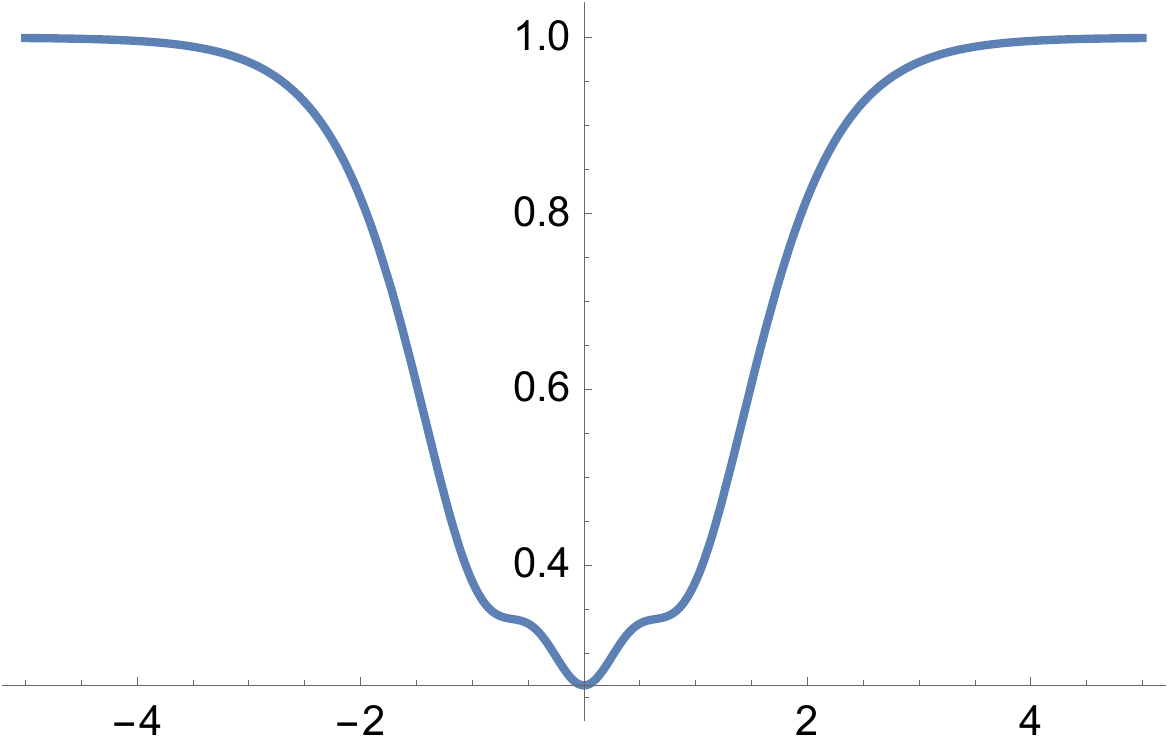} \hspace{0.5cm}\includegraphics[height=5.2cm]{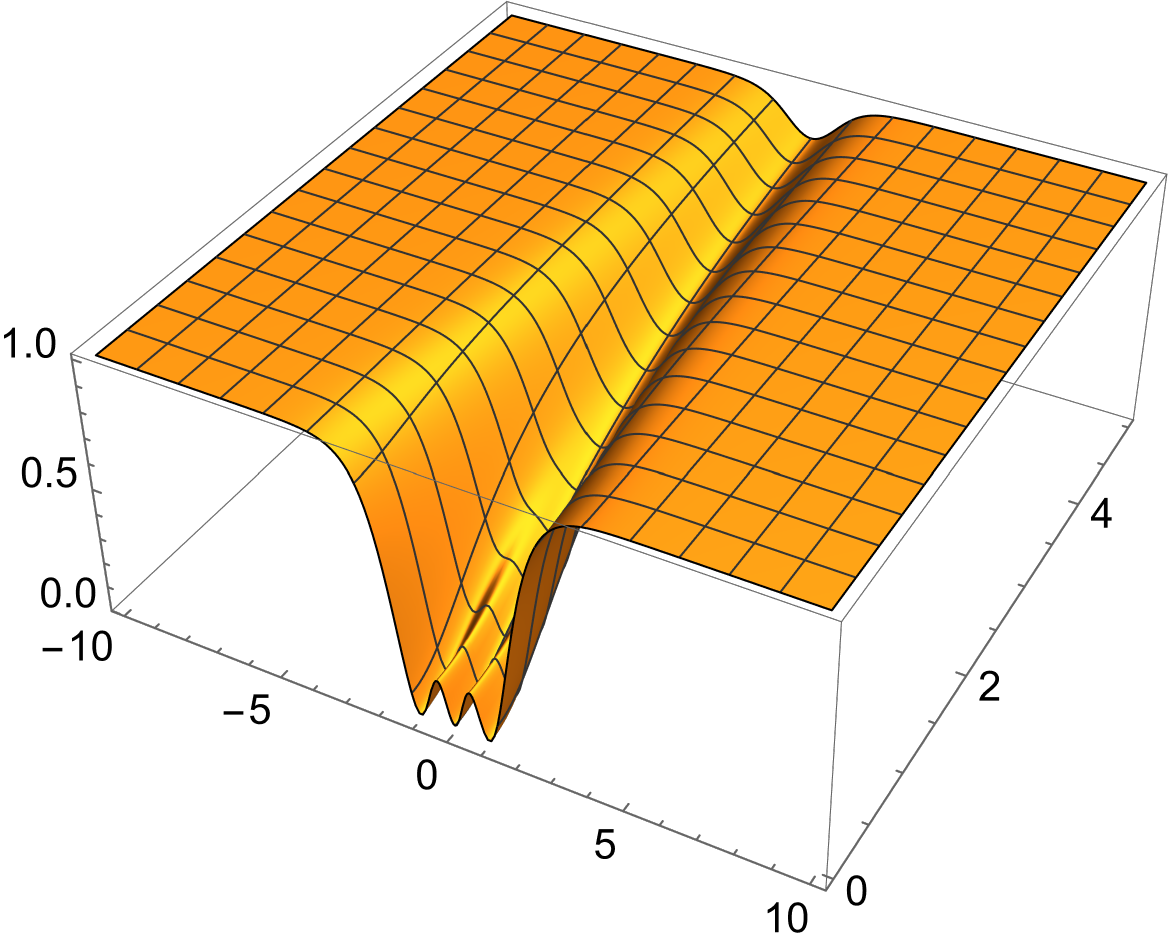} }
\caption{ Total density of probability of scattering spinor waves for $q=1$ (left) 3D plot of the total scattering density probability as function of $y$ and $q$ (right). Note that the probability density of spinor waves is depleted around the Kink center.}
\end{figure}

\end{enumerate}

\subsection{Fermionic quanta emmerging from spinor Kink fluctuations}
To finish this Section we expand first
the classical spinor field in terms of the one-particle states:
\begin{eqnarray*}
\Sigma(y,\tau) &=& B_0 \left( \begin{array}{c}{\rm sech}^N y \\ 0  \end{array} \right)+ \sum_{l=1}^{N-1} \, \Big[ B_l  \left(\begin{array}{c} \psi_1^{(l)}(y) \\ \psi_2^{(l)}(y) \end{array}\right)  e^{-i \Omega_F^{(l)} g t}+C_l^*\left(\begin{array}{c} \phi_1^{*(l)}(y) \\ \phi_2^{*(l)}(y) \end{array}\right)  e^{i \Omega_F^{(l)} g t} \Big] \\ &+& \int \frac{dq}{\sqrt{4\pi \Omega_F(q)}}\left[ B(q) \left(\begin{array}{c} \psi_1(y;q) \\ \psi_2(y;q) \end{array}\right)  e^{-i \Omega_F(q) g t}+  C^*(q) 
\left(\begin{array}{c} \phi^*_1(y;q) \\ \phi^*_2(y;q) \end{array}\right)  e^{i \Omega_F(q) g t} \right] 
\end{eqnarray*}
We stress that the eigenspinors $\psi$ of the Dirac-Kink operator and of its $g$ to $-g$ transformed $\phi$ have been taken as a complete system in the space of spinor fields.

 The next step is the promotion of the coefficients to Fermi operators demanding  anticommutation rules between them to establish the canonical quantization procedure:
\begin{eqnarray*}
&& \{\hat{B}_0, \hat{B}_0^\dagger\}=1 \qquad , \qquad \{ \hat{B}_{l_1}, \hat{B}^\dagger_{l_2}\}=\delta_{l_1 l_2} \qquad , \qquad \{ \hat{C}_{l_1}, \hat{C}^\dagger_{l_2}\}=\delta_{l_1 l_2} \\ && \{\hat{B}(q_1), \hat{B}^\dagger(q_2)\}=  \delta(q_1-q_2)= \{\hat{C}(q_1), \hat{C}^\dagger(q_2)\} \, 
\, , \, \, \, 
\{ \hat{\Sigma}(y_1,\tau), i \hat{\Sigma}^{\dagger}(y_2,\tau)\}=i  \delta(y_1-y_2)
\end{eqnarray*}
and all the remaining anti-commutators are set to zero.
The Fermi Kink ground state and in general the Fermionic Fock space follows describing electron/positron multiparticle states with Fermi statistics built in:
\begin{eqnarray*}
&& {}_F\langle K\vert \hat{B}_0 \vert K \rangle_F =0 \quad , \quad {}_F\langle K\vert \hat{B}_l \vert K \rangle_F =0= {}_F\langle K\vert \hat{C}_l \vert K \rangle_F  \, \, \, , \, \, \, \forall l=1,2,\cdots, N \\ && \hspace{4cm} {}_F\langle K\vert \hat{B}(q) \vert K \rangle_F =0= {}_F\langle K\vert \hat{C}(q) \vert K \rangle_F  \, \, \, , \, \, \, \forall q\in \mathbb{R}
\\ && \prod_{a=1}^{n_a^-}\hat{B}^\dagger_{l_a}\prod_{b=1}^{n_b^+} \hat{C}^\dagger_{l_b}\prod_{c=1}^{n_b^-} \hat{B}^\dagger(q_c)\prod_{d=1}^{n_d^+} \hat{C}^\dagger(q_d)\vert K \rangle_F= \vert \prod_{a=1}^{n_a^-} 1_{l_a}^- \prod_{b=1}^{n_b^+} 1_{l_b}^+  \prod_{c=1}^{n_c^-} 1^-(q_c)  \prod_{d=1}^{n_d^+} 1^+(q_d); K\rangle_F
\end{eqnarray*}

 As a practical computation we show the normal ordered Fermi number operator, all the annihilation operators placed at the right of the creation operators denoted by the $:\hat{F}:$ symbol.
\begin{eqnarray*}
:\hat{F}:&=& \int \, dy \, \frac{1}{2}\left[\hat{\Sigma}^{\dagger}(y,\tau), \hat{\Sigma}(y,\tau)\right] \\ 
: \hat{F} : &=& \frac{1}{2}[\hat{B}_0^\dagger, \hat{B}_0]+\sum_{l=1}^{N-1} (\hat{B}_l^\dagger \hat{B}_l-\hat{C}_l^\dagger\hat{C}_l)+\int dq \rho_F(q) (\hat{B}^\dagger(q)\hat{B}(q)-\hat{C}^\dagger(q)\hat{C}(q))\\ &=&\hat{N}_0-\frac{1}{2}+\sum_{l=1}^{N_1} (\hat{N}_l^--\hat{N}_l^+)+\int d q \rho_F(q) (\hat{N}^-(q)-\hat{N}^+(q))
\end{eqnarray*}
Due to the unpaired zero mode we see that the expectation value of this operator in any state of the Fermionic Fock space is fractionary. There is a one-to-one correspondence  electron and positron spinors:
\[
\sigma_1 \left(\begin{array}{c} \psi_1(y;q) \\ \psi_2(y;q) \end{array}\right)=\left(\begin{array}{c} \phi_1(y;q) \\ \phi_2(y;q) \end{array}\right) \quad {\rm if} \, \, \, \Omega_F >0
\]
Therefore because the $\sigma_1$ matrix maps the positive eigenspinors of $h_{DK}(g)$ into those of $h_{DK}(-g)$, not only the states in the discrete spectra are paired (except the zero mode) but also the spectral densities are identical in the continuous spectra.

The most important observable however is the quantum energy coming from the spinor Kink fluctuations
\[
\hat{K}_D= g \int \,dy \, \, \hat{\Sigma}^\dagger(y,\tau) i \frac{\partial}{\partial \tau} \hat{\Sigma}(y,\tau)  \, \, .
\]
Due to the orthogonality and completeness of he one-particle spinor waves and bound states we find
\begin{eqnarray}
\hat{K}_D &=& g \Big( \sum_{l=1}^N \Omega_F^{(l)} \left(\hat{B}_l^\dagger \hat{B}_l - \hat{C}_l \hat{C}_l
^\dagger\right)+ \int \, dq \, \Omega_F(q) \rho_F(q)\left(\hat{B}^\dagger(q)\hat{B}{q}-\hat{C}(q)\hat{C}^\dagger(q)\right)\Big)\nonumber \\ &=&  g \left( \sum_{l=1}^N \Omega_F^{(l)} \left(\hat{B}_l^\dagger \hat{B}_l + \hat{C}_l^\dagger \hat{C}_l
\right)+ \int \, dq \, \Omega_F(q)\rho_F(q) \left(\hat{B}^\dagger(q)\hat{B}{q}+\hat{C}^\dagger(q)\hat{C}(q)\right)\right)\nonumber \\ &-& g \left( \sum_{l=1}^N \Omega_F^{(l)}\delta_{ll} + \int dq \, \Omega_F(q) \rho_F(q) \lq\lq \delta"(0)\right) \, \, .  \label{fkcc} 
\end{eqnarray}

\section{ Bose/Fermi one-loop Kink mass shifts.}
We enter now in the central task in this paper. Since that we are going to compute the one-loop kink mass shifts coming either from Bose or from Fermi quanta it is convenient to resurface $\hbar$ in the formalism. In the system of units where only the speed of light in vacuum is set to one, $c=1$, we write the Jackiw-Rebbi action in the form:
\begin{equation}
 S[\Phi, \Psi] = \int_{\mathbb{R}^{1,1}} dt dx \left\{\frac{1}{2}\left(\frac{\partial \Phi}{\partial t}\right)^2-\frac{1}{2}\left(\frac{\partial \Phi}{\partial x}\right)^2-\frac{\lambda^2}{2}\left(\Phi^2-v^2\right)^2-i \bar{\Psi}\left(\beta \frac{\partial}{\partial t}+\alpha\frac{\partial}{\partial x}\right)\Psi-g \bar{\Psi}\Phi\Psi \right\} \, \, . \label{JRactfd}
\end{equation}
The physical dimension of the action is thus $[S]=M L$ whereas the dimensions of fields and couplings are:
\[
[\Phi^2]=[v^2]=ML \quad , \quad [\bar{\Psi}\Psi]=M \quad , \quad [\lambda]=[g]=M^{-\frac{1}{2}}L^{-\frac{3}{2}} \, \, .
\]
In terms of the non dimensional fields, coupling and coordinates 
\[
\phi=\frac{1}{v} \Phi\quad , \quad \psi=\frac{1}{\sqrt{\lambda v^3}}\Psi, \quad , \quad \nu=\frac{g}{\lambda} \quad , \quad \tau=(\lambda v) t \quad , \quad y=(\lambda v) x 
\]
the classical Lagrangian density of the JR model reads:

\[
\frac{{\cal L}}{\lambda^2 v^4}= \left\{\frac{1}{2} \left(\frac{\partial \phi}{\partial \tau}\right)^2-\frac{1}{2} \left(\frac{\partial \phi}{\partial y}\right)^2-\frac{1}{2}(\phi^2-1)^2 +i \bar{\psi}\left(\beta\frac{\partial}{\partial \tau}+\alpha\frac{\partial}{\partial y}\right) \psi-\nu \bar{\psi} \phi \psi  \right\}
\]
We perform next the $\hbar$-expansion: 
\begin{enumerate}
\item Firstly, in the \underline{vacuum sector}
\[
\phi(\tau, y)= 1+ \frac{\sqrt{\hbar}}{v} H(\tau,y) \quad , \quad \psi(\tau,y)=\frac{\sqrt{\hbar}}{v} \chi(\tau,y)  \, \, .
\]
In terms of the Higgs field $H(\tau, y)$ the Lagrangian density becomes
\begin{eqnarray}
&& \hspace{-1cm}\frac{{\cal L}}{\lambda^2 v^4}=\frac{\hbar}{2 v^2} \Big( H(\tau,y)\left(-\frac{\partial^2}{\partial \tau^2}+
\frac{\partial^2}{\partial y^2}\right) H(\tau,y)+4 H^2(\tau,y)\Big)+ 2 \frac{\sqrt{\hbar^3}}{v^3} H^3(\tau,y)+ \frac{1}{2}\frac{\hbar^2}{v^4} H^4(\tau,y)\nonumber \\ && \hspace{-0.8cm} +i \frac{\hbar}{v^2}\Big(\bar{\chi}(\tau,y) \left(\beta \frac{\partial}{\partial \tau}+\beta\alpha\frac{\partial}{\partial y}\right)\chi(\tau,y)-\nu \bar{\chi}(\tau,y)\chi(\tau,y) \Big)-\nu \frac{\sqrt{\hbar^3}}{v^3}\bar{\chi}(\tau,y)H(\tau,y)\chi(\tau,y) \label{hexpvac}
\end{eqnarray}
The associated non-linear Euler-Lagrange equations become
\begin{eqnarray}
&& \hspace{-1.4cm}\frac{\hbar}{v^2}\Big[\left( \frac{\partial^2}{\partial \tau^2}-\frac{\partial^2}{\partial y^2}+4 \right)H(\tau,y)+\frac{6\sqrt{\hbar}}{v}H^2(\tau, y)+\frac{\hbar}{v^2} H^3(\tau,y)- \nu \frac{\sqrt{ \hbar}}{v}\bar{\chi}(\tau,y,y)\chi(\tau,y)\Big]=0 \label{elwkb1} \\ && \hspace{-1.2cm} \frac{\hbar}{v^2}\Big[\left(i\beta\frac{\partial}{\partial\tau}+i\beta\alpha\frac{\partial}{\partial y}-\nu\right)\chi(\tau,y)+\nu \frac{\sqrt{\hbar}}{v}\chi(\tau,y) H(\tau,y)\Big]=0 \label{elwkb2} 
\end{eqnarray}

Look now at the field equation (\ref{elwkb1}) (for $\chi=0)$ and select the Higgs field as:
\begin{equation}
H(\tau,y)= A(\tau) \sqrt{\frac{3}{2}}\frac{\sinh y}{\cosh^2y}+\rho(\tau,y) \, \,  \label{shscat}
\end{equation}
Here $A(\tau)$ is the time-dependent amplitude of the bosonic shape mode $f_{\sqrt{3}}(y)$ and $\rho(\tau,y)$ collects the bosonic .radiation (scattering) states. Entering this expression in the field equations we find at first-order
the following equation linking the bosonic vibrational kink mode with the bose scattering modes:
\[
f_{\sqrt{3}}(y) (\ddot{A}(\tau)+3 A(\tau))+\ddot{\rho}(\tau,y)-\rho^{\prime\prime}(y)- \frac{3}{\cosh^2 y}\rho(\tau,y)=-6 \tanh y A^2(\tau) f^2_{\sqrt{3}}(y) \, \, .
\]
In references \cite{Merabet} and \cite{Blanco} it is shown by solving this equation that the bosonic shape mode decays by emmiting
bosonic radiation with the funny peculiarity that the scattered bosons off the kink have twice the frequency of the shape mode. A finer analysis requires adiabatic switching of the interactions such that the asymptotic quantum states \lq\lq in\rq\rq and \lq\lq out\rq\rq might be well defined. A similar strategy to estimate the possible emission of fermions from the bosonic shape mode starts with the plugging of the Higgs configuration (\ref{shscat}) in equation (\ref{elwkb2}). In this problem the analysis is much more delicate. In (\cite{Blanco}) is explained how a Bogoliubov-Valatin transformation is needed to identify the spaces \lq\lq in\rq\rq and \lq\lq out\rq\rq of initial and final fermions spaces in the kink background in such a way that conservation of the Fermi number is ensured in the process of fermionic emission.  Also powerful numerical methods are developed to compute the quantum emission of fermions. {\footnote{Once that fermions enter in the Kink bachground not only fermionic radiation is induced from the fermionic shape mode but also bosonic radiation arises from vibrational fermi modes of kink fluctuations. Plugging a fermionic shape mode in the term $\bar{\chi}\chi$ of equation (\ref{elwkb1}) bosonic emission of radiation emerges.}}

From the QFT perspective, the conventional wisdom from the diagrammatica representation of quantum perturbation theory reveals us that there is the boson propagator plus bosonic tri-valent and four-valent vertices respectively at order $\frac{\hbar}{v^2}$, $ \frac{\sqrt{\hbar^3}}{v^3}$ and $\frac{\hbar^2}{v^4}$ in the WKB expansion. The fermion propagator as well as one bose-fermi trivalent vertex also appear at the appropriate order in the expansion. The Bose mass $m^2_B=4 \lambda^2 v^2$ and the Fermi mass $m_F=g v$ also 
enter in the formula (\ref{hexpvac}).

\item Secondly, we move to the \underline{kink sector}.
\[
\phi(\tau, y)= \tanh y + \frac{\sqrt{\hbar}}{v} H(\tau,y) \quad , \quad \psi(\tau,y)=\frac{\sqrt{\hbar}}{v} \chi(\tau,y)  \, \, .
\]
The Jackiw-Rebbi action in the kink sector after the shifting of the kink fluctuations becomes
\begin{eqnarray}
&&\ S_K^{JR}[H, \chi]=\int \, d\tau dy \,\, \frac{{\cal L}}{\lambda^2 v^4}= \frac{1}{2}\lim_{T\to \infty } \int_{-T/2}^{T/2} d\tau \int_{-\infty}^\infty dy \, \frac{1}{\cosh^4 y} + \nonumber\\ && + \frac{\hbar}{2 v^2} \int _{-\infty}^\infty \, d\tau \int_{-\infty}^\infty \, dy \,\left\{ H(\tau,y) \left(-\frac{\partial^2}{\partial \tau^2}+\frac{\partial^2}{\partial y^2}\right) H(\tau, y) - \left(2- \frac{3}{\cosh^2 y}\right)H^2(\tau,y)\right\}+\nonumber \\ &&  -\int _{-\infty}^\infty \, d\tau \int_{-\infty}^\infty \, dy \,\left\{  \frac{2\sqrt{h^3}}{v^3}\tanh y \, H^3(\tau,y)+\frac{ \hbar^2}{2 v^4} H^4(\tau,y)\right\}+\nonumber\\ && + i \frac{\hbar}{v^2} \int _{-\infty}^\infty \, d\tau \int_{-\infty}^\infty \, dy \,\left\{ \bar{\chi}(\tau,y)\left(\beta\frac{\partial}{\partial \tau}+\beta\alpha\frac{\partial}{\partial y}\right)\chi(\tau,y)-\nu \bar{\chi}(\tau,y) (\tanh y) \, \chi(\tau,y) -\right. \nonumber\\ && -\left. \nu \frac{\sqrt{\hbar}}{v} \bar{\chi}(\tau,y) H(\tau,y) \chi(\tau,y)\right\} \label{hbarJR}
\end{eqnarray}

The subsequent non-linear field PDE's read as follows:
\begin{eqnarray}
 && \frac{\hbar}{v^2}\left[ \left(\frac{\partial^2}{\partial \tau^2}-\frac{\partial^2}{\partial y^2}+4 -\frac{6}{\cosh^2 y}\right)H(\tau,y)  \nonumber \right. + \\  && \hspace{2.cm} +
 \left. \frac{6 \sqrt{\hbar}}{v} \tanh y \, H^2(\tau,y)+ \frac{\hbar}{v^2}H^3(\tau,y)-\nu \bar{\chi}(\tau,y) \chi(\tau,y) \right]=0 \label{elkwkb1} \\ && \frac{\hbar}{v^2} \left[i\left(\beta\frac{\partial}{\partial \tau}+
 \beta\alpha \frac{\partial}{\partial y}\right) \chi(\tau,y) - \nu \tanh y \, \chi (\tau,y)-\nu \frac{\sqrt{\hbar}}{v} H(\tau,y) \chi(\tau,y)\right]=0 \label{elkwkb2}
\end{eqnarray}

In the $\hbar$-expansion  (\ref{hbarJR}) of the JR action the parameter $\bar{h}=\frac{\hbar}{v^2}$ measures the Planck constant with respect to the natural (physical) action of the system $v^2$. If the vacuum expectation value is large, $v^2 >> \hbar$, the parameter of the asymptotic WKB expansion $\bar{h}=\frac{\hbar}{v^2}$ is small. Because the WKB expansion is an asymptotic series, a large $v$ offers a very good approximation at low orders. Therefore, few terms accurately approximate the JR action in the semi-classical regime. We observe that the term independent of $\hbar$ is the classical kink energy (divergent in infinite time). There are two kinds of terms proportional to $\frac{\hbar}{v^2}$ 
governing respectively the propagations of bosonic and fermionic fluctuations over the kink background. In the bosonic case a variable mass arises as a P$\ddot{\rm o}$sch-Teller well with a threshold of the continuous spectrum which is precisely the mass of the freely propagating bosons. There are two bound states: the lower bound with zero eigenvalue, corresponding to a boson travelling without cost of energy with the kink center of mass. The second bound has positive eigenvalue but lesser than the scattering threshold. This state correspond to a boson trapped by the kink that vibrates and changes its shape with respect to the kink at rest. A variable mass also influence the propagation of fermionic fluctuations. Because $m_F(y) = g v \tanh y$   bound states of fermions trapped by the kink may arise, besides the scattering of fermions off the kink background, recall sub-Section \S.3.2. There are also two terms, one bosonic and one fermionic, proportional to $\bar{h}^{3/2}=\frac{\sqrt{\hbar^3}}{v^3}$  giving rise to two trivalent vertices. The first one is purely bosonic but point dependent. In fact is attractive over the left and repulsive on the right half-lines or viceversa. The other trivalent vertex incorporates two fermions and one boson and it is standard, i.e., its strength is identical along the whole real line. Finally, a fourvalent purely bosonic vertex arise in the order $\frac{\hbar^2}{v^4}$ which is also standard.

\end{enumerate}

\subsection{One-loop kink mass shifts due to bosonic fluctuations}
Before of describing the novelties in the computation of fermionic one-loop kink mass shifts in the JR model, we summarize the computation of the bosonic one-loop kink mass shifts. Notice that we only needs the terms proportional to $\frac{\hbar}{v^2}$
in the expansions (\ref{hexpvac}) and (\ref{hbarJR}) for the calculation of this magnitude at one-loop.

Standard lore in the computation of the one-loop shift to the Kink classical energy in the pure $\lambda \phi^4$ model is based in the \underline{Dashen-Hasslacher-Neveu paradigm}, see \cite{DHN}. Reminding that $m_B=2\lambda v^2$, the formula for the one-loop Kink mass reads, 
\begin{equation}
\Delta M_B^{\rm DHN}=- \hbar m_B \left[\frac{\sqrt{3}}{2}+\frac{1}{2\pi}\int_0^\infty\, dq \, (\rho(q)-l)\sqrt{q^2+4}-\frac{3}{\pi}(1-\int_0^\infty \, dq \, \frac{1}{\sqrt{q^2+4}})\right] \, \, . \label{DHNshift}
\end{equation}
see the References \cite{AIJMG}-\cite{GAGTS} and papers quoted therein.
In the derivation of formula (\ref{DHNshift}) we have assummed that the system is defined on a very long interval of (non dimensional) length $l=\lambda L$. Periodic boundary conditions are imposed on scattering waves. In formula (\ref{DHNshift}) there are three ingredients:
\begin{enumerate}

\item \underline{Kink Casimir Energy}. The divergence arising from all the Bosonic quantum states being unoccupied, both in the discrete and continuous
spectra, is partially tammed by subtracting the analogous divergence due to the absence of occupied  quanta in the vacuum.
The energy of the state where one quanta is trapped by the kink is $\sqrt{3}$ and the contribution of scattering states is expressed in terms of the subtraction of thes pectral density of kink fluctuations minus the spectral density of ground state fluctuations:
\[
\rho(q)= \frac{l}{2\pi}+\frac{1}{2\pi}\frac{\partial \delta}{\partial q} \, \quad , \quad \, \rho(q)-\rho_0(q)=\frac{1}{2\pi}\frac{\partial \delta}{\partial q} 
\]
which in turn is determined from the total phase shift $\delta(q)$ induced by the kink on the scalar fluctuations. After this zero point renormalization still a logarithmic diverence is left.

\item \underline{Mass renormalization}. Cure of this divergence is achieved by standard normal ordering 
procedures that introduce the one-lopp mass renormalization counter-term:
\[
\frac{3}{\pi}\int_0^\infty \, dq \, \frac{1}{\sqrt{q^2+4} } \,\, .
\]
To understand how this integral arises we first recall that the Feynman propagator is the difference of the Wick chronological product and the normal product of two field operators defined at different points in space-time:
\begin{eqnarray}
T(\hat{\phi}(x^\mu)\hat{\phi}(y^\mu))&=& : \hat{\phi}(x^\mu)\hat{\phi}(y^\mu) : +\Delta_{FB}(x^\mu-y^\mu) \label{chronp}\\ \Delta_{FB}(x^\mu-y^\mu)&=& -\frac{i}{(2\pi)^2} \int \int_{\tilde{\mathbb{R}}^{1,1}}dq_0dq_1 \frac{e^{iq_\mu(x^\mu-y^\mu)}}{q_\mu q^\mu-4+i \varepsilon} \label{feyprop}
\end{eqnarray} 
The chronological product is defined as follows:
\[
T(\hat{\phi}(x^\mu)\hat{\phi}(y^\mu))=\left\{\begin{array}{c} \hat{\phi}(x^\mu)\hat{\phi}(y^\mu) \, \, \, {\rm if} \, \, \, x^0 > y^0 \\ \hat{\phi}(y^\mu)\hat{\phi} (x^\mu) \, \, \, {\rm if} \, \, \, x^0 < y^0 \end{array}\right.
\]
The double colon in $:\hat{A}(\hat{\phi}):$ denotes that the operator $\hat{A}$ is normal ordered. This means that being $\hat{\phi}$ a linear superposition of creation and annilation operators, in $\hat{A}$ all the annihilation operators are on the right of all the creation operators. This prescription forbids spurious ultraviolet divergent processes.

The most dangerous sigularities occur in products of fields  defined at the same point in Minkowski space. The basic structure is the product of coinciding two-fields. In this case formulas (\ref{chronp}) and (\ref{feyprop}) become 
\[
  \hat{\phi}^2(x^\mu)= : \hat{\phi}^2(x^\mu): +\Delta_{FB} (0) \qquad , \qquad \Delta_{FB}(0)=\frac{1}{4 \pi} \int_{\mathbb{R}} dq_1 \frac{dq_1}{\sqrt{q_1^2+4}}
\]
because the notion of time-ordering is lost.
To proceed from the normal ordering of $\hat{\phi}(x^\mu)\hat{\phi}(x^\mu)$ to the normal ordering of any functional of the quantum fields the procedure is provided by the Wick's Theorem, which expresses any functional of the quantum fields as a sum from zero to all possible \lq\lq contractions\rq\rq  of two fields. Applied to the energy density Wick's theorem offers the normal ordered functional as the formal series:
\begin{eqnarray}
 U(\hat{\phi}(y^\mu)) & = &   : \left(  {\rm exp}[\frac{\hbar m_B}{2} \Delta_{FB}(0)\frac{\delta^2}{\delta \phi^2}]U(\hat{\phi}(y^\mu) \right) : \nonumber \\ &\simeq &  :U(\hat{\phi}(y^\mu)):+\frac{\hbar m_B}{2} \Delta_{FB}(0):\frac{\delta^2 U}{\delta\phi^2}(\hat{\phi}(y^\mu)):+ {\cal O}(\hbar^2) \, \, , \, \, \, m_B=2\lambda v
 \label{wick}
 \end{eqnarray}
In  formula (\ref{wick}) we observe that stopping at one-loop order there is no need of going farther in the asymptotic $\hbar$-expansion. Why only the quanta propagating over the vacuum are needed in this analysis? The reason is that both the vacuum and the kink are eigenstates not of the number operators but of the annihilatio operators, i.e., coherent states like those of displaced harmonic oscillators. Moreover,
expectation values of normal ordered operators at coherent states, see Reference \cite{Cahill}, reproduce the classical profiles:
\begin{eqnarray*}
&& \hat{\phi}(y^\mu) \vert V\rangle =  \vert V \rangle \qquad , \qquad \hat{\phi}(y^\mu) \vert K \rangle = \phi_K(y) \vert K \rangle \\ && \langle V \vert :F(\hat{\phi}(y^\mu): \vert V \rangle =F(1) \quad , \quad \langle K \vert :F(\hat{\phi}(y^\mu): \vert K \rangle =F(\phi_K(y)) \\ && \langle V \vert : \frac{\delta^2 U}{\delta \phi^2}(\hat{\phi}):\vert V \rangle= \frac{\delta^2 U}{\delta \phi^2}(1)=4 \quad , \quad \langle K \vert : \frac{\delta^2 U}{\delta \phi^2}(\hat{\phi}):\vert K \rangle= \frac{\delta^2 U}{\delta \phi^2}(\phi_K(y))=4-\frac{6}{\cosh^2 y}
\end{eqnarray*}

Therefore, from formula (\ref{wick}) we read the contribution to the one-loop shift of the Kink mass from the mass renormalization counterterms

\begin{eqnarray*}
 \Delta M_R &=& \frac{\hbar m_B}{2}\Delta_{FB}(0) \int_{-\infty}^\infty \, dy \, \left(\frac{\delta^2 U}{\delta \phi^2}(\phi_K)-\frac{\delta^2 U}{\delta \phi^2}(1) \right)\\ &=& \frac{\hbar m_B}{4\pi}\int_{0}^\infty \, \frac{dq}{\sqrt{q^2+4}}\langle V(y) \rangle = -\frac{\hbar m_B}{4\pi}\int_{0}^\infty \, \frac{dq}{\sqrt{q^2+4}} \int_{-\infty}^\infty \, dy \, \frac{6}{\cosh^2 y}\\ && = -\frac{3 \hbar m_B}{\pi}\int_{0}^\infty \, \frac{dq}{\sqrt{q^2+4}} 
\end{eqnarray*}

\begin{figure}[ht]
\centerline{\includegraphics[height=3.5cm]{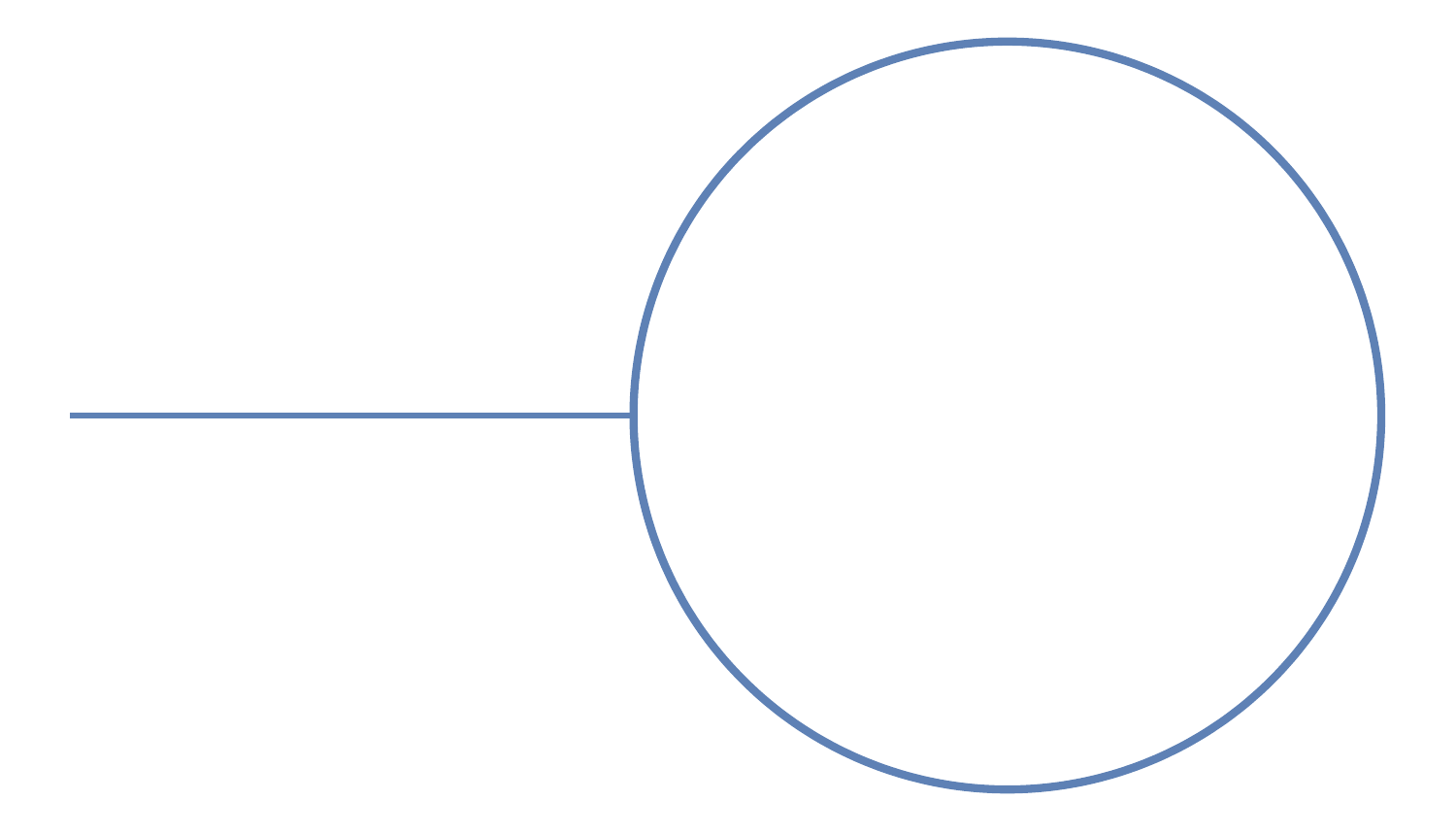}
 \hspace{2cm} \includegraphics[height=3.5cm]{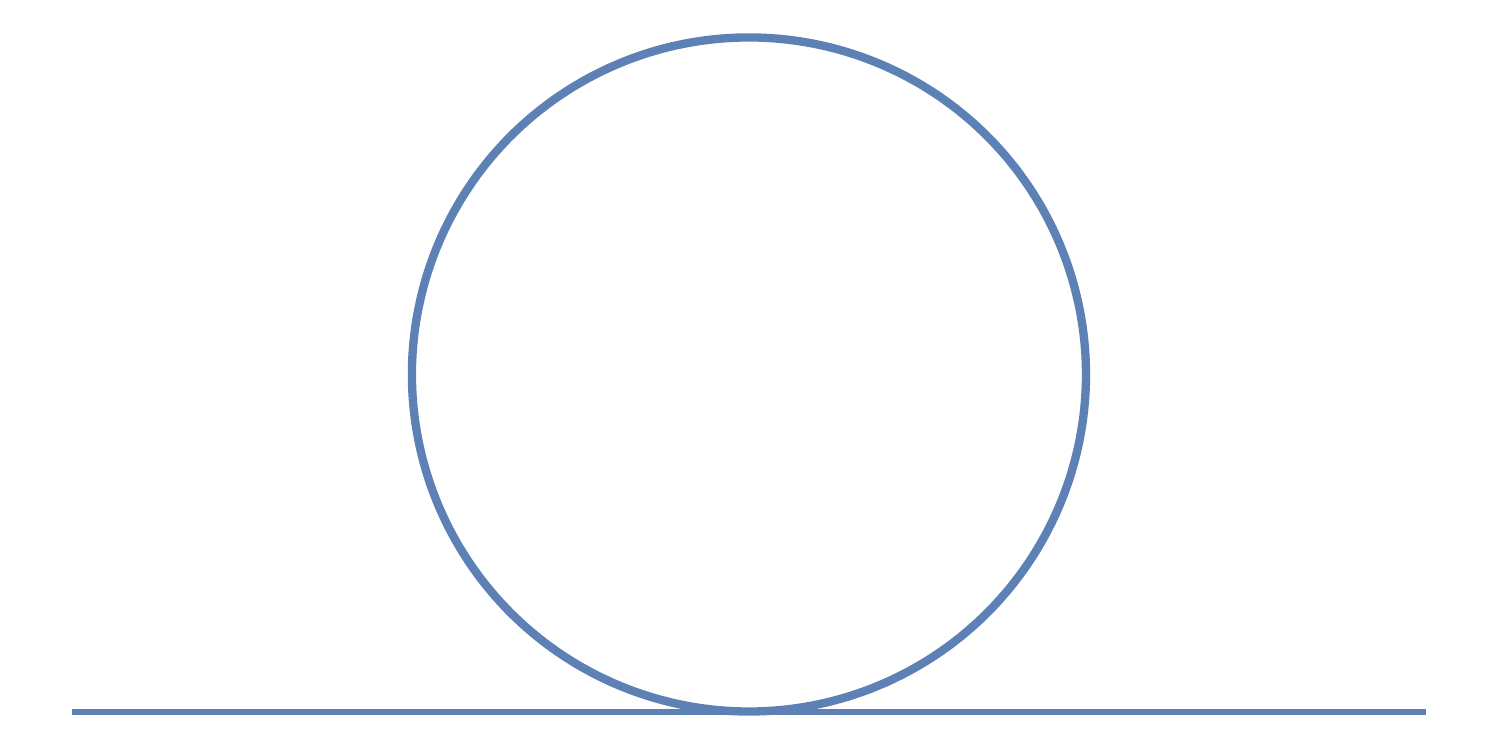}}
\caption{Divergent one-loop graphs due to creation/annihilation of bosonic quanta. The Feynman rules assign to the loops in the graphs the Green's function $\Delta_{FB}(0)$ when the momenta in the external legs are null. The factor in front comes from $-\int_{-\infty}^\infty \, dy \,\frac{6}{\cosh^2 y}=-12$}
\end{figure}

\item \underline{Mode number versus energy cutoffs}. Finally, a very subtle finite renormalization term $-\frac{3}{\pi}$ arises when one regularizes the theory, following 
DHN, by using a cutoff in the number of modes rather than in the energy. Note that the \lq\lq number\rq\rq of fluctuation modes 
in the Kink and vacuum sectors are different and this term responds to this fact. This effect is captured by the phase shift at very large momentum

\begin{eqnarray*}
&& \hspace{-1cm}\delta(q)=-i \log[\frac{(1-i q)}{(1+ i q)}\cdot\frac{(2-i q)}{(2+i q)}] \, \, ; \, \, \,  \frac{1}{2\pi} \int_0^\Lambda \, dq \, \frac{d \delta}{d q}(q)\cdot \sqrt{q^2+4} = \frac{1}{2\pi} \delta(\Lambda)\Lambda -\frac{1}{2 \pi} \int_0^\Lambda \, dq \, \frac{q \delta(q) }{\sqrt{q^2+4}} \\ && \hspace{-0.8cm} \frac{1}{2 \pi} \delta(\Lambda) \Lambda \simeq_{\Lambda\to \infty} -\frac{i}{2 \pi}\log[1+ \frac{1}{2 i \Lambda}\int_{-\infty}^\infty \, dy \, V(y)]\Lambda =-\frac{1}{4 \pi}\langle V(y)\rangle  
\end{eqnarray*}

In sum,,
\begin{eqnarray}
\Delta M_B^{{\rm DHN}} &=& - \hbar m_B\left[\frac{\sqrt{3}}{2}-\frac{3}{\pi}\left(1 + \int \, dq \, \{ \sqrt{q^2+4}\frac{2+q^2}{4+ 5 q^2+q^4}- \frac{1}{\sqrt{q^2+4}}\}\right)\right]\nonumber \\ &=&  \hbar m_B (\frac{1}{2 \sqrt{3}}-\frac{3}{\pi})=- \hbar m_B \cdot 0.666265  \quad , \quad m_B = 2 \lambda v \label{olmsb}
\end{eqnarray}

\item \underline{Cahill-Comtet-Glauber formula}

If the Schr$\ddot{\rm o}$dinger operator governing the one-loop kink fluctuations corresponds to reflectionless potentials, Cahill,Comtet and Glauber derived a formula allowing to express the one-loop kink mass shift in terms of the eigenvalues of the discrete spectrum only,
see \cite{CCG}. Applied to the $\lambda \phi^4$-kink, with room for two bound states, one zero and one shape modes with energies 
respectively 
$\omega_0=0$ and $\omega_1=\sqrt{3}$, the CCG formula reads:
\begin{equation}
\Delta M_B^{{\rm CCG}}=-2\hbar \lambda v \left(\frac{2 (\sin \theta_0 -\theta_0 \cos \theta_0)+2(\sin \theta_1 -\theta_1 \cos \theta_1) }{\pi}\right)
\end{equation}
where $\theta_0={\rm arccos} \frac{\omega_0}{2}=\frac{\pi}{2}$ and $\theta_1= {\rm arccos} \frac{\omega_1}{2}=\frac{\pi}{6}$. Note that the factor $2$ inside brackets obeys to the square root of the threshold energy of the continuum. One easily checks that 
\[
\Delta M_B^{{\rm CCG}}=- 2\hbar \lambda v (\frac{1}{2 \sqrt{3}}-\frac{3}{\pi})
\]
and the CCG, if applicable, and DHN formulae give the same answer.

\end{enumerate}

\subsection{Fermionic one-loop kink mass shifts: the Fermi Cahill-Comtet-Glauber formula}

Our goal in this sub-Section is to try the computation of one-loop kink mass shifts induced by Fermionic fluctuations. With inspiration in the paper by Casahorran, \cite{Casa}, we propose the modification of the CCG and DHN formulas needed to take into account 
the effect of Fermionic fluctuations in one-loop kink mass shifts within the Jackiw-Rebbi model. To put it in other way we are going to compute the one-loop backreaction of fermions over kinks in the JR model. We consider the case $\nu=N$, $N\in {\mathbb N}^*$, such that the two potential wells governing the up and down components of the spinor fluctuations are transparent. Thus, the Cahill-Comtet-Glauber method works and we generalize the 
\lq\lq Fermionic\rq\rq  CCG formula in the following manner
\begin{equation} 
\Delta M_F^{\rm CCG}=\hbar g v \frac{2 N}{\pi}
   \left(\frac{1}{2}+\sum _{j=1}^{N-1}\left(\sin \alpha_j -\alpha_j \cos \alpha_j \right)\right) \quad , \quad \alpha_j ={\rm arccos} \frac{\sqrt{(2N-j)j}}{N}\label{ccgf}  
\end{equation}
There are two conceptual differences with respect to the effect of one-loop Bosonic fluctuations:
\begin{enumerate}

\item In QFT is well known that closed loops of Fermions carry associated a minus sign. This is due to Fermi Statistics and related to the fact tat unoccupied fermionic states in the vacuum decrease the energy.Therefore the contribution of Fermions to one-loop kink mass shifts is positive unlike the contribution of Bosons which is negative.

\item In the Jackiw-Rebbi model the scalar field is real. Consequently, the bosonic quanta associated are their own antiparticles.   The spinor fields, however, are {\it complex} Dirac spinors encompassing two components which may fluctuate. Therefore, in the JR/QFT model the quanta coming from quantization of the spinor field are both fermions and antifermions. Moreover, there is quasi-symmetry particle-antiparticle. This symmetry is incomplete because  the zero mode may be occupied by fermions, no anti-fermions. Therefore, the fermions contribute to the one-loop 
kink mass shifts not only with different sign but also twice per massive fermionic bound state due to the antiparticle contribution.
The summand $1/2$ is due to the fact that there is only one zero mode where a fermionic particle travel with the kink. No fermionic  anti-particle travel freely with the kink. 

\end{enumerate}
Application of formula (\ref{ccgf}) allows us to compute promptly the Fermionic one-loop kink mass shifts for the lower integer 
ratios bewteen the Yukawa coupling $g$ and the scalar quartic self-coupling $\lambda$:

\begin{eqnarray*}
\Delta M_F^{\rm CCG}(1)&=&  \hbar g v \frac{1}{\pi}= 0.31831 \cdot \hbar \lambda v \\ 
\Delta M_F^{\rm CCG}(2)&=&  \hbar g  v (-\frac{1}{\sqrt{3}}+\frac{4}{\pi})= 0.695889 \cdot 2 \hbar \lambda v\\ \Delta M_F^{\rm CCG}(3)&=&\hbar g v\frac{\left(9-4
   \sqrt{2} \cos^{-1}\left(\frac{2
   \sqrt{2}}{3}\right)-2
   \sqrt{5} \sec^{-1}\left(\frac{3}{\sqrt{5}}\right)\right)}{\pi}= 1.21408\cdot 3\hbar \lambda v\\
   \Delta M_F^{\rm CCG}(4)&=& \hbar g v \left(-\frac{2}{\sqrt{3}}-\frac{2
   \left(-8+\sqrt{7} \sec^{-1}\left(\frac{4}{\sqrt{7}}\right)+\sqrt{15} \sec^{-1}\left(\frac{4}{\sqrt{1 5}}\right)\right)}{\pi }\right)= 1.88682 \cdot 4 \hbar  \lambda v
\end{eqnarray*}
It is clear that the bigger $g$ with respect to $\lambda$ the stronger is the fermionic one-loop kink mass shift with respect to the bosonic one-loop kink mass shift, because in the JR model the number of fermionic-Kink bound states grows bigger and bigger while the bosonic-Kink bound states remains fixed, always two.

\subsection{Fermionic one-loop kink mass shift: the Fermi Dashen-Hasslacher-Neveu formula}

Because the square of the Dirac operator in the kink background is a diagonal $2\times 2$-matricial differential operator of second order such that the matrix elements are transparent P$\ddot{\rm o}$sch-Teller Hamiltonians it is expected that the DHN formula may be 
applied to account for the fermionic kink fluctuations measuring the backreaction of fermions on the kink at one-loop order. In the DHN spirit the Fermionic 
contribution to the one-loop kink mas shift reads
\begin{eqnarray}
\Delta M_F^{\rm DHN}(N)&=&\hbar g v\left(- \sum_{j=1}^{N-1} \sqrt{(2N-j)j}+\frac{2}{\pi}\int_0^\infty \, dq \, \sqrt{q^2+N^2}\left(\sum_{j=1}^{N-1} \frac{j}{j^2+q^2} + \frac{1}{2} \frac{N}{N^2+q^2}\right)\nonumber \right. \\ &+& \left. \frac{N^2}{\pi}(1-\int_0^\infty \, dq \, \frac{1}{\sqrt{q^2+N^2}} ) \right)  \label{DHNF}
\end{eqnarray}
The rationale behind the different terms and their coefficients is explained as follows:
\begin{enumerate}

\item First, there is the contribution of all the positive bound states being unoccupied. With respect to the analogous contributions in the Bosonic DHN formula a factor of $2$ arises that is due to the fact that the Fermi bound state spectrum is doubled:  one bound state for the \lq\lq electron\rq\rq  and one bound state for the \lq\lq positron\rq\rq for each positive eigenvalue. The zero eigenvalue is attached only to the \lq\lq electron\rq\rq but it does not contribute to the shift at one-loop order. A global minus sign also appear here because Fermi statistics.

\item The next summand in the upper row of formula (\ref{DHNF}) responds to the fermionic kink Casimir energy. Observe that there is a factor of two and a minus sign with respect to the analogous bosonic kink Casimir energy. This magnitude collects the contribution of unnocupied scattered off the kink quantum states to the kink backreaction minus the analogous zero point energy of the vacuum.. The spectral density of the Dirac operator in the kink background with respect to the vacuum spectral density acts as integration measure of the square root of the eigenvalues in If $\nu=N$, the effective spectral density encompass the relative spectral densities of \lq\lq electrons\rq\rq and \lq\lq positrons\rq\rq scattered by the kink background,
\begin{eqnarray}
\rho_C^F(q, \slashed{\partial})&=&\rho^F_K(q)-\rho^F_V(q) \quad , \quad \rho^F_K(q)= 2(\frac{g L}{2\pi} +\frac{1}{2\pi} \frac{\partial\delta^F}{\partial q}) \quad , \quad \rho^F_V(q)= 2\frac{g L}{2\pi} \nonumber \\ \rho_C^F(q, \slashed{\partial})&=&\frac{2}{\pi}\left(\sum_{j=1}^{N-1} \frac{j}{j^2+q^2}+ \frac{1}{2}\frac{N}{N^2+q^2}\right) \label{specdenkf}
\end{eqnarray}
Here $\delta^F$ is the  phase shift triggered by the potential on the upper components of \lq\lq electron\rq\rq and \lq\lq positron\rq\rq spinors
\[
V_N^{\mp}(y)=\left(\begin{array}{cc} N^2- N(N\pm 1) \sech^2 y & 0 \\ 0 & N^2-N(N\mp 1) \sech^2 y \end{array}\right)
\]
Note that a minus sign has been introduced in the fermionic kink Casimir energy. A fine point: the family of scattering states with
spectral density $\frac{N^2}{N^2+q^2}$ only contributes to the electron kink fluctuations because for imaginary momentum $q=i N$ 
gives rise to the electron zero mode  and there is no positron zero mode.{\footnote{We ommit to consider the half bound state $j=0$ because there is also one half-bound state in the vacuum sector which cancels the contribution of its counterpart in the kink sector}}

\item There is a very subtle contribution that is due to the difference in the numbers of fluctuation modes in the kink sector with respect to the number of modes in vacuum. This correction is the term in the left of the second row in formula (\ref{DHNF}). It is known from the DHN papers that using a cutoff in the number of modes rather than a cutoff in the energy as regularization method produces a difference given by the phase shift at very high incident momentum.
In Born approximation this is captured as the Fourier transform of the potential at zero momentum transfer. In our spinor system we must count both these differences for $V_N^{11}$ and $V_{N}^{22}$ (and take the \lq\lq square\rq\rq root, the factor $1/2$ in formula (\ref{specdenkf})). Therefore, we find
\begin{equation}
-\frac{1}{4\pi}\left(\langle V_N^{11}(y) +V_{N}^{22}(y)\rangle \right)= -\frac{1}{4\pi}\int_{-\infty}^\infty 2 N^2 \sech^2 y dy=-\frac{N^2}{\pi}
\end{equation}
Again Fermi statistics requires multiply by minus $1$.

\item Sum of these three terms is still logarithmically divergent despite of having subtracted the vacuum energy. Bose quanta mass renormalization due to closed loops of fermions comes to the rescue. The (modulus of) the Feynmann amplitude for the creation of a Fermi particle/antiparticle pair from the annihilation of a Boson with zero bimomentum and the posterior reversed process reads, see Figure 8 (left), 
\begin{eqnarray*}
 \Delta_{FF}(0)&=&  - i\frac{N^2}{2}\int_{\mathbb{R}^{1,1}} \frac{dq^0 dq^1}{(2\pi)^2} {\rm tr}\left[\frac{i}{-\slashed{q}+N}\cdot \frac{i}{\slashed{q}+N}\right]\\ &=& i N^2\int_{-\infty}^\infty \frac{dq^0}{2\pi}\int_{-\infty}^\infty\frac{d q^1}{2\pi}\left[\frac{1}{q^0q^0-q^1q^1-N^2}\right] \, ,
\end{eqnarray*} 
see Figure 8 (right).
Thus, one only needs to replace $\Delta_{FB}(0)$ by $\Delta_{FF}(0)$ in formula (\ref{wick}) to account for computing the impact of Fermions in the Boson mass renormalization at one-loop, see the term in the right of the second row in formula (\ref{DHNF}).

\vspace{0.2cm}
\begin{figure}[ht]
\centerline{\includegraphics[height=2.3cm]{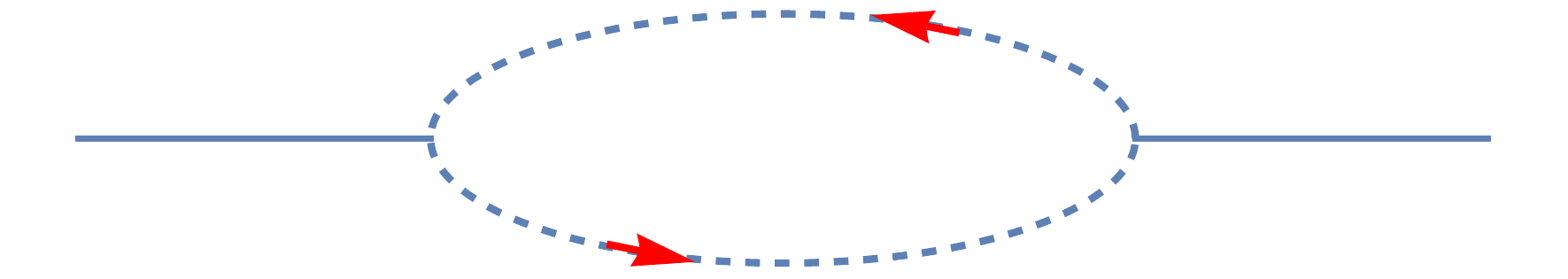}\hspace{0.22cm} 
\includegraphics[height=5.5cm]{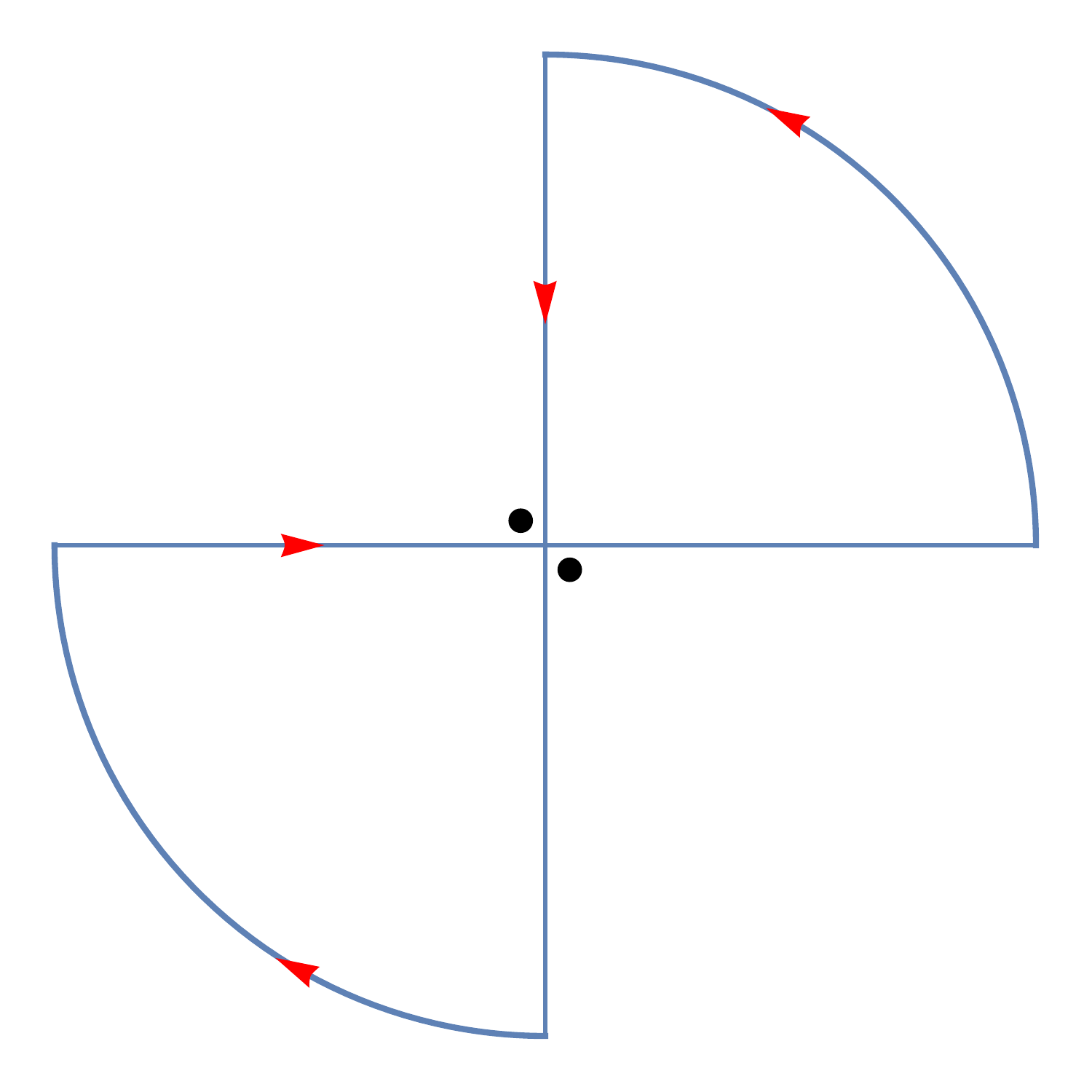}}
\caption {{} (left) \, The bosonic two-point Green's function with a closed loop of Fermions. Unlike the identical graph with a closed loop of bosons this graph diverges logarithmically in the ultraviolet and needs to be renormalized. On the contrary, tadpole graphs with Fermionic closed loops add to nothing due to the trace in the Clifford algebra. (right) \, The black dots are the poles of the integrand shifted from the abscissa axis by the standard $\pm i \varepsilon$}
\end{figure}

We perform first the integral in $q^0$:
\[
\int_{-\infty}^\infty \, \frac{d q^0}{2\pi}\cdot \frac{1}{q^0q^0-A^2} \quad , \quad A=\sqrt{q^1q^1+N^2} \, \, .
\]
The integrand presents two poles $q^0=\pm (A-i \varepsilon)$ living respectively on the fourth and second quadrants in the $q^0$
complex plane. We choose a countour along the following stages: 1) Travel along the real $q^0$ axis: ${\rm Im} q^0=0)$ 2) Follow through the arc of the circumference
$({\rm Re}q^0)^2+ ({\rm Im}q^0)^2=R^2$ in the first quadrant, i.e., start at $({\rm Re}q^0=R, {\rm Im} q^0=0)$ and ends in $({\rm Re}q^0=0, {\rm Im}q^0=R)$. 3) Run the imaginary axis, ${\rm Re}q^0 =0$ from ${\rm Im}q^0=R$ up to ${\rm Im}q^0=-R$. 4) Close the countour travelling the arc of the circumference of radius $R$ in the third quadrant.

\vspace{0.2cm}
\
There are no poles of the integrand surrounded by this integration countour. According the Cauchy theorem the integral is zero. 
In the $R\to \infty$ limit the integral along both circumference arcs is zero because the integrand decreases as $\frac{1}{q^0q^0}$.
Therefore, we conclude that 
\begin{eqnarray*}
&& \int_{-\infty}^\infty dq^0 = - \int_{i \infty}^{-i \infty} dq^0 \, =i \int_{-\infty}^\infty d Q \quad , \quad Q=i q^0 \\ && 2 \int_0^\infty dQ \, \frac{1}{Q^2+A^2}= \frac{\pi}{\sqrt{q^1q^1+N^2}}
\end{eqnarray*}
and finally,
\begin{equation}
\Delta_{FF}(0)=-\frac{N^2}{\pi} \int_0^\infty \frac{d q}{\sqrt{q^2+N^2}} \, \, .
\end{equation}
\end{enumerate}
Clearly the addition of these four terms leads to the Fermionic  DHN formula (\ref{DHNF}).

With the help of Mathematica one obtains for the $N=1,2,3,4$ cases the results:

\begin{eqnarray*}
\Delta_F^{\rm DHN}(1)&=& \hbar g v \frac{1}{\pi}\left(\lim_{\Lambda\to \infty} \arcsinh{\Lambda} +1 -\lim_{\Lambda\to \infty}\arcsinh \Lambda\right) = 0.31381 \cdot \hbar \lambda v   \\
\Delta_F^{\rm DHN}(2)&=&\left(\frac{4}{\pi}-\frac{1}{\sqrt{3}}\right) \hbar g v=0.695889\cdot 2 \hbar \,\lambda v \\ \\
  \Delta M_F^{\rm DHN}(3)&=&-\frac{\hbar g v}{\pi}\left(-9+(2
   \sqrt{2}+\sqrt{5})\pi \right. \\ 
    &-& \left. 4 \sqrt{2} \tan
   ^{-1}\left(\frac{7}{4
   \sqrt{2}}\right)-2 \sqrt{5}
   \tan ^{-1}\left(\frac{1}{4
   \sqrt{5}}\right)-2 \sqrt{5}
   \tan
   ^{-1}\left(\frac{2}{\sqrt{5
   }}\right)-4 \sqrt{2} \cot
   ^{-1}\left(2
   \sqrt{2}\right)\right)
   \\ &=& 1.21408 \cdot 3 \hbar \, \lambda v\\
 \Delta M_F^{\rm DHN}(4)&=&\frac{\hbar g v}{3} \left(-2
   \sqrt{3}-3 \sqrt{7}-3
   \sqrt{15}\right. \nonumber \\ &+& \left.\frac{6
   \left(8-\sqrt{7} \tan
   ^{-1}\left(\frac{1}{3
   \sqrt{7}}\right)+\sqrt{7}
   \tan
   ^{-1}\left(\frac{3}{\sqrt{7
   }}\right)+\sqrt{15} \tan
   ^{-1}\left(\frac{7}{\sqrt{1
   5}}\right)+\sqrt{15} \cot
   ^{-1}\left(\sqrt{15}\right)
   \right)}{\pi }\right)  \nonumber \\ &=& 1.88682 \cdot 4 \hbar \,\lambda v
\end{eqnarray*}
in perfect (astonishing) agreement with the results achieved by means of the CCG formula.

In sum, the addition of the effects of the scalar and spinor Kink fluctuations on the Kink mass, $\Delta M = \Delta M_B(2)+\Delta M_F(N)$, in the JR model.  is negative if $\frac{g}{\lambda}=1$ and positive if  $\frac{g}{\lambda}>1$. In the first case the back reaction induced by one-loop bosonic and fermionic fluctuations makes the Kink lighter and heavier in all the other cases.

\section{Feynman/Berezin functional integrals: the Jackiw-Rebbi system}
Justification of the computations in the previous  subsections may be alternatively based in Feynman's sum over histories conceptual approach 
in QFT. Because our results have been reached in the WKB approximation the Feynman quantization procedure is particularly appropriated to accomplish our goal. Applied to the Jackiw-Rebbi system Feynmann formally interpret the quantum \lq\lq partition\rq\rq function as the \lq\lq sum\rq\rq over all the Euclidean classical evolutions between identical, initial, $ \bar{\tau}=i \tau=0$, and final inverse temperature, $\bar{\tau}=i \tau\in (0,\bar{T}=\frac{1}{\beta})$, configurations plus integration over all the final configurations:
\begin{equation}
{\rm Tr}_{{\rm L}^2} \left[ {\rm exp}(- \frac{1}{\hbar}\frac{\hat{H}\bar{T}}{\lambda v})\right]=K(\bar{T}) \int_{{\rm L}^2} {\cal D}\phi(\bar{\tau},y) {\cal D} \psi^\dagger (\bar{\tau},y) {\cal D}\psi(\bar{\tau},y)\cdot 
{\rm exp}\left[ -\frac{1}{\hbar} S_{\rm JR}^{\rm E}(\phi, \psi^\dagger,\psi)\right] \quad , \quad \bar{\tau}=i \tau
\end{equation}
Here ${\rm L^2}$ is the infinite dimensional space of square integrable scalar and spinor fields or their completion by plane waves and/or their Fourier transforms. ${\cal D}$ is the symbol for Euclidean Feynmann-Berezin integration measures in this space. It is important 
to notice that, being the classical limit of Fermi fields Grassmann fields, the integration measure over Grassmann spinors are Berezin measures. $K(\bar{T})$ is an infinite normalization function of the inverse temperature $\bar{T}$.

The WKB-semi-classical-stationary phase approximation, the steepest descent between saddle points method in the Euclidean version, to this infinite dimensional integral captures the one-loop phenomena we are interested in. Introducing now in the action the scalar and spinor fluctuations around the kink, the saddle point, in the form
\[
\phi(\bar{\tau},y)= \tanh  y +\frac{\sqrt{\hbar}}{v} \eta (\bar{\tau}, y)  \quad  , \quad \psi(\bar{\tau},y)= \frac{\sqrt{\hbar}}{v} \psi(\bar{\tau},y) \, \, ,
\]
we rewrite the $\hbar$-expansion of the action
\begin{eqnarray}
\frac{S_{\rm JR}^{\rm E}}{\hbar}&=&\frac{4}{3}\frac{ v^2}{\hbar}\bar{T} +\frac{1}{2}\int_0^{\bar{T}} d \bar{\tau} \int_{-\infty}^\infty dy \left\{\eta(\bar{\tau }, y)\left(\frac{\partial^2}{\partial \bar{ \tau}^2}+\frac{\partial^2}{\partial y^2}- 4 +\frac{6}{\cosh^2 y}\right)\eta(\bar{\tau},y)\right. \nonumber \\ &+&  \left. 2  \frac{\sqrt{\hbar}}{v} \cdot\tanh y \, \, \eta^3(\bar{\tau}, y)+ \frac{\hbar}{2 v^2} \cdot \eta^4(\bar{\tau}, y) +  \bar{\psi}^+(\bar{\tau},y) \left(-\frac{\partial}{\partial \bar{\tau}}+i \alpha \frac{\partial }{\partial y}-\nu \beta \tanh y \right)\psi^+(\bar{\tau}, y)\nonumber \right. \\ &+& \bar{\psi}^-(\bar{\tau})\left(\frac{\partial}{\partial\bar{\tau}}+i\alpha\frac{\partial}{\partial y}-\nu\beta \tanh y\right) \psi^-(\bar{\tau}) - \frac{\sqrt{\hbar}}{v} \bar{\psi}^+(\bar{\tau},y) \eta(\bar{\tau},y) \psi^+(\bar{\tau},y) \\ &-& \left. \frac{\sqrt{\hbar}}{v} \bar{\psi}^-(\bar{\tau},y) \eta(\bar{\tau},y) \psi^-(\bar{\tau},y) \right\} \nonumber 
\end{eqnarray}
where $\psi^+$, respectively $\psi^-$, includes the spinor field terms that, after quantization, creates \lq\lq electrons\rq\rq, respectively, creates \lq\lq positrons\rq\rq.
Only the terms independent on $\hbar$ are kept in the WKB approximation to the $\hbar$-expansion of the Euclidean action. These terms, in particular in the Euclidean JR action,
are quadratic in the fields. 

Therefore, we are compelled to perform integrals of Gaussians of two types in infinite dimensional spaces of either scalar or Grassmann spinor fields. We pause to remind us how these integrals are computed in simple settings. First choose the real scalar field in two points of the Minkowski space: $\phi_1=\phi(\tau_1,y_1)$, $\phi_2(\tau_2,x_2)$. In the quantum partition function integrals of the form
\[
I_S= \frac{1}{\pi} \int_{-\infty}^\infty d\phi_1 \int_{-\infty}^\infty d\phi_2 e^{-a_1 \phi_1^2-a_2 \phi_2^2}=\frac{1}{\sqrt{a_1a_2}} \, \quad {\rm if } \, \, a_1>0 \, \, , \, \, a_2 >0 \, 
\]
enter. In this formula we find that
\[
I_S= \frac{1}{{\rm det}^{\frac{1}{2}} A_d} \quad {\rm where} \quad A_d=\left(\begin{array}{cc} a_1 & 0 \\ 0 & a_2 \end{array}\right) \, .
\]
Defining now $A=O A_d O^T$ such that $O^T=O^{-1}$ is an orthogonal matrix and introducing the new variables
\[
\left(\begin{array}{c} \eta_1 \\ \eta_2\end{array}\right)=O\cdot\left(\begin{array}{c} \phi_1 \\ \phi_2\end{array}\right) 
\]
one checks that the Gaussian integral formula is also valid for non diagonal matrices:
\[
I_S= \frac{1}{\pi} \int_{-\infty}^\infty d\eta_1 \int_{-\infty}^\infty d\eta_2 \, e^{- \eta^T A \eta}= \frac{1}{{\rm det}^{\frac{1}{2}}A} 
\]
because ${\rm det}A_d={\rm det} A$ and the Jacobian of the transformation from the $\phi$- to the $\eta$-variables is one. The consequence is that the integration \lq\lq volumes\rq\rq in the $\phi$- and $\eta$-varibles are identical.

It remains to analize what happens with the Grassmann variables. Focusing on a complex Grassmann variable such that 
\[
B^*B+B B^*=0 \quad , \quad B^2= B^{*2}=0 \quad \Rightarrow  \quad {\rm exp}[-a B^*B]=1-a B^*B \quad , \quad a\in\mathbb{R}^+
\]
the Berezin integration measure
\[
\int dB^* =0 \quad , \quad \int dB=0 \quad , \quad \int dB^*dB=1 
\]
means that: $ I_G= -\int dB^* dB {\rm exp}[-a B^*B]=a$. For a Grassmann complex spinor field we find
\begin{equation}
I_G =\int d\psi^\dagger d\psi {\rm exp}[\psi^\dagger A \psi]= {\rm det} A \, \, . \label{berint}
\end{equation}
 
Therefore, generalization of these concepts to the functional integral version of the partition function of the Jackiw-Rebbi model
at the one-loop order prompts the following result:
\begin{eqnarray}
&& K(\bar{T}) \int_{L^2} {\cal D}\phi {\cal D}\bar{\psi}^+ {\cal D}\psi^+ {\cal D}\bar{\psi}^-{\cal D}\psi^- {\rm exp}[-\frac{1}{\hbar}S_{\rm JR}^{\rm E}(\phi,\bar{\psi}^+, \psi^+, \bar{\psi}^-, \psi^-) ] \label{WKBjrpf} \\ && \hspace{2cm} \simeq {\rm exp}[-\frac{4}{3}\frac{v^2}{\hbar}\bar{T}]\cdot K(\bar{T}) \frac{{\rm Det} {\rm \slashed{D}_E} {\rm Det} \slashed{D}_E^\dagger}{{\rm Det}^{1/2} \Box_E}+ {\cal O}(\sqrt{\hbar}) \label{WKBjrpf} \, \, . \nonumber
\end{eqnarray}
after expanding the spinor fields in terms of the eigenfunctions of $\slashed{D}_E$ or $\slashed{D}_E^\dagger$, and the scalar fields in terms of those of  $\Box_E$. The integration measure is then defined as products of Gaussians in the expansion coefficients, respectively complex Grassmann and reals, times the corresponding eigenvalues. We denote now the determinants of differential operators by means of capital letters in order to distinguish them from determinants of ordinary matrices and the hard task is to make sense of this formal result. 

Formula (\ref{WKBjrpf}) suggest the choice
\[
K(\bar{T})= \frac{{\rm Det}^{1/2} \Box^0_{\rm E}}{{\rm Det} \slashed{D}_{\rm E}^0\cdot{\rm Det}\slashed{D}_E^{0\dagger}}
\]
for the infinite normalization factor with the purpose of refering the semi-classical action in the kink sector to its counterpart in the vacuum sector.

The infinite determinants of the Dirac and Klein-Gordon operators extended to include potential enegrgy densities
\begin{eqnarray*}
\slashed{D}_{\rm E}&=&-\frac{\partial}{\partial \bar{\tau}}+i \alpha \frac{\partial}{\partial y}- N \beta \tanh y  \quad , \quad \slashed{D}_{\rm E}^0=-\frac{\partial}{\partial \bar{\tau}}+i \alpha \frac{\partial}{\partial y}- N \beta  \\ \Box_E &=& \frac{\partial^2}{\partial \bar{\tau}^2}+\frac{\partial^2}{\partial y^2}-4+ \frac{6}{\cosh^2 y} \quad \quad , \quad \quad \Box_E^0 = \frac{\partial^2}{\partial \bar{\tau}^2}+\frac{\partial^2}{\partial y^2}-4
\end{eqnarray*}
may be regularized à la Ray-Singer in the mathematical framework of spectral zeta function theory. We shall argue here following physical ideas to make sense of infinite determinants, although the spectral zeta function spirit will be behind this analysis even if the spectrum is mixed of discrete and continuous sets of eigenvalues.

\subsection{The Bosonic Determinant of the Klein-Gordon operator}
 
 To start with, we discuss the spectrum of $\Box_{\rm E}$ 
 assuming Periodic Boundary conditions in the inverse temperature $\bar{T}$
\[
\Box_E f_\lambda (\bar{\tau},y)= \lambda^2 f_\lambda(\bar{\tau},y) \quad , \quad f_\lambda(0, y)=f_\lambda (\bar{T},y)
\]
The generic form of the eigenfunctions complying with this PBC is consequently
\[
f_{\lambda_n}(\bar{\tau},y) =\sin( \frac{ \pi n}{\bar{T}}\bar{\tau}) g(y) \quad , \quad n\in \mathbb{Z}^+ \, \, ,
\]
which are eigenfunctions of $\Box_{\rm E}$ only if 
\begin{eqnarray*}
&& \left( \frac{d^2}{d y^2}-4+\frac{6}{\cosh^2 y}\right)g_q(y)= -\omega^2(q) g_q(y) \\
&& g_q(y) = e^{i q y}(3 \tanh^2 y-3 i q \tanh y -1-q^2) \quad , \quad \omega^2(q)=q^2+4 \, \, , \, {\rm if} \, \, q \in \mathbb{R} \\ && g_i(y)= \frac{\sinh y}{\cosh^2 y} \quad , \quad \omega^2(i)=3 
\end{eqnarray*}
In sum, the eigenfunctions and eigenvalues of $\Box_{\rm E} $ forming a \lq\lq basis\rq\rq in $L^2(\mathbb{R})$ are the following
\begin{eqnarray*}
&& f_{\lambda_{n (q)}}(y)=\sin (\frac{n \pi}{\bar{T}}\bar{\tau})\, \cdot \, e^{i q y} P_2(\tanh y , q) \quad , \quad \lambda^2_{n (q)}=-(\frac{\pi^2}{\bar{T}^2}n^2 + q^2+4) \\ && f_{\lambda_{n(i)}}(y)= \sin (\frac{\pi  n}{\bar {T}} \bar{\tau})\, \cdot \, e^{-y} \, P_2(\tanh y,i)  \, \, \quad ,  \, \, \quad \lambda^2_{n(i)}=-(\frac{ \pi^2}{{\cal T}^2} n^2 + 3 ) \qquad .
\end{eqnarray*}
There are scattering waves, Higgs quanta propagating through Kink wells, when $q$ is real. There is also a bound state at imaginary momentum $q=i$ that obeys to a Higgs quantum trapped in the kink well. We shall normalize the scattering waves, in the standard way in QFT, on a finite interval of very long length $l=\lambda L$ by means of the spectral density, both for $\Box_{\rm E}$ and $\Box_{\rm E}^0$
\[
\rho(q)= \frac{l}{2\pi}+\frac{1}{2\pi}\frac{\partial \delta}{\partial q}=\frac{l}{2\pi}-\frac{1}{\pi}\sum_{j=1}^2 \, \frac{j}{j^2 + q^2}  \quad , \quad , \quad \quad \rho_0(q)=\frac{l}{2 \pi} 
\]
where $\delta(q)$ is the phase shift induced by the Kink well. Besides the bound state at $q=i$ there is a zero mode
at $q=2 i$. The existence of this zero mode is due to the freedom to fix along the real line the Kink center of mass. Thus, the only contribution arise at ${\cal O}(\sqrt{\hbar})$-order . Alternatively one may evaluate the zero mode via a Faddeev-Popov determinant. Accordingly, we shall compute the determinant on the orthogonal subspace to the zero mode in $L^2(\mathbb{R})$. We write now the spectral zeta functions of $\Box_E^0$ and $\Box_E$ as functions of the complex parameter $s$:
\begin{eqnarray*}
\zeta_{\Box_E^0}(s)&=& \sum_{n\in\mathbb{Z}^+}\frac{l}{2\pi}\left(\pi\frac{n}{\bar{T}}\right)^{-2 s}\int_{-\infty}^\infty \, dq \, \left(1+\frac{q^2+4}{\pi^2 n^2}\bar{T}^2\right)^{-s}\\ \zeta_{\Box_E}^\perp (s)&=& \sum_{n\in\mathbb{Z}}^+\left(\pi\frac{n}{\bar{T}}\right)^{-2s}\left[\left(1+\frac{3}{\pi^2 n^2}\bar{T}^2\right)^{-s}+\right.\\ && \hspace{2.4cm}+ \left.\int_{-\infty}^\infty dq \left(\frac{l}{2\pi}-\frac{1}{\pi}\sum_{j=1}^2 \frac{j}{j^2+q^2}\right)\left(1+\frac{q^2+4}{\pi^2n^2}\bar{T}^2\right)^{-s}\right]
\end{eqnarray*}
which is a meromorphic function in the $s$-complex plane. The Ray-Singer proposal for the regularization of an operator $A$ with infinite eigenvalues reads
\[
{\rm Det} A = {\rm exp}[-\frac{d \zeta_A}{d s}(0)] \, .
\]
We expect that, due to the choice of $K(\bar{T})$ the one-loop correction incorporates a factor
\[
\left( \frac{{\rm Det} \Box_E^0 }{ {\rm Det}^\perp \Box_E}\right)^{1/2}
\]
in the Bosonic part of the effective action. The notation ${\rm Det}^\perp$ means that the detrminant is computed in the subspace orthogonal to the zero mode. According to the Ray-Singer formula we find:
\[
\frac{{\rm Det} \Box_E^0 }{ {\rm Det}^\perp \Box_E}= {\rm exp}\left[\frac{d \zeta^\perp_{\Box_E}}{d s}(0)-\frac{d \zeta_{\rm \Box_E^0}}{d s}(0)\right]
\]
From
\[
\frac{d}{d s}\left[\left(\frac{\pi}{\bar{T}}n\right)^{-2s}\left(1+\frac{\omega^2\bar{T}^2}{\pi^2 n^2}\right)^{-s}\right]\Big\vert_{s=0}= -\log\left(\frac{\pi}{\bar{T}}n\right)^2- \log\left(1+ \frac{\omega^2\bar{T}^2}{\pi^2 n^2}\right)
\]
a (not too long) calculation shows that:
\begin{eqnarray}
-\frac{d \zeta_{\Box_E^0}}{d s}(0)&=& \sum_{n\in\mathbb{Z}^+}\left[\log\left(\frac{\pi n}{\bar{T}}\right)^2+\frac{l}{2\pi}\int_{-\infty}^\infty dq \log \left(1+ \frac{q^2+4}{\pi^2 n^2}\bar{T}^2\right)\right] \label{zetfdet1}\\ \frac{d \zeta_{\Box_E}}{d s}(0)&=&- \sum_{n\in\mathbb{Z}^+}\left[\log\left(\frac{\pi n}{\bar{T}}\right)^2+\frac{l}{2\pi}\int_{-\infty}^\infty dq \log \left(1+ \frac{q^2+4}{\pi^2 n^2}\bar{T}^2\right)\right]-\nonumber \\ &-& \sum_{n\in\mathbb{Z}^+}\left[ \log\left(1+ \frac{3 \bar{T}^2}{4 \pi^2 n^2}\right)-\frac{1}{\pi}\int dq \left(\sum_{j=1}^2 \frac{j}{j^2+q^2}\right) \log \left(1+\frac{(q^2+4)\bar{T}^2}{ \pi^2 n^2}\right) \right] \label{zetfdet2}\, \, .
\end{eqnarray}
In the very low temperature regime, when $\bar{T} \to \infty$ is very large, the sums in formulas (\ref{zetfdet1}-\ref{zetfdet2}) become integrals:
\begin{equation}
\sum_{n\in \mathbb{Z}^+}\, \log\left(1+\frac{\omega^2}{ \pi^2 n^2}{\bar{T}}^2\right) \, \longrightarrow \,\frac{\bar{T}}{\pi} \int_{0}^\infty \, dz \, \log\left(1+\frac{\omega^2}{z^2}\right) \quad , \quad z^2 \simeq _{\bar{T}\to \infty}\frac{\pi^2 n^2}{\bar{T}^2}
\end{equation}
Therefore, because
\[
\int_{0}^\infty \, \frac{dz}{\pi}\log \left(1+ \frac{\omega^2}{z^2}\right)=\vert \omega \vert
\]
the square root of the regularized ratio of determinants reads{\footnote{The infinite kinetic factor $\sum_{n\in\mathbb{Z}^+}\log(\frac{\pi^2n^2}{\bar{T}^2})$ has been absorbed in the normalization of the Euclidean integration measure .}}
\[
\left(\frac{{\rm Det}\Box_E^0}{{\rm Det}^\perp \Box_E}\right)^{\frac{1}{2}}={\rm exp}\left[-\frac{\sqrt{3}}{2}\, \bar{T}+\frac{\bar{T}}{2\pi}\int_{-\infty}^\infty \, dq \, \sqrt{q^2+4}\, \sum_{j=1}^2 \frac{j}{j^2+q^2}\right] \, \, .
\]
Thus, the infinite normalization due to multiplication by the determinant of $\Box_{\rm E}^0$ subtracts $\rho_0(q)$ from $\rho(q)$
and the summation $ \sum_n\log(\frac{ \pi^2}{\bar{T}^2}n^2+4)$ from the summations arising in the definition of ${\rm Det}\Box_E$. Moreover, the infinite product of $-1$ appearing in the spectrum of $\Box_{\rm E}$ is cancelled by the identical infinite factor arising in the spectrum of $\Box_{\rm E}^0$.
Finally, the effective partition function due to Bosonic fluctuations up to one-loop order is found to be:
\begin{equation}
{\rm Tr}_{\rm L^2(\mathbb{R})} \, \, {\rm exp}[-\frac{S_{\rm JR}^{\rm E}}{\hbar}] \simeq {\rm exp} \left\{-\frac{1}{\hbar} \left( \frac{4}{3}v^2 - \frac{\hbar}{2} (\int_{-\infty}^\infty \, \frac{dq}{\pi}\sqrt{q^2+4}\sum_{j=1}^2 \frac{j}{j^2+q^2}-\sqrt{3})+ {\cal O}(\hbar^{3/2})\right)\bar{ T}\right\} \label{ltpartfJR} \, \, .
\end{equation}
Note that the $\hbar$-independent contribution in the exponent is the Kink Casimir energy. Only the contribution of the renormalization mass counter-terms are not included in this formula, as well as the finite renormalization due to the use of mode number regularization. In sum, up to one-loop renormalization, the Bosonic  one-loop kink mass shifts obtained in canonical or Feynman integration quantization are identical.

\subsection{ The Fermionic Determinant of the Dirac operator}

The contribution of Fermionic Kink fluctuations up to the one-loop order of the quantum Kink mass shift is encoded in the quotient of determinants: 
\[
\frac{{\rm Det} \slashed{D}_{\rm E}\cdot{\rm Det} \slashed{D}^\dagger_E}{{\rm Det} \slashed{D}_{\rm E}^0\cdot {\rm Det} \slashed{D}_E^{0\dagger}} \, \, . 
\]
We shall focus in the case where
the $\nu=\frac{g}{\lambda}=N$ parameter, involving the ratio of Yukawa and scalar self-interaction couplings, is a positive natural number. Otherwise formulas are very complicated encompassing Euler Gamma and Gauss Hypergeometric functions. We now mention a sublety when Dirac spinors enter the game. While in non-linear Klein-Gordon equations it is enough to replace real by imaginary time (temperature inverse) to deal with Euclidean field theory, in the case of the Dirac equation it would be necessary replace also the Clifford algebra 
by Euclidean gamma matrices. We shall stick to a mixed situation: we will consider imaginary time (no) propagation but we shall keep
the $\alpha$ and $\beta$ Dirac matrices without change. By taking this point of view our Dirac equation is  a difussion equation
like the equations arising in the stochastic quantization of supersymmetric field theories, so closely related to Fermionic functional integrals, see \cite{Parisi}.

To cope with a sensible definition of the quotient of determinants we start with the Dirac spectral problem
\begin{eqnarray*}
&& \slashed{D}_{\rm E}\, \psi_\varepsilon^+(\bar{\tau},y)=\varepsilon ^+\, \psi_\varepsilon^+(\bar{\tau},y) \quad , \quad \slashed{D}^\dagger_{\rm E}\, \psi_\varepsilon^-(\bar{\tau},y)=\varepsilon ^-\, \psi_\varepsilon^-(\bar{\tau},y)\\ && \psi_\varepsilon^\pm(0,y)=- \psi_\varepsilon^\pm(\bar{T},y) 
\end{eqnarray*}
with anti-periodic boundary conditions in the inteval $[0,T]$ over the spinor fields. Because 
\[
\slashed{D}_{\rm E}=\left(\begin{array}{cc} -\frac{\partial}{\partial\bar{\tau}}& 0 \\ 0 & -\frac{\partial}{\partial\bar{\tau}}\end{array}\right)+ \left(\begin{array}{cc} 0 & \frac{\partial}{\partial y}+N \, \tanh y \\ -\frac{\partial}{\partial y}+ N \tanh y \end{array}\right)
\]
\[
\slashed{D}^\dagger_{\rm E}=\left(\begin{array}{cc} \frac{\partial}{\partial\bar{\tau}}& 0 \\ 0 & \frac{\partial}{\partial\bar{\tau}}\end{array}\right)+ \left(\begin{array}{cc} 0 & \frac{\partial}{\partial y}+N \, \tanh y \\ -\frac{\partial}{\partial y}+ N \tanh y \end{array}\right)
\]
it suffices to remind the spectrum of $H_D=-i \alpha \frac{\partial}{\partial y}+N \, \tanh y \, \beta $:
\begin{eqnarray*}
&& \hspace{3 cm} H_D\left(\begin{array}{c}\psi_1(y;\omega) \\ \psi_2(y;\omega) \end{array} \right)= \omega \left(\begin{array}{c}\psi_1(y;\omega )\\ \psi_2(y;\omega) \end{array} \right) \\\
&& \hspace{-0.7cm} {\rm Discrete} : \omega_j= \sqrt{(2N-j)j}\, \, , \, \, j=1,2, \cdots, N-1 \, \, \, , \, \, \, {\rm doubly} \, {\rm degenerate}
\, \, , \, \, \omega_0=0 \, \, \, , \, \, \, {\rm unpaired} \\ && \hspace{-0.7cm}{\rm Continuous}: \omega(q^2) = \sqrt{q^2+N^2} \, \, , \, \, {\rm doubly} \, {\rm degenerate} \, \, , \,\rho(q)= \frac{l}{2 \pi} -\frac{1}{2\pi}\left(\sum_{l=1}^{N-1}\frac{2 l}{l^2+q^2}+ \frac{N}{N^2+q^2} \right)
\end{eqnarray*}

Given that $H_D$ is $\bar{\tau}$-independent it is convenient (and possible) to plug the separation ansatz in the spectral problems of $\slashed{D}_{\rm E}$
and its adjoint, demanding also  anti-periodic boundary contitions in the inverse temperature:
\begin{equation}
\psi^\pm_{\varepsilon_{n\omega}}(\bar{\tau},y)={\rm exp}\left[ \frac{ i (2n+1)\pi}{\bar{T}}\bar{\tau}\right]\left(\begin{array}{c} \psi^\pm_1(y;\omega)\\ \psi^\pm_2(y;\omega)\end{array}\right) \quad , \quad n\in \mathbb{Z}
\end{equation}
Therefore, the spectra of $\slashed{D}_{\rm E}$ and $\slashed{D}_{\rm E}^\dagger$ are respectively:
\begin{eqnarray*}
\varepsilon^+_{n \omega_j}&=& (\frac{i(2 n+1)\pi}{\bar{T}}+\omega_j) \quad , \quad \varepsilon^-_{n \omega_j}= (\frac{-i(2 n+1)\pi}{\bar{T}}+\omega_j)\\ \varepsilon^+_{n\omega(q)}&=& (\frac{i(2 n+1)\pi}{\bar{T}}+\omega(q)) \quad , \quad \varepsilon^-_{n\omega(q)}= (\frac{-i(2 n+1)\pi}{\bar{T}}+\omega(q))
\end{eqnarray*}
Note that there is no zero mode because $\varepsilon_{00}=\frac{\pi}{\bar{T}}$. 

Given that 
\[
{\rm Det}\slashed{D}_E \cdot {\rm Det} \slashed{D}_E^\dagger= {\rm Det} (\slashed{D}_E^\dagger \slashed{D}_E) \quad , \quad {\rm Det}\slashed{D}^0_E \cdot {\rm Det} \slashed{D}_E^{0\dagger}= {\rm Det} (\slashed{D}_E^{0\dagger }\slashed{D}^0_E)
\]
is compelling to compute $\zeta_{\slashed{D}_E^\dagger\slashed{D}_E}(s)$ and $\zeta_{\slashed{D}_E^{0\dagger}\slashed{D}^0_E}(s)$ as an intermediate step in finding the Ray-Singer regularized determinants.

The spectral zeta functions of products of Dirac operators  times its adjoints in the Jackiw-Rebbi model
are the following:
\begin{eqnarray*}
\zeta_{\slashed{D}_E^{0\dagger} \slashed{D}_E^0}(s)&=& \sum_{n\in\mathbb{Z}}\frac{l}{2\pi}\left(2\frac{(2n+1)\pi}{\bar{T}}\right)^{-2 s}\int_{-\infty}^\infty \, dq \, \left(1+\frac{q^2+N^2}{(2n+1)^2\pi^2}\bar{T}^2\right)^{-s}\\ \zeta_{\slashed{D}_E^\dagger\slashed{D}_E} (s)&=& \sum_{n\in\mathbb{Z}}\left(2\frac{(2n+1)\pi}{\bar{T}}\right)^{-2s}\left[\sum_{l=1}^{N-1}\left(1+\frac{(2N-l)l}{(2n+1)^2\pi^2}\bar{T}^2\right)^{-s}+\right.\\ && + \left.\int_{-\infty}^\infty dq \left(\frac{l}{2\pi}-\frac{1}{2\pi}\left(\sum_{j=1}^{N-1} \frac{2j}{j^2+q^2}+\frac{N}{N^2+q^2}\right)\right)\left(1+\frac{q^2+N^2}{(2n+1)^2\pi^2}\bar{T}^2\right)^{-s}\right]
\end{eqnarray*}
Calculation of the quotient of determinants according to the Ray-Singer prescription requires knowledge of  the derivatives of the spectral zeta functions in the origin of the complex $s$-plane:
\begin{eqnarray*}
\frac{d \zeta_{\slashed{D}_E^{0\dagger}\slashed{D}_E^0}}{d s}(0)&=& -\sum_{n\in\mathbb{Z}}\left[ \log\left(\frac{(2 n+1)\pi}{\bar{T}}\right)^2+\frac{l}{2\pi}\int_{-\infty}^\infty dq \log \left(1+ \frac{q^2+N^2}{(2n+1)\pi^2}\bar{T}^2\right)\right]-\log 4 \\ - \frac{d \zeta_{\slashed{D}_E^\dagger\slashed{D}_E}}{d s}(0)&=& \sum_{n\in\mathbb{Z}}\left[\log\left(\frac{(2n+1)\pi}{\bar{T}}\right)^2+\frac{l}{2\pi}\int_{-\infty}^\infty dq \log \left(1+ \frac{q^2+N^2}{(2 n+1)^2\pi^2}\bar{T}^2\right)\right]+\log 4\\ && \hspace{-2.5cm} +\sum_{n\in\mathbb{Z}}\left[ \log\left(\frac{(2 n+1)\pi}{\bar{T}}\right)^2 + \log \sum_{l=1}^{N-1}\left(1+ \frac{(2 N-l)l \bar{T}^2}{(2 n+1)^2 \pi^2}\right)-\right.\\ && \left. -\frac{1}{\pi}\int dq \left(\sum_{j=1}^{N-1} \frac{j}{j^2+q^2}+\frac{1}{2}\frac{N}{N^2+q^2}\right) \log \left(1+\frac{(q^2+N^2)\bar{T}^2}{(2 n+1)^2 \pi^2}\right) \right]+\log 4 \, \, .
\end{eqnarray*}
leading to:
\begin{eqnarray*}
\frac{{\rm Det} \slashed{D}_E^\dagger \slashed{D}_E}{{\rm Det} \slashed{D}_E^{0\dagger} \slashed{D}^0_E}&=& {\rm exp}[ - \frac{d \zeta_{\slashed{D}_E^\dagger\slashed{D}_E}}{d s}(0)+ \frac{d \zeta_{\slashed{D}_E^{0\dagger}\slashed{D}_E^0}}{d s}(0)]={\rm exp}\left[ \sum_{l=1}^{N-1}\log \prod_{n=-\infty}^\infty \left(1+\frac{(2N-l)l\bar{T}^2}{(2 n+1)^2\pi^2}\right)-\right. \\ &-& \left.\frac{1}{\pi}\int_{-\infty}^\infty dq \left( \sum_{j=1}^{N-1}\frac{j}{j^2+q^2}+\frac{1}{2}\frac{N}{N^2+q^2}\right)\log \left(\prod_{n=-\infty}^\infty \left(1+\frac{q^2+N^2}{(2 n+1)^2 \pi^2}\bar{T}^2\right)\right)\right]
\end{eqnarray*}
In this formula the one-oop impact of the vacuum fermionic fluctuation has been subtracted and the kinematical factors have been absorbed in the integration measure. The next task is the evaluation of the infinite products. For a single mode of frequency $\omega$
the result is particularly significative at very low temperature $\bar{T}\to\infty$:
\begin{equation}
\prod_{n=-\infty}^\infty 4\left(1+\frac{\omega^2 \bar{T}^2}{(2n+1)^2\pi^2}\right) =4\cosh^2(\frac{\omega \bar{T}}{2})\simeq_{\lim_{\bar{T}\to \infty}} e^{\omega \bar{T}} \label{lowTI}
\end{equation}
From the formula (\ref{lowTI}) for a single oscillation mode we infer the following expression for the quotient of the Fermionic determinants at  high $T$:
\begin{equation}
\frac{{\rm Det}\slashed{D}_E^\dagger\slashed{D}_E}{{\rm Det} \slashed{D}_E^{0\dagger}\slashed{D}_E^0}\simeq _{\bar{T}>> 1} \,{\rm exp}\Big[\bar{T}\Big( \sum_{l=1}^{N-1}\sqrt{(2 N-l)l}-\frac{2}{\pi }\int_{0}^\infty dq \left(\sum_{j=1}^{N-1}\frac{j}{j^2+N^2}+\frac{1}{2}\frac{N}{N^2+q^2}\right)\sqrt{q^2+N^2}\Big) \Big]
\end{equation}

In sum, the effective partition function in the Jackiw-Rebbi model up to one-loop order including the impact of both Bosonic  (scalar) and Fermionic  (spinor) Kink fluctuations reads:
\begin{eqnarray}
&&{\rm Tr}_{\rm L^2(\mathbb{R})\sqcup L^2(S)} \, \, {\rm exp}[-\frac{S_{\rm JR}^{\rm E}}{\hbar}] \simeq  {\rm exp} \left\{-\frac{1}{\hbar} \left( \frac{4}{3}v^2 - \frac{\hbar}{2}\left (\int_{-\infty}^\infty \, \frac{dq}{\pi}\sqrt{q^2+4}\, \sum_{j=1}^2 \frac{j}{j^2+q^2}-\sqrt{3} \right. \right. \right. \label{fltpartfJR} \\ &+2& \left. \left. \left. \sum_{j=1}^{N-1}\sqrt{(2N-j)j}-\frac{4}{\pi} \int_0^\infty \, dq \, \left(\sum_{l=1}^{N-1}\frac{l}{l^2+q^2}+\frac{N}{2(N^2+q^2)}\right)\sqrt{q^2+N^2}\right)+ {\cal O}(\hbar^{3/2})\right)\bar{T}\right\} \nonumber \, \, .
\end{eqnarray}

 From this formula (\ref{fltpartfJR}) is not difficult to derive the Bose/Fermi DHN prescription for the one-loop Kink mass before of the incorporation of one-loop mass renormalization and choosing the mode number regularization scheme.
 Fermionic Kink fluctuations with respect to Bosonic Kink fluctuations contribute to the Kink Casimir energy with two differences: 1) Fermions add to the classical Kink mass at one-loop order whereas Bosons subtract. This is due to the difference in quantum statistics. 2) A single Fermionic fluctuation mode contributes twice as compared with a single Bosonic fluctuation mode. The reason is that there are electron and positrons fluctuating over the Kink.  

In order to reach a finer understanding about how Fermi statistics emerges we shall focus now in the contribution to the effective action of a single fermionic mode
\[
4 \cosh^2 \frac{\omega \bar{T}}{2}=\left( e^{-\frac{\omega \bar{T}}{2}}+ e^{\frac{\omega T}{2}}\right)^2=e^{-\omega \bar{T}}\left( 1+2 \, e^{\omega \bar{T}}+ e^{2 \omega \bar{T}} \right)
\]
The polynomial between brackets encodes the following facts: 1) \boxed{1}. This is the vacuum state. Neither the electron nor the positron occupy this state. 2) $\boxed{2 e^{\omega \bar{T}}}$. There are two one-particle states. Either the electron or the positron occupy the $\omega$-mode.
3) $\boxed{e^{2 \omega \bar{T}}}$. This is a two-particle state. Both the electron and the positron occupy the $\omega$-state.

Next we consider that only two fluctuation modes $\omega_1 < \omega_2$ contribute to the partition function
\begin{eqnarray*}
&& 4 \cosh^2 \frac{\omega_1 \bar{T}}{2}\cdot 4 \cosh^2 \frac{\omega_2\bar{T}}{2}=\left(e^{-\frac{\omega_1 \bar{T}}{2}}+e^{\frac{\omega_1 \bar{T}}{2}}\right)^2 \cdot \left(e^{-\frac{\omega_2 \bar{T}}{2}}+e^{\frac{\omega_2 \bar{T}}{2}}\right)^2 
= e^{-\bar{T}(\omega_1+\omega_2)}\times \\ &&\times\left (1+2(e^{\bar{T}\omega_1}+e^{\bar{T}\omega_2})+ e^{2 \bar{T}\omega_1}+e^{2 \bar{T}\omega_2}+
4 e^{\bar{T}(\omega_1+\omega_2)}+ 2(e^{\bar{T}(2\omega_1+\omega_2)}+ e^{\bar{T}(\omega_1+2\omega_2)})+ e^{2\bar{T}(\omega_1+\omega_2)} \right)
\end{eqnarray*}
The partition function shows that there is a vacuum state with neither $\omega_1$ nor $\omega_2$ occupied by any fermions. There are four one-particle states where either $\omega_1$ or $\omega_2$ are occupied by either one electron or one positron. There are six two-particle states. The alternatives are: One electron and one positron occuppy either $\omega_1$ or $\omega_2$. Either two electrons or
two positrons occupy respectively $\omega_1$ and $\omega_2$. Finally, one electron occupy $\omega_1$ and one positron occupy $\omega_2$ 
or viceversa. There are also four three particle states: One electron-positron pair occupy $\omega_1$ and either one electron or a positron occupy $\omega_2$ plus identical configurations exchanging $\omega_1$  by $\omega_2$. There is a single four-particle state where both $\omega_1$ and $\omega_2$ are occupied by single electron-positron pairs. This no more than the partition function of a gas of two molecule kinds, each molecule having two degrees of freedom, and complying with {\it Fermi-Dirac statistics}. Promotion of this idea to the one-loop partition function accounting for fermionic kink fluctuations in the Jackiw-Rebbi model follows much more cumbersome but essentially identical combinatorics, see e.g. Reference
\cite{Rajaraman}, Chapter 9.

\section{Outlook on future research}

Bidimensional relativistic Quantum Field Theories encompassing fermions and anti-fermions as well as kink topological defects are by no means only of theoretical interest but are realized in Nature in the realm of Condensed Matter Physics. The quantum dynamics is qualitatively described by the Jackiw-Rebbi model in e.g. 
graphene nanotubes, see \cite{JakubskyJPA14}, or trans polyacetilene molecules  \cite{Schrieffer} where electrons and holes move in one dimensional chains  with a linear dispersion relation between energy and momentum while kink-like defects also develops in these linear materials. Therefore, it is natural to investigate other models where the scalar (bosons,phonons) self-interaction is governed by other energy densities that the
$\lambda\phi^4$-polynomial of the JR model. Moreover, systems with more than one scalar fields, or several branches of phonons, are rather interesting. Of course, one must select mathematical expressions and the appropriate number of fields providing room for other types of kink topological defects, see e.g. \cite{AIJMG}. The Yukawa coupling of fermions to these exotic kinks will give rise, at one-loop order,
to Dirac Hamiltonians whose square are not P$\ddot{\rm o}$sch-Teller operators but more general matrix Schr$\ddot{\rm o}$dinger operators. This means that the spectra of kink spinor fluctuations are not exactly known and only heat kernel/zeta function regularization methods are available to compute one-loop mass shifts. This approach was developed first in the paper \cite{BGNV} in the computation of the one-loop Susy kink mass shift. A similar procedure was applied latter, see \cite{AIJMG}, to calculate kink masses in one-field models where the purely scalar fluctuations are governed by Schr$\ddot{\rm o}$dinger operators such that its spectra are not analytically known. To extend this technology to  kinks in $N$-field scalar models is not difficult and several interesting cases may be found in \cite{GAGTS}. Adding spinor fields to these classes of models will be interesting.  We propose to perform a parallel development to what has been done
in this paper on the Jackiw-Rebbi model. Moreover, the crucial ingredient, the computation of the spectral zeta function of the square of the Dirac operator from the Mellin transform of the Gilkey-DeWitt heat kernel expansion, seems to be plausible.  

Another scenario where the interaction between Fermions and Topological defects at one-loop order deserves deep investigation is the Abelian Higgs model. Despite of being a very difficult problem even in the purely bosonic setting. The difficulties in the Abelian Higgs model are due both to the leap in dimensions to three-dimensional space-time and the addition of one Abelian gauge field. Information about the one-loop BPS-vortex mass shifts is available, see \cite{GAGTS}, \cite{AIMGTM0}, \cite{AIGFMGMT} and \cite{AIMGTM}. The computations were achieved with a mixture  of the Gilkey-DeWitt heat kernel expansion and symbolic calculations in a Mathematica environment appropriate to solve sophisticated recurrence relations. Our new proposal is to evaluate the impact of including Fermions
in the Abelian Higgs model at one-loop level. A first attempt in this direction was tried by Jackiw and Rossi \cite{JRo} describing the
zero modes of circularly symmetric BPS vortices of Fermionic type. Expectations of finding in this model the phenomenon of Fermion fractionization previously discovered by Jackiw and Rebbi were partially verified in \cite{AFGV}. In this paper it was shown that the spectral assymmetry is indeed non null in the Jackiw-Rossi model as required by Fermi fractionization but unlike in the Jackiw-Rebbi the spectral eta function is not a topological invariant. 
The Jackiw-Rossi model, however, is difficult to deal with because one hermitian
Yukawa coupling of one complex scalar field to spinor fields requires the simultaneous consideration of the spinor field and its complex conjugate. A better situation with respect to compute one-loop vortex mass shifts is ${\cal N}=2$ supersymmetric Abelian gauge theory in $(2+1)$-Minkowski space-time. In this model Vassilevich \cite{DVass} achieved a brilliant computation of the quantum correction of the vortex mass shift having into account that the bosonic and fermionic fluctuations cancel but specific boundary conditions preserving SUSY partially leave a trace. The challenge is to escape in this model from supersymmetry by, e.g., varying the couplings and trying the computation of one-loop vortex mass shifts due to bosonic and fermionic fluctuations. Of course, heat-kernel/zeta-function regularization methods plus Mathematica computational help will be necessary.

During the last ten years conceptual advances have been reached in the understanding of the low energy dynamics of BPS vortices in the Abelian Higgs model. Starting from improved knowledge of zero modes as well as discover of bounded fluctuations of BPS-vortex fluctuations, see \cite{AIJMG1} and \cite{AIJMG2}, it has been studied how the low energy effective dynamics of BPS-vortices departs from the Hitchin-Manton-Samols geodesic dynamics in the vortex moduli space, see \cite{AGMG} and \cite{AMGW}. Another suggestive challenge is to extend the Abelian Higgs model by the incorporation of fermions and then to study the modification induced in the effective theory, i.e, to promote the theoretical developments achieved in Reference \cite{Guilarte} about the effective dynamics
of kink-fermion systems to the realm of BPS topological vortices.


\begin{thebibliography}{AA99}

\bibitem{JakubskyJPA14} V. Jakubsky, S. Kuru, and J. Negro, {\sl Carbon Nanotubes in an Inhomogeneous Transverse Magnetic Field: Exactly Solvable Model}, J. Phys. A: Math. Theor. {\bf 47} (2014) 115307

\bibitem{GuilarteIJTP11} J. Mateos Guilarte and J.M. Mu$\tilde{\rm n}$oz Casta$\tilde{\rm n}$eda,{\sl Double-Delta Potentials: One-Dimensional Scattering}, Int. J. Theor.Phys. {\bf 50} (2011) 2227-2241

\bibitem{Coleman} S.F. Coleman, {\sl Quantum Field Theory}, World Scientific, New Jersey, 2018

\bibitem{Jackiw} R. Jackiw and C. Rebbi, {\sl Solitons with fermion number $\frac{1}{2}$}, Phys. Rev. {\bf D13}, 3398 (1976)

\bibitem{Schrieffer} R. Jackiw and R. Schrieffer,{\sl Solitons with fermionic number $\frac{1}{2}$ in condensed matter physics and relativistic field theory}, Nuclear Physics {\bf B190}[FS 3](1981) 253-332

\bibitem{Niemi1986} A. Niemi and G.W. Semenoff, {\sl Fermion number fractionazitation in quantum field theory}, Physics Reports {\bf 135} (1986)99

\bibitem{Fresneda} A.  Alonso-Izquierdo,R. Fresneda, J. Mateos Guilarte, and D. Vassilevich, {\sl Soliton fermionic number from the heat kernel expansion}, European Physical Journal {\bf C79} (2019) 525

\bibitem{Kirsten} K. Kirsten, {\sl  Basic zeta functions and some applications in physics}, in {\bf A window into zeta functions and modular phisics},  MSRI publications, {\bf 57}, {\bf Cambridge University Press}

\bibitem{Rebhan} A. Rehban and P. van Nieuwnhuizen, {\sl No saturation of the quantum Bogomolny bound by two-dimensional ${\cal N}=1$ solitons}, Nucl. Phys. {\bf B508} (1997) 449

\bibitem{Shifman} M. Shifman, A.Vainsthein and M. Voloshin, {\sl Anomalies and quantum corrections to solitons in two-dimensional theories with minimal supersymmetries}, Phys. Rev. {\bf D59} (1999) 045016

\bibitem{Graham} N.Graham and R. Jaffe, {\sl Energy, central charge, and the BPS bound for $1+1$-dimensional supersymmetric solitons}, Nucl. Phys. {\bf B544} (1999) 432

\bibitem{Wimmer} R. Wimmer, {\sl Quantization of supersymmetric solitons}, Ph. D. Thesis, Wien University, 2001, arXiv: hep-th/0109119

\bibitem{Vachaspati} Y.Chu and T. Vachaspati, {\sl Fermions on one or fewer Kinks}, Phys. Rev. {\bf D77}(2008)025006

\bibitem{Bazeia} D. Bazeia, A. Mohammadi, {\sl Fermionic bound states in distinct kinlike backgrounds}, European Physical Journal {\bf C77} (2017)203

\bibitem{Rubakov} V.A. Rubakov, {\sl Classical theory of gauge fields}, Princeton University Press, Princeton, New Jersey; 2002

\bibitem{Rubakov1} V.A. Rubakov, {\sl Monopole catalysis of proton decay}, Rep. Prog. Phys. {\bf 41} (1978) 137

\bibitem{Morse} P. M. Morse and H. Fesbach, {\sl Methods in Mathematical Physics, Volume 1}, Mc Graw Hill, New York, 1953, pp. 139-144 

\bibitem{Drazin} P. G. Drazin, R. S. Johnson, {\sl Solitons: an introduction}, Cambridge University Press, Cambridge U.K., 1989

\bibitem{Abramowitz} M. Abramowitz, I. Stegun, {\sl Handbook of mathematical functions}, Dover, New York, 1972, tenth printing

\bibitem{Segal} G. Segal, {\sl Lectures on Conformal Field Theory}, Easter Term, Department of Pure Mathematics, Cambridge University, 1993.

\bibitem{Merabet} N.S. Manton, H. Merabet , {\sl $\phi^4$-Kinks. Gradient flow and dynamics }, Npnlinearity, {\bf 10} (1997) 3-18

\bibitem{Blanco} S. Alameda-Calvo, J.J. Blanco-Pillado, A. Gacia Martin-Caro, {\sl Quantum fermion emission frome xcited kinks}, JHEP {\bf 02} (2026) 226

\bibitem{DHN} R. Dashen, B. Hasslacher and A. Neveu, {\sl Non-perturbative methods and extended hadron models in field theory} Phys. Rev. {\bf D10} (1974) 4130

\bibitem{FK} L.D. Faddeev and V.E. Korepin, {\sl Quantum theory of solitons}, Physics Reports {\bf 42} (1978) 1-78

\bibitem{AIJMG} A. Alonso-Izquierdo, J. Mateos Guilarte, {\sl One-loop kink mass shifts: A computational approach}, Nuclear Physics {\bf B852[PM]} (2011) 696-735

\bibitem{GAGTS} J. Mateos Guilarte, A. Alonso-Izquierdo, W. Garcia Fuertes, M. de la Torre Mayado, M.J. Senosiain, {\sl Quantum fluctuations around low dimensional topological defects}, Proceedings of Science, PoS, {\bf ISFT209: 013}

\bibitem{Cahill} K. E. Cahill, {\sl Extended particles and solitons}, Physics Letters {\bf 53B} (1976) 174-176

\bibitem{CCG} K. Cahill, A. Comtet, R. Glauber, {\sl Mass formulas for static solitons}, Physics Letters {\bf 64 B} (1976) 283-285

\bibitem{Casa} J. Casahorran, {\sl The quantum mass formula for topologically extended objects in bidimensional theories}, International Journal of Modern Physics {\bf A05} (1990) 363

\bibitem{Guilarte} J.Mateos Guilarte, {\sl Low energy dynamics of vibrating kinks}, Front. Phys. {\bf 13} (2026)1647949, arXiv: 2510.14883 hep:th

\bibitem{Parisi} G. Parisi, N. Sourlas, {\sl Supersymmetric field theories and stochastic differential equations}, Nucl. Phys. {\bf B206} (1982) 321 

\bibitem{BGNV} M. Bordag, A. Goldhaber, P. van Nieuwenhuizen, D. Vassilevich, {\sl Heat kernel and zeta function regularization for the mass of the SUSY kink}, Phys. Rev. {\bf D66}(2002) 125014

\bibitem{AIMGTM0}  A. Alonso-Izquierdo,  W. Garcia Fuertes,J. Mateos Guilarte, M. de la Torre Mayado, {\sl Quantum corrections to the mass of self-dual vortices}, Phys. Rev. {\bf D70}061702(R) (2004)

\bibitem{AIGFMGTM} A. Alonso-Izquierdo, W. Garcia Fuertes, J. Mateos Guilarte, M. de la Torre Mayado, {\sl Quantum oscillations of self-dual Abrikosov-Nielsen-Olesen vortices}, Phys. Rev. {\bf D71} (2005) 125010

\bibitem{AIMGTM} A. Alonso-Izquierdo, J. Mateos Guilarte, M. de la Torre Mayado, {\sl Quantum magnetic flux lines, BPS vortex zero modes, and one-loop string tension shifts}, Phys. Rev.{\bf D94} (2016) 045008

\bibitem{JRo} R. Jackiw, P. Rossi, {\sl Zero modes of the vortex-fermion system}, Nucl. Phys. {\bf B190} (1981) 681

\bibitem{AFGV} C. Almeida, A. Alonso-Izquierdo, R. Fresneda, J. Mateos Guilarte, D. Vassilevich, {\sl Non-topological fractional fermion number in the Jackiw-Rossi model}, Phys. Rev. {\bf D103}, 125015 (2021)

\bibitem{DVass} D. Vassilevich, {\sl Quantum corrections to the mass of the supersymmetric vortex}, Phys. Rev. {\bf D68} (2003) 045005

\bibitem{AIJMG1} A. Alonso-Izquierdo, W. Garcia Fuertes, J. Mateos Guilarte, {\sl A note on BPS vortex bound states}, Phys. Lett. {B753} (2019) 29

\bibitem{AIJMG2} A. Alonso-Izquierdo, W. Garcia Fuertes, J. Mateos Guilarte, {\sl Dissecting zero modes and bound states on BPS vortices in Ginzburg-Landau superconductors}, JHEP05(2016) 074

\bibitem{AGMG} A. Alonso-Izquierdo, W. Garcia Fuertes, N. S. Manton, J. Mateos Guilarte, {\sl Spectral flow of vortex shape modes over the BPS 2-vortex moduli  space}, JHEP01(2024)020

\bibitem{AMGW} A. Alonso-Izquierdo, N. S. Manton, J. Mateos Guilarte, A. Wereszczynski, {\sl Collective coordinate models for 2-vortexshape mode dynamics} Phys. Rev.{\bf D110} (2024) 085006

\bibitem{Rajaraman} R. Rajaraman, {\sl Solitons and Instantons, An Introduction to Solitons and Instantons in Quantum Field Theory}, North Holland , 1982

\end{thebibliography}
\end{document}